\documentclass[
    aps,
    prx,
    reprint,
    amsmath,
    amssymb,
    superscriptaddress,
    longbibliography
]{revtex4-2}

\usepackage[T1]{fontenc}
\usepackage{libertinus}
\usepackage[libertine]{newtxmath}
\usepackage{microtype}

\usepackage{mathtools}
\usepackage{siunitx}

\usepackage{graphicx}
\usepackage{booktabs}
\usepackage{dcolumn}
\usepackage{subcaption}

\usepackage{bm}
\usepackage{braket}

\usepackage{xcolor}

\usepackage[colorlinks=true,linkcolor=blue,citecolor=blue,urlcolor=blue]{hyperref}

\usepackage{tikz}
\usetikzlibrary{calc,arrows.meta,positioning,backgrounds}

\usepackage[font=small,labelfont=bf]{caption}
\AtBeginDocument{%
  \setlength{\abovedisplayskip}{6pt plus 2pt minus 2pt}%
  \setlength{\belowdisplayskip}{6pt plus 2pt minus 2pt}%
  \setlength{\abovedisplayshortskip}{4pt plus 2pt minus 2pt}%
  \setlength{\belowdisplayshortskip}{4pt plus 2pt minus 2pt}%
}

\definecolor{pqcbox}{RGB}{225,205,225}
\definecolor{pqcoutline}{RGB}{160,120,160}
\definecolor{bellblob}{RGB}{210,230,245}
\definecolor{classwire}{RGB}{128,0,128}
\definecolor{boxgray}{RGB}{235,235,235}
\definecolor{advblue}{RGB}{120,140,170}
\definecolor{textdark}{RGB}{30,30,30}
\definecolor{textpurple}{RGB}{92,48,108}

\usepackage{xcolor}

\definecolor{NitishColor}{RGB}{0,0,200} 

\begin{document}

\title{Quantum-Resistant Quantum Teleportation}

\author{Xin Jin}
\altaffiliation{These two authors contributed equally to this work.}
\affiliation{Department of Computer Science, University of Pittsburgh, Pittsburgh, Pennsylvania 15260, USA}

\author{Nitish Kumar Chandra}
\altaffiliation{These two authors contributed equally to this work.}
\affiliation{Department of Informatics and Networked Systems, University of Pittsburgh, Pittsburgh, Pennsylvania 15260, USA}

\author{Mohadeseh Azari}
\affiliation{Department of Informatics and Networked Systems, University of Pittsburgh, Pittsburgh, Pennsylvania 15260, USA}

\author{Jinglei Cheng}
\affiliation{Department of Computer Science, University of Pittsburgh, Pittsburgh, Pennsylvania 15260, USA}

\author{Zilin Shen}
\affiliation{Purdue University, West Lafayette, Indiana 47907, USA}

\author{Kaushik P. Seshadreesan}
\affiliation{Department of Informatics and Networked Systems, University of Pittsburgh, Pittsburgh, Pennsylvania 15260, USA}

\author{Junyu Liu}
\email{junyuliu@pitt.edu}
\affiliation{Department of Computer Science, University of Pittsburgh, Pittsburgh, Pennsylvania 15260, USA}

\date{\today}

\begin{abstract}
Quantum teleportation requires classical control bits transmitted via a classical communication channel, which are vulnerable to quantum adversaries. We propose a quantum-resistant quantum teleportation (QRQT) framework protected by post-quantum cryptography (PQC), with a comprehensive security analysis. By applying PQC to protect the classical control bits, QRQT renders quantum teleportation fully quantum-resistant and eliminates the classical attack surface. Our analysis reveals that quantum memory is a hidden bottleneck linking physical and computational security. Its finite coherence time simultaneously limits the communication distance, constrains the tolerable PQC overhead, and restricts the adversary’s attack window.
Under realistic parameters (1~ms coherence, fiber-optic propagation), the maximum secure teleportation distance ranges from $\sim$191~km (FrodoKEM-1344) to $\sim$199~km (Kyber512), quantifying the security-distance tradeoff imposed by the selected PQC scheme. We show a joint classical--quantum attack probability $P_{\mathrm{joint}}(t)$ exhibits a non-monotonic, Bell-shaped profile due to the opposing time dependencies of classical cryptanalysis and quantum decoherence, thereby establishing a bounded optimal attack window after which the probability of adversarial success decays exponentially.
Complementing the attack probability analysis we further analyze how leakage of the classical correction bits affects the security of quantum teleportation, where confidentiality relies on keeping these bits secret until the receiver completes the recovery. To capture realistic patterns of time dependent exposure, we consider four stochastic leakage scenarios: independent exponential, sequential, burst, and correlated leakage, when the eavesdropper manages to compromise the receiver’s quantum system. We also account for decoherence in the entangled resource by modeling amplitude damping on the receiver's half of the shared Bell pair.  For each scenario, we derive closed form expressions for the average Holevo quantity and the average teleportation fidelity as functions of time. Based on the Holevo quantity, our information theoretic analysis provides measurement independent upper bounds on the information extractable by an adversary, which can help assess leakage risks in quantum communication and assist in developing protocols that are robust against such attacks.

\end{abstract}

\maketitle

\section{Introduction}
\label{sec:introduction}


Quantum teleportation is a core communication primitive for quantum networks, enabling quantum states to be transferred, coordinated, and processed across remote nodes~\cite{Dahlberg2019,Valivarthi2020}. Introduced by Bennett \emph{et al.} in 1993, teleportation enables the transfer of an unknown quantum state using shared entanglement and classical communication, without direct physical transmission of the quantum system itself~\cite{Bennett1993}. Teleportation fundamentally builds block for long distance entanglement distribution, quantum repeater protocols, and distributed quantum computing architectures~\cite{Kimble2008, wehner2018quantum, RevModPhys.83.33, PhysRevA.59.4249, Van2014quantum,Hermans2022}. As quantum networking advances from isolated demonstrations to scalable systems, teleportation should be viewed not merely as a canonical protocol of quantum information theory, but as an operational primitive whose security is essential to quantum network infrastructure~\cite{Gangopadhyay2022,Tserkis2020,yam2025practicalquantumteleportationfiniteenergy,Chandra2026}.

While shared entanglement supplies the quantum resource for teleportation, state transfer is completed only when the associated classical information reaches the receiver. Alice (sender) performs a joint measurement on the unknown input qubit and her share of an entangled pair, while Bob (receiver) holds the other share. This measurement produces two classical outcomes that must be communicated to Bob~\cite{Nielsen2012}. Bob then uses this information to determine the corresponding Pauli correction to recover the original state. The classical channel therefore carries the measurement outcomes that determine the recovery operation at the receiver. Any delay, corruption, or compromise of this classical information directly affects the completion of the protocol, making the correction channel a critical part of both the functionality and the security of quantum teleportation~\cite{Parakh2022,Ma2012}.

This reliance on classical information introduces a fundamental security asymmetry: while the quantum correlations are established without transmitting information, its successful and trustworthy completion also depends on a classical channel that can be vulnerable ~\cite{Cane2020,Liu2024}. In practice, the entangled quantum channel is often treated as the main object of protection, whereas the classical correction channel is commonly assumed to be authenticated or secured using conventional public key cryptography~\cite{Renner2023}. Such assumptions become problematic in the presence of quantum capable adversaries, because widely used public key schemes such as Rivest-Shamir-Adleman (RSA)~\cite{10.1145/359340.359342} and the Digital Signature Algorithm (DSA)~\cite{1994} derive their security from integer factorization and the discrete logarithm problem, both are vulnerable to Shor’s algorithm~\cite{Shor1999}. The classical correction channel thus becomes a natural attack surface: it carries the information required to complete state recovery, its protection is commonly delegated to public-key schemes that are broken by Shor's algorithm. Protecting only the quantum states is therefore insufficient, securing teleportation end-to-end requires quantum-resistant authentication of the classical correction path as well~\cite{jin2025quantumresistantnetworksusingpostquantum}.

Several existing approaches can be used to authenticate or secure the classical correction channel, but they can be operationally cumbersome or infrastructure intensive in scalable teleportation networks. Wegman-Carter authentication~\cite{Wegman1981} provides information theoretic message authentication, it requires pre-shared secret keys whose distribution and periodic refresh become increasingly burdensome as the number of nodes and teleportation rounds grows. Quantum key distribution (QKD), a method for establishing shared secret keys using quantum states, offers another possibility for securing the classical channel~\cite{Bennett2014}. However, it still relies on an initially authenticated classical channel, and in practice it requires each protected link to be paired with dedicated QKD hardware and supporting infrastructure~\cite{Scarani2009,Portmann2022}. Quantum digital signatures provide an alternative approach~\cite{Amiri2016}, but their practical deployment remains constrained by limitations in transmission distance, signature overhead, and multi-party scalability. As a result, a practical gap remains between solutions that provide strong security but require substantial infrastructure and coordination overhead, and those that would offer the scalable public-key functionality needed for large-scale quantum communication systems~\cite{Aquina2025,Baseri2025}.


Post quantum cryptography (PQC) provides a practical response to the problem of securing the classical correction channel while preserving the deployment advantages of public key cryptography. Since launching its PQC standardization effort in 2016, NIST has advanced PQC from a research direction to a deployable standards framework~\cite{Chen2016}. In 2022, NIST selected CRYSTALS Kyber for key encapsulation and CRYSTALS Dilithium and SPHINCS+ among its first post quantum digital signature candidates, and in 2024 finalized FIPS 203, FIPS 204, and FIPS 205, which specify algorithms derived from these schemes~\cite{Bos2018,Ducas2018}. Kyber and Dilithium are lattice based constructions built from module lattice assumptions related to Learning With Errors (LWE) and Short Integer Solution, while SPHINCS+ provides algorithmic diversity through a stateless hash based design~\cite{Bernstein2019,Moody2020}. This progress makes it timely to revisit teleportation security from a post quantum perspective, particularly for the classical correction channel, where public key protection is especially well suited to networked settings because it can secure protocol critical classical information without requiring pairwise pre-shared secret keys or dedicated quantum security infrastructure~\cite{Wang2021,Scarani2009}, while still relying on standard public key management and authentication mechanisms.

Building on recent progress in post quantum cryptography, we propose the \textit{Quantum-Resistant Quantum Teleportation (QRQT)} framework, which applies PQC to the classical control channel of quantum teleportation (See Fig.~\ref{fig:qrqt}). QRQT leaves the underlying teleportation protocol and entanglement resource unchanged, but protects the Bell measurement outcomes that must be delivered to the receiver for state reconstruction. It therefore secures the protocol critical classical correction path, ensuring that the information required for Pauli correction remains protected against quantum enabled adversaries.

Integrating PQC into teleportation is not a purely cryptographic substitution. The use of PQC-protected classical communication introduces additional processing and transmission delay before Bob can obtain the measurement outcomes and apply the required Pauli correction. During this interval, the receiver's qubit must remain stored in quantum memory, so the success of the protocol is also constrained by coherence time~\cite{BarGill2013,Zhong2015,Heshami2016,Hucul2014}. The PQC latency therefore translates into a physical resource constraint: the longer the classical correction path takes to complete, the greater the risk that decoherence will destroy the stored quantum state before recovery is possible. Computational security on the classical channel thus becomes directly coupled to physical reliability on the quantum side.

We consider an adversary whose goal is to intercept and recover the transmitted quantum state, requiring simultaneous access to both the classical correction outcomes and the stored quantum state. This creates a time-dependent joint constraint on the adversary: a longer transmission window provides more time to break the PQC-authenticated classical channel, but also increases the probability that the quantum state has decohered and is no longer recoverable. Conversely, a shorter window preserves the quantum state but limits the time available for classical cryptanalysis. Successful interception therefore requires the adversary to compromise both channels within the same coherence window.

To assess QRQT security comprehensively, we develop a hybrid threat model that jointly captures adversarial behavior on both the classical and quantum channels. On the classical side, we model attack success probabilities against LWE-based cryptosystems underlying standardized post-quantum encryption schemes. On the quantum side, we consider an active interception strategy in which an adversary attempts to substitute, extract, or temporarily retain entangled qubits intended for the receiver through a SWAP-type interaction; related interception and entanglement-tampering scenarios have been studied in the contexts of unreliable entanglement assistance, malicious entanglement in quantum networks, and active eavesdropping on quantum communication channels~\cite{Genovese2001,Lederman2024,Shaban2024}. This formulation enables us to quantify attack success under realistic temporal constraints, thereby capturing the interplay between PQC-induced latency and decoherence in finite quantum memory.

To analyze this coupled security problem, we use two complementary frameworks. The first is an explicit attack analysis that captures adversarial behavior on both the classical and quantum channels, allowing us to quantify attack success probabilities under finite computation time, communication latency, and quantum memory constraints. The second component is an information theoretic analysis based on the Holevo quantity, which quantifies the information an adversary can extract from partial correction data together with access to Bob's quantum state. Together, the two components allow us to distinguish between concrete protocol compromise and residual information exposure, thereby providing a more complete characterization of QRQT security.



\begin{figure*}[htbp] 
    \centering
    
    \begin{subfigure}[b]{0.48\textwidth}
        \centering
        \includegraphics[width=\textwidth]{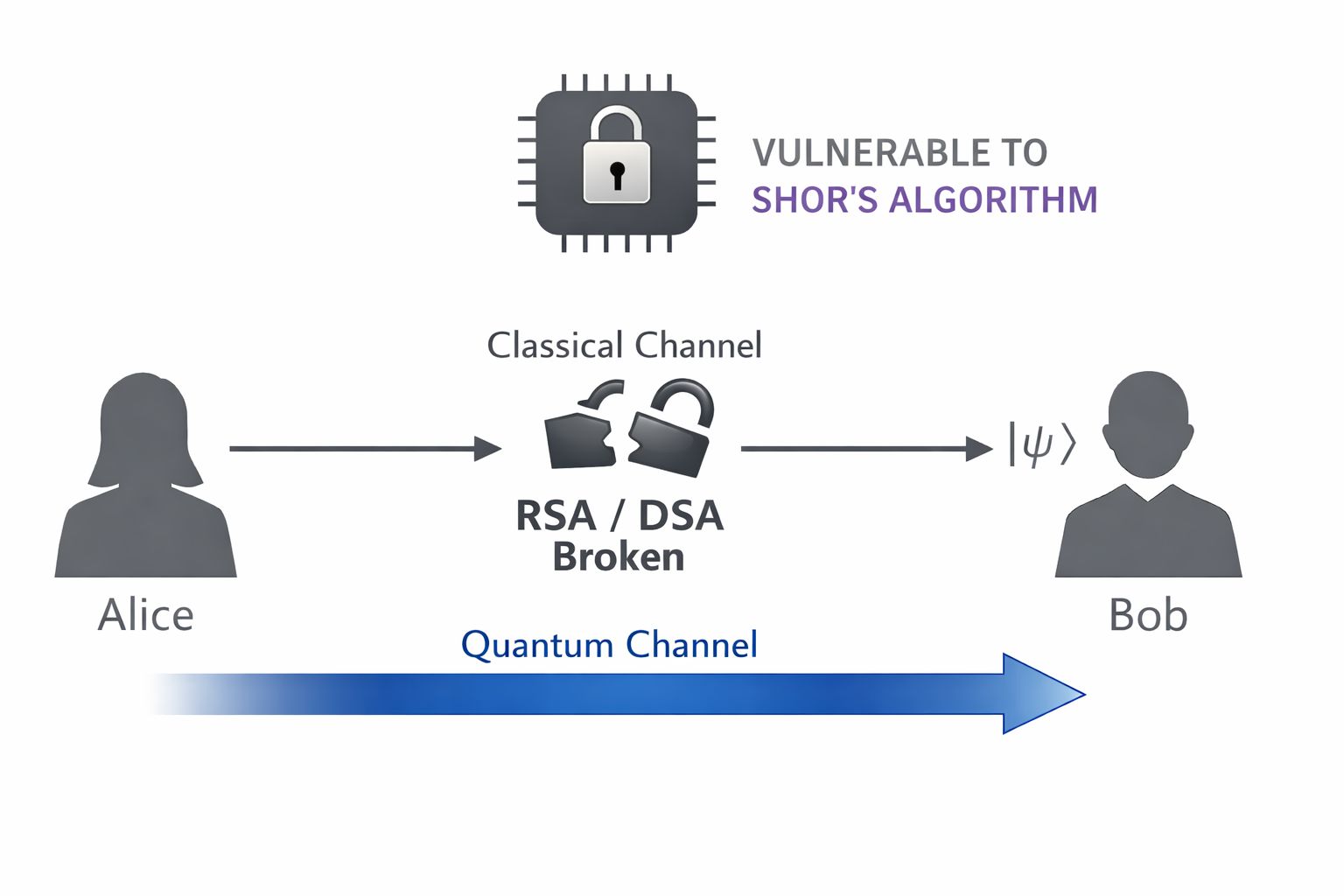} 
       \caption{Traditional Quantum Teleportation}
        \label{fig:top_left}
    \end{subfigure}
    \hfill 
    \begin{subfigure}[b]{0.48\textwidth}
        \centering
        \includegraphics[width=\textwidth]{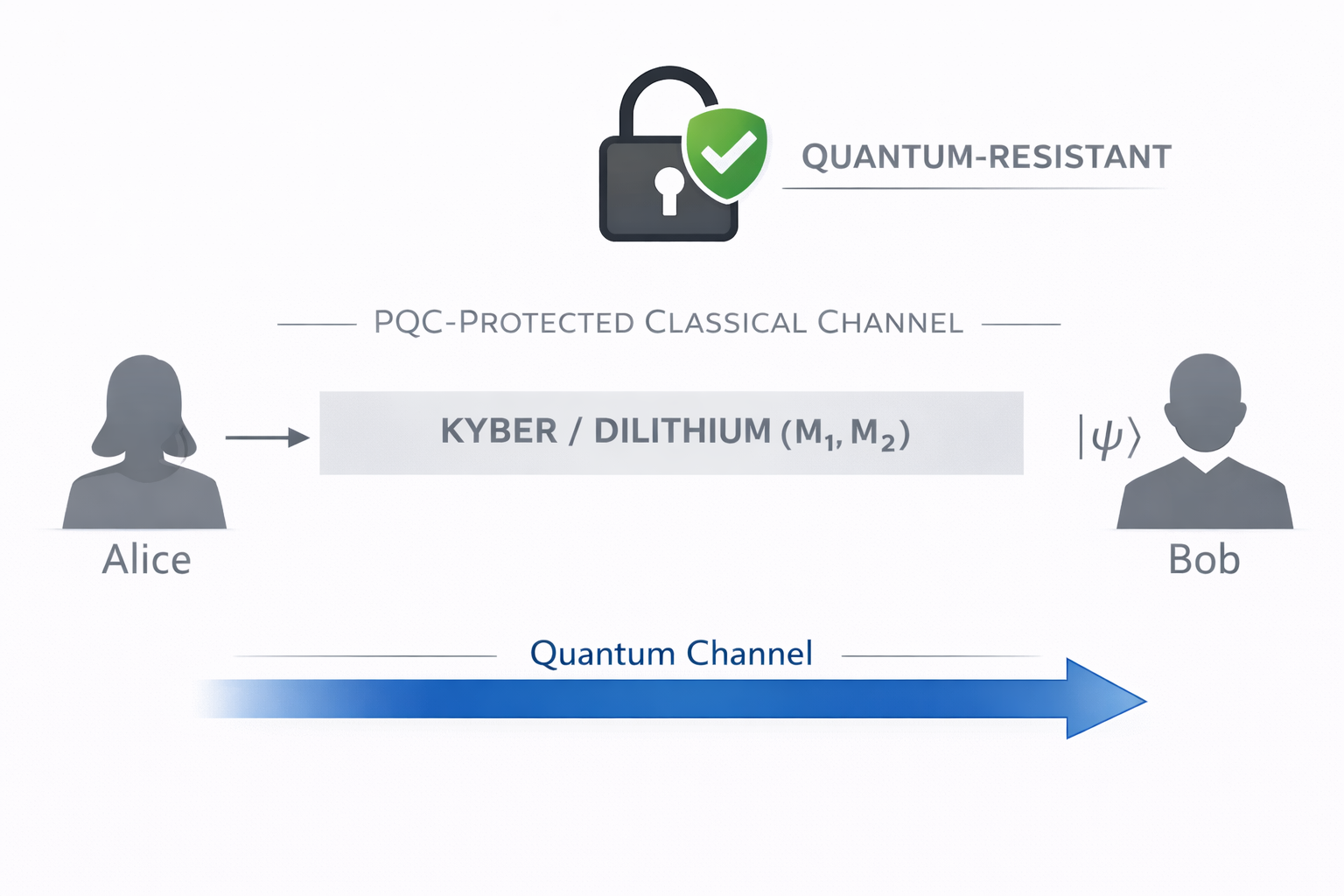}
        \caption{QRQT Framework}
        \label{fig:top_right}
    \end{subfigure}

    \vspace{1em} 

    \begin{subfigure}[b]{0.48\textwidth}
        \centering
        \includegraphics[width=\textwidth]{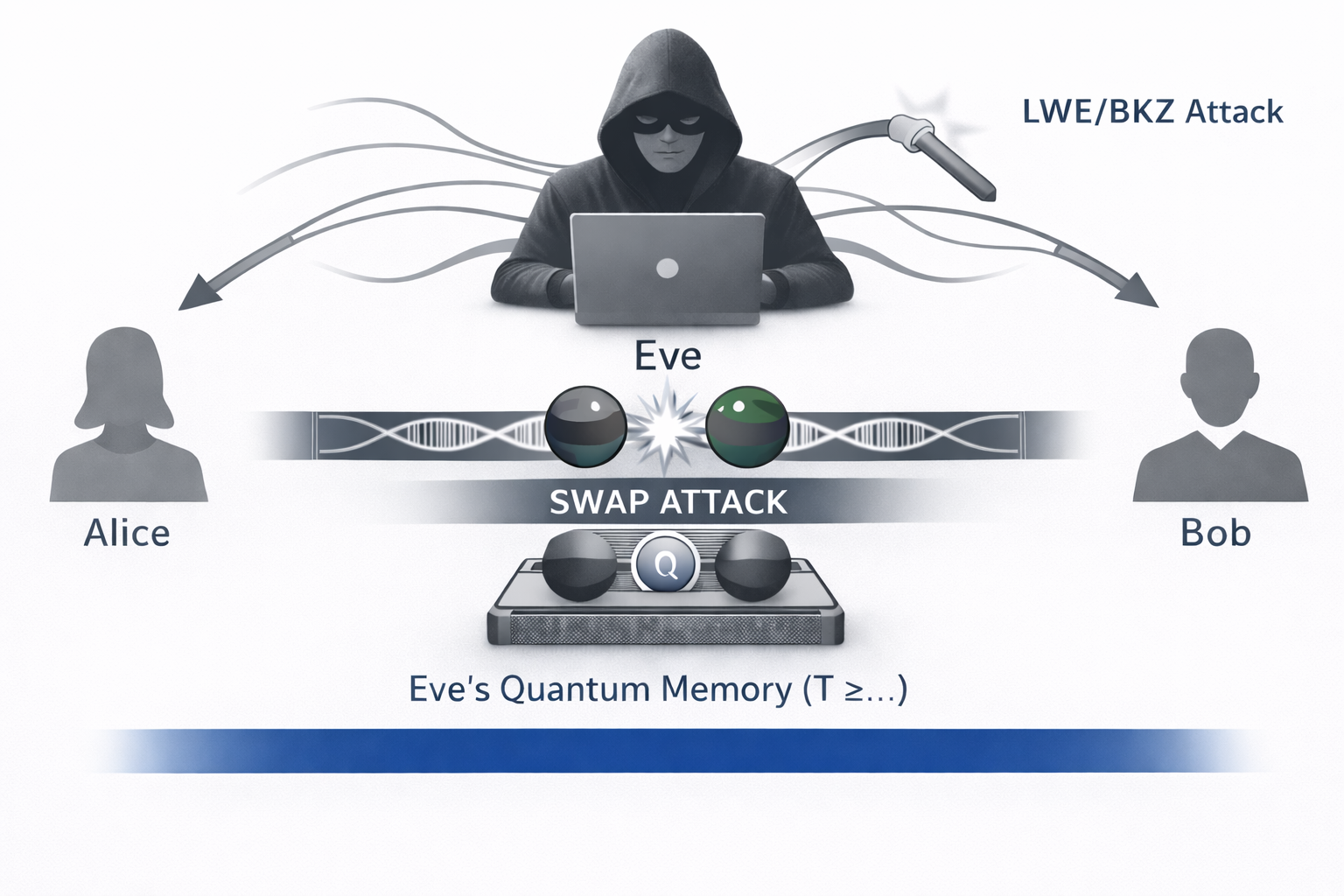}
        \caption{Joint Threat Model}
        \label{fig:bottom_left}
    \end{subfigure}
    \hfill
    \begin{subfigure}[b]{0.48\textwidth}
        \centering
        \includegraphics[width=\textwidth]{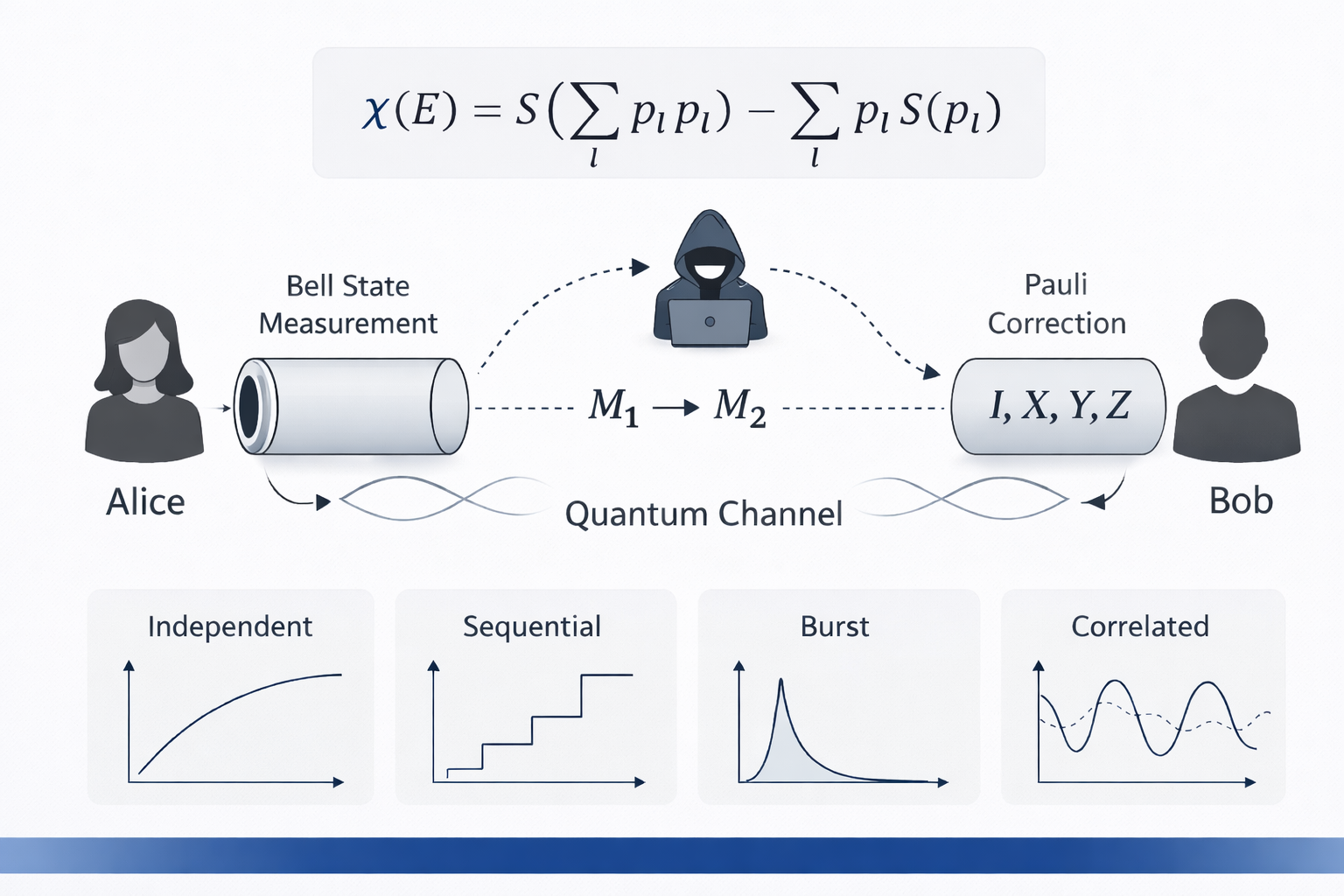}
        \caption{Information Theoretic Analysis}
        \label{fig:bottom_right}
    \end{subfigure}
    
    \caption{Overview of the QRQT framework. (a)~Standard teleportation with classical public-key protection, vulnerable to Shor's algorithm. (b)~QRQT replaces the classical link with PQC; Bob holds the qubit in quantum memory during PQC processing. (c)~Hybrid threat model combining lattice attacks on the classical channel and SWAP interception on the quantum channel. (d)~Holevo-bound analysis of information leakage under decoherence and classical key compromise.}
    \label{fig:overview}
\end{figure*}

The main contributions of this work are as follows:
\begin{itemize}

\item We propose the \textit{Quantum-Resistant Quantum Teleportation (QRQT)} framework (See Fig.~\ref{fig:overview}), which protects the classical correction channel of quantum teleportation using post-quantum cryptography, and identify the resulting joint security tradeoff between cryptographic protection and quantum memory constraints. Specifically, because the receiver must store the entangled qubit while awaiting PQC-protected correction bits, communication distance, tolerable cryptographic overhead, and the adversary's attack window become fundamentally coupled. This contribution highlights that the classical attack success probability grows with computation time, whereas the quantum interception probability decays as the entangled state decoheres, giving rise to a joint security landscape that cannot be captured by analyzing either channel in isolation.

\item If decoherence occurs before the correction bits are received, the original quantum state cannot be faithfully reconstructed, undermining both fidelity and practical feasibility. The interaction between cryptographic latency and quantum coherence remains largely unexplored, raising a fundamental question: under what temporal and physical conditions can QRQT remain both secure and operational? Motivated by this constraint, we quantitatively analyze the feasibility of QRQT by evaluating multiple standardized PQC algorithms and estimating the minimum coherence time required for successful state reconstruction across varying communication distances and security parameters.



    \item We formulate a hybrid threat model that jointly captures lattice-based attacks on the classical channel and SWAP-based interception on the quantum channel, and characterize the resulting time-dependent joint attack probability.
\item
We further present an information theoretic analysis based on the Holevo bound and introduce four physically motivated stochastic leakage models, namely independent, sequential, burst, and correlated leakage, that capture common qualitative patterns of information exposure arising in statistical failure and communication error processes. This framework enables us to quantify the residual information available to an adversary and to characterize how partial, delayed, and correlated leakage lead to time dependent degradation of the teleported quantum state.


\end{itemize}

\definecolor{textdark}{RGB}{48,54,60}
\definecolor{boxgray}{RGB}{238,240,242}

\definecolor{bellfill}{RGB}{220,231,223}
\definecolor{bellborder}{RGB}{122,145,130}

\definecolor{pqcfill}{RGB}{226,232,244}
\definecolor{pqcborder}{RGB}{96,118,156}
\definecolor{pqctext}{RGB}{78,100,138}

\definecolor{advcol}{RGB}{91,128,134}

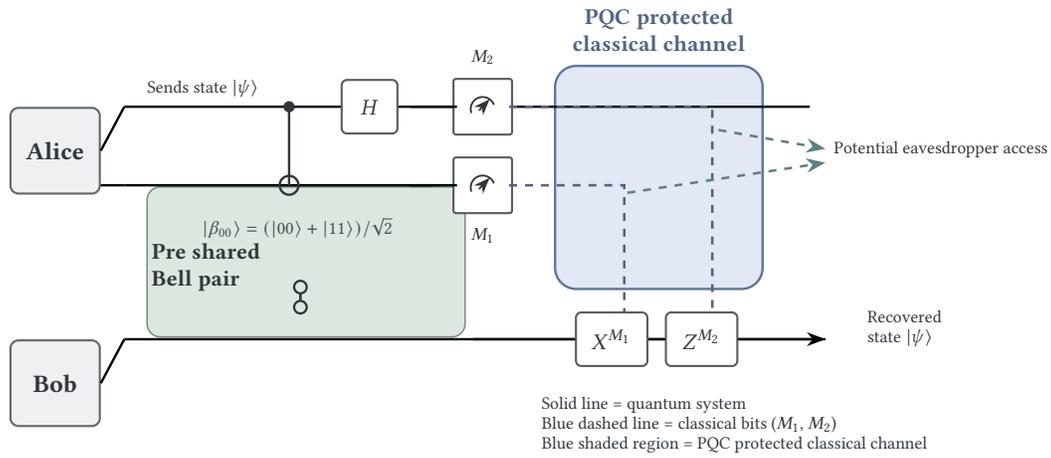
\begin{figure*}[htbp]
\centering

\begin{tikzpicture}[
    x=0.88cm,
    y=0.88cm,
    scale=0.95,
    transform shape,
    >=Stealth,
    thick,
    userbox/.style={
        draw=textdark!75,
        rounded corners=3pt,
        fill=boxgray,
        minimum width=1.25cm,
        minimum height=1.20cm,
        align=center
    },
    gate/.style={
        draw=textdark!80,
        rounded corners=2pt,
        minimum width=0.90cm,
        minimum height=0.75cm,
        fill=white,
        inner sep=1.2pt
    },
    meas/.style={
        draw=textdark!80,
        rounded corners=1pt,
        minimum width=0.82cm,
        minimum height=0.78cm,
        fill=white
    },
    classpath/.style={
        draw=pqctext,
        dashed,
        line width=0.95pt,
        ->,
        >=Stealth
    },
    advpath/.style={
        draw=advcol,
        dashed,
        line width=0.95pt,
        ->,
        >=Stealth
    }
]

\node[userbox, text=textdark, font=\bfseries\normalsize] (alice) at (0,1.75) {Alice};
\node[userbox, text=textdark, font=\bfseries\normalsize] (bob)   at (0,-1.95) {Bob};

\draw[black, line width=0.9pt] (1.10,2.45) -- (12.00,2.45);
\draw[black, line width=0.9pt] (1.10,1.20) -- (6.00,1.20);
\draw[black, line width=0.9pt] (1.10,-1.25) -- (12.25,-1.25);

\node[above, text=textdark, font=\scriptsize] at (2.35,2.47) {Sends state $|\psi\rangle$};

\draw[black, line width=0.9pt] (alice.east) -- (1.10,2.45);
\draw[black, line width=0.9pt] (alice.east |- 1.10,1.20) -- (1.10,1.20);
\draw[black, line width=0.9pt] (bob.east) -- (1.10,-1.25);

\begin{scope}[on background layer]
\node[
    rectangle,
    rounded corners=5pt,
    fill=bellfill,
    draw=bellborder,
    line width=0.6pt,
    minimum width=4.45cm,
    minimum height=2.10cm,
    anchor=west,
    inner sep=0pt
] (bell) at (1.45,-0.02) {};
\end{scope}

\node[align=left, text=textdark, font=\bfseries\small] at (2.40,-0.10) {Pre shared\\Bell pair};
\node[text=textdark!90, font=\scriptsize] at (3.85,0.52) {$|\beta_{00}\rangle=(|00\rangle+|11\rangle)/\sqrt{2}$};

\draw[line width=0.85pt, draw=textdark] (3.90,-0.42) circle (0.10);
\draw[line width=0.85pt, draw=textdark] (3.90,-0.74) circle (0.10);
\draw[line width=0.95pt, draw=textdark] (3.90,-0.52) -- (3.90,-0.64);

\node[gate, minimum width=0.78cm, text=textdark, font=\small] (H) at (5.00,2.45) {$H$};

\filldraw[textdark] (3.72,2.45) circle (1.8pt);
\draw[textdark, line width=0.9pt] (3.72,2.45) -- (3.72,1.20);
\draw[textdark, line width=0.9pt] (3.72,1.20) circle (0.15);

\node[meas] (M2box) at (6.80,2.45) {};
\node[meas] (M1box) at (6.80,1.20) {};

\draw[line width=0.7pt, draw=textdark] (6.62,2.35) arc[start angle=200,end angle=20,radius=0.19];
\draw[->, line width=0.7pt, draw=textdark] (6.74,2.39) -- (6.91,2.57);

\draw[line width=0.7pt, draw=textdark] (6.62,1.10) arc[start angle=200,end angle=20,radius=0.19];
\draw[->, line width=0.7pt, draw=textdark] (6.74,1.14) -- (6.91,1.32);

\node[above=2pt, text=textdark, font=\scriptsize] at (M2box.north) {$M_2$};
\node[below=2pt, text=textdark, font=\scriptsize] at (M1box.south) {$M_1$};

\draw[black, line width=0.9pt] (5.65,2.45) -- (6.39,2.45);
\draw[black, line width=0.9pt] (5.65,1.20) -- (6.39,1.20);

\begin{scope}[on background layer]
\path[
    fill=pqcfill,
    draw=pqcborder,
    line width=0.9pt,
    rounded corners=8pt
] (7.95,3.10) rectangle (11.25,-0.45);
\end{scope}

\node[align=center, text=pqctext, font=\bfseries\small] at (9.60,3.68) {PQC protected\\classical channel};

\draw[classpath] (7.22,2.45) -- (10.45,2.45) -- (10.45,-1.25);
\draw[classpath] (7.22,1.20) -- (9.05,1.20) -- (9.05,-1.25);

\node[gate, minimum width=1.00cm, text=textdark, font=\small] (Xcorr) at (8.85,-1.25) {$X^{M_1}$};
\node[gate, minimum width=1.00cm, text=textdark, font=\small] (Zcorr) at (10.28,-1.25) {$Z^{M_2}$};

\draw[black, line width=0.9pt] (6.65,-1.25) -- (Xcorr.west);
\draw[black, line width=0.9pt] (Xcorr.east) -- (Zcorr.west);
\draw[->, black, line width=0.9pt] (Zcorr.east) -- (12.25,-1.25);

\node[align=left, anchor=west, text=textdark, font=\scriptsize] at (12.78,-1.08) {Recovered\\state $|\psi\rangle$};

\node[right, text=textdark, font=\scriptsize] (advtext) at (12.25,1.78) {Potential eavesdropper access};
\draw[advpath] (10.55,2.08) -- (12.25,1.78);
\draw[advpath] (9.10,1.08) -- (12.25,1.56);

\node[align=left, anchor=north west, text=textdark, font=\scriptsize] at (7.60,-2.02) {
Solid line = quantum system\\
Blue dashed line = classical bits ($M_1$, $M_2$)\\
Blue shaded region = PQC protected classical channel
};

\end{tikzpicture}

\caption{\textbf{Quantum Resistant Quantum Teleportation (QRQT) framework.}
Alice performs a Bell state measurement on her input qubit $|\psi\rangle$ and her half of the Bell pair $|\beta_{00}\rangle=\tfrac{1}{\sqrt{2}}(|00\rangle+|11\rangle)$.
A \textsc{cnot} gate followed by a Hadamard gate and computational basis measurements yield two classical outcomes $M_1$ and $M_2$.
These control bits are sent to Bob through a \emph{PQC} protected classical channel, while Bob retains his entangled qubit in quantum memory.
Upon receiving the bits, Bob applies Pauli corrections $X^{M_1}$ and $Z^{M_2}$ to recover $|\psi\rangle$.}
\label{fig:qrqt}

\end{figure*}

The remainder of the paper is organized as follows. Section~\ref{sec:preliminaries} reviews the background theory relevant to this work. Sections~\ref{sec:qrqt} and~\ref{sec:joint-threat} present the QRQT framework, analyze memory lifetime limitations, and develop the hybrid threat model along with the corresponding attack success probability. Sections~\ref{sec:info-theoretic-security}, \ref{sec:holevo-damped}, and~\ref{stochastic} present an information theoretic analysis of the teleportation protocol under an amplitude damping noise model for different scenarios, with particular focus on the Holevo quantity, the fidelity of the quantum state, and their time dependent behavior under four stochastic models of classical leakage. Finally, Section~\ref{sec:conclusion} concludes the paper and discusses future directions for integrating PQC protected teleportation into scalable quantum networks.

\section{Preliminaries}
\label{sec:preliminaries}

 In this section, we review the fundamental concepts used throughout the paper. We first review the quantum teleportation protocol, examine how noise degrades quantum information through the amplitude damping channel, discuss the role of Pauli corrections as an encryption mechanism, and introduce the von Neumann entropy and the Holevo quantity used in the security analysis. We further discuss  the lattice-based cryptographic hardness assumption underlying post-quantum security and the teleportation fidelity metric.

\subsection{Quantum Teleportation Protocol}
\label{subsec:teleportation}

Quantum teleportation~\cite{Bennett1993,Nielsen2012} is a fundamental protocol in quantum information processing that enables the transfer of an arbitrary unknown quantum state between two spatially separated parties using shared entanglement and classical communication (see Fig.~\ref{fig:qrqt}). The protocol relies on two essential resources:
\begin{itemize}
    \item a pre-shared entangled state, typically a maximally entangled Bell pair, and
    \item a classical communication channel used to convey measurement outcomes.
\end{itemize}

Let the input state to be teleported be a single qubit
\[
\ket{\psi} = \alpha \ket{0} + \beta \ket{1},
\]
with \( |\alpha|^2 + |\beta|^2 = 1 \). The teleportation procedure proceeds as follows:
\begin{enumerate}
    \item Alice performs a joint Bell basis measurement on the input qubit and her half of the shared entangled pair, projecting the combined system onto one of four Bell states.
    \item The measurement outcome is encoded into two classical bits \((m_1,m_2)\in\{0,1\}^2\). (Throughout this paper, $m_1 \equiv M_1$ and $m_2 \equiv M_2$. These symbols denote the same classical bits and are used interchangeably in some sections for notational convenience and clarity.)
    \item Alice transmits these classical bits to Bob over a classical communication channel.
    \item Upon receiving the measurement outcomes, Bob applies the conditional Pauli operation
    \[
    U_{m_1m_2} = Z^{m_2}X^{m_1}
    \]
    to his qubit, thereby recovering the original state \(\ket{\psi}\).
\end{enumerate}

In the ideal setting, assuming perfect entanglement, noiseless quantum operations, and error free classical communication, the teleportation protocol reconstructs the input state at Bob's location with unit fidelity.

\subsection{Quantum Noise and the Amplitude Damping Channel}

In realistic settings, quantum systems inevitably interact with their surrounding environment, leading to decoherence and the loss of quantum information. Such noise processes are mathematically described by completely positive trace preserving maps, which are commonly represented using the Kraus operator formalism~\cite{Nielsen2012}.

A particularly relevant noise model for many physical qubit implementations is the amplitude damping channel, which captures energy relaxation mechanisms such as spontaneous emission in optical platforms or energy decay in superconducting qubits~\cite{Klimov2018,Khatri2020}. The amplitude damping channel acting on a single qubit is characterized by the Kraus operators
\begin{equation}
E_0 =
\begin{pmatrix}
1 & 0 \\
0 & \sqrt{1-\gamma}
\end{pmatrix},
\qquad
E_1 =
\begin{pmatrix}
0 & \sqrt{\gamma} \\
0 & 0
\end{pmatrix},
\end{equation}
where \(\gamma\in[0,1]\) denotes the probability of energy decay.

When a qubit with density matrix \(\rho\) undergoes amplitude damping, its evolution is described by
\begin{equation}
\mathcal{E}_{\mathrm{AD}}(\rho) = E_0 \rho E_0^\dagger + E_1 \rho E_1^\dagger.
\end{equation}

In this work, we use the amplitude damping channel as a representative model for such decoherence processes, which lead to degradation of the teleported quantum state.

\subsection{Pauli Corrections and Encryption of Quantum Information}

The Pauli operators \(\{I,X,Y,Z\}\) form a complete orthonormal basis for single qubit operators. In the quantum teleportation protocol, the specific Pauli correction applied by Bob is determined by the outcome of Alice's Bell state measurement. The required unitary operation takes the form
\begin{equation}
U_{M_1M_2} = Z^{M_2}X^{M_1},
\end{equation}
where \((M_1,M_2)\in\{0,1\}^2\) denote the classical bits communicated by Alice.

This conditional correction reverses the random Pauli operator induced by the Bell measurement and enables Bob to recover the original quantum state. Prior to receiving the classical information, however, Bob's qubit remains correlated with Alice's measurement outcome. When averaged over the unknown correction bits, Bob's reduced state is maximally mixed and therefore reveals no information about the input state.

From an information theoretic perspective, the absence of the correction bits effectively hides the teleported quantum information. Any observer without access to the classical outcomes perceives Bob's qubit as carrying no extractable information about the input state. In this sense, the teleportation protocol provides a form of physical layer encryption, with the classical correction bits acting as a decryption key~\cite{Boykin2003}. Consequently, partial or complete leakage of the correction bits undermines this protection and can allow an adversary to infer information about the original quantum state.

\subsection{Von Neumann Entropy and Quantum Ensembles}

The von Neumann entropy~\cite{Nielsen2012} is the standard measure of entropy for quantum states and generalizes classical Shannon entropy to the quantum setting. For a state described by a density matrix \(\rho\), it is defined as
\begin{equation}
S(\rho) = -\mathrm{Tr}\!\left(\rho \log_2 \rho\right),
\end{equation}
and quantifies the mixedness of the state.

In quantum information theory, an ensemble of states is represented as \(\mathcal{E}=\{p_i,\rho_i\}\), where \(\rho_i\) is prepared with probability \(p_i\). The corresponding average state is
\begin{equation}
\rho_{\mathrm{avg}}=\sum_i p_i \rho_i.
\end{equation}
The Holevo quantity is then given by
\begin{equation}
\chi(\mathcal{E})
=
S\!\left(\sum_i p_i \rho_i\right)
-
\sum_i p_i S(\rho_i), \label{eq:holevo-def}
\end{equation}
which upper bounds the accessible classical information that can be extracted from the ensemble by any measurement.

In the present setting, the Holevo quantity provides an information theoretic upper bound on how much Eve can learn from partial classical knowledge together with access to Bob's quantum system. In later sections, we use this framework to analyze residual information exposure under different stochastic leakage models and relate it to the information contained in the teleported quantum state.

\subsection{Lattice-Based Cryptography and the LWE Problem}
\label{subsec:lwe-prelim}

The classical channel protection employed in this work relies on post quantum cryptographic schemes whose security reduces to lattice problems, that is, computational problems defined over highly regular geometric point sets in high dimensional spaces that are believed to be hard to solve efficiently. The central hardness assumption is the \textit{Learning With Errors (LWE)} problem~\cite{Regev2009}. In this problem, one is given a public matrix $A\in\mathbb{Z}_q^{n\times m}$ and noisy samples of the form
\begin{equation}
    \mathbf{p} = A^{\top}\mathbf{sk} + \mathbf{e} \pmod{q},
\end{equation}
and the goal is to recover the hidden vector $\mathbf{sk}$.

More specifically, $D_{\mathbb{Z},\sigma_e q}$ is a secret vector that plays the role of the hidden key, $q$ is a modulus that fixes the arithmetic range, $m$ is the number of samples, and $\mathbf{e}\in D_{\mathbb{Z},\sigma_e q}$ is an error vector drawn from a discrete Gaussian distribution with standard deviation $\sigma_e\cdot q$. Here, the public matrix $A$ is known to everyone and provides the structured linear data from which the samples are formed, while the notation $\mathbb{Z}_q$ means that all arithmetic is performed modulo $q$, so values wrap around after reaching the modulus. The error vector $\mathbf{e}$ represents deliberately added random noise, with smaller errors more likely than larger ones.

Intuitively, LWE can be viewed as a noisy system of linear equations. Without the error term, the secret vector could in principle be recovered using standard linear algebra. The added noise, however, makes this recovery task computationally difficult, and this difficulty is what gives LWE based cryptographic schemes their security. Standardized post quantum schemes such as CRYSTALS Kyber and FrodoKEM derive their security from module LWE and plain LWE variants, respectively~\cite{Bos2018,Ducas2018}. Here, plain LWE refers to the original form of the problem, whereas module LWE introduces additional algebraic structure that improves efficiency while maintaining the same general hardness foundation.

The best known classical attacks against LWE proceed through BKZ (Block Korkine Zolotarev) lattice reduction~\cite{Chen2011,Albrecht2015}. In broad terms, lattice reduction attempts to replace a given lattice basis by another basis that spans the same lattice but consists of shorter and more nearly orthogonal vectors, since such a basis is more useful for cryptanalysis. A lattice basis $\mathbf{B}=(\mathbf{b}_1,\dots,\mathbf{b}_m)$ is simply a set of vectors whose integer linear combinations generate all points of the lattice. BKZ works on this basis and iteratively improves it.

More precisely, BKZ applies an exact shortest vector oracle within successive blocks of dimension $\beta$ to produce a reduced basis whose Gram Schmidt vectors $\tilde{\mathbf{b}}_i$ decrease in length at a predictable rate. The block dimension $\beta$ controls the strength of the reduction: a larger block size generally yields a better reduced basis, but at a significantly higher computational cost. The shortest vector oracle is an idealized subroutine that finds the shortest nonzero vector inside a given block, while the Gram Schmidt vectors are an orthogonalized version of the basis vectors that make the geometry of the reduced basis easier to analyze.

The quality of the reduced basis is commonly characterized by the root Hermite factor $\delta_{\mathrm{root}}$, defined so that the shortest reduced vector satisfies
\begin{equation}
\|\mathbf{b}_1\|\approx\delta_{\mathrm{root}}^{m}\,(\det\Lambda)^{1/m},
\end{equation}
where $\Lambda$ is the lattice and $m$ is its rank~\cite{Schnorr1994}. Here, $\det\Lambda$ is the determinant of the lattice, which measures the effective volume of a fundamental cell, and the rank $m$ is the number of basis vectors. Smaller values of $\delta_{\mathrm{root}}$ correspond to stronger reduction and therefore to more effective attacks.

To model the structure of a BKZ reduced basis analytically, one often adopts the \textit{Geometric Series Assumption (GSA)}, which is a standard heuristic description of how the orthogonalized basis vectors shrink under reduction. Under this assumption, the Gram Schmidt norms decay geometrically:
\begin{equation}
\|\tilde{\mathbf{b}}_i\| = \|\tilde{\mathbf{b}}_1\|\,\delta_{\mathrm{root}}^{-2(i-1)},
\label{eq:gsa_decay_prelim}
\end{equation}
which provides a convenient analytic approximation to the output of BKZ reduction~\cite{Schnorr1994}. In other words, the GSA assumes that the orthogonalized vectors decrease in a regular geometric pattern, making the behavior of BKZ easier to model mathematically.

The computational cost of BKZ is commonly parameterized by $\delta_{\mathrm{root}}$ through the empirical relation
\begin{equation}
\log_2 T_{\mathrm{BKZ}} = \frac{a}{\log_2 \delta_{\mathrm{root}}} - b,
\label{eq:bkz_time_prelim}
\end{equation}
where $a$ and $b$ are empirical scaling constants encoding the adversary's hardware throughput and algorithmic efficiency~\cite{Albrecht2015}. This relation captures the practical tradeoff between attack quality and runtime: achieving a smaller root Hermite factor, and hence a stronger reduction, generally requires substantially more computation.

Once lattice reduction has been carried out, the secret is recovered by Nearest Planes enumeration~\cite{Lindner2011}. At each Gram Schmidt layer $i$, the algorithm searches over integer coefficients within a radius $d_i$ and succeeds with probability governed by
\begin{equation}
\operatorname{erf}\bigl(d_i\|\tilde{\mathbf{b}}_i\|\sqrt{\pi}/(2s)\bigr),
\end{equation}
where $s$ is the LWE noise standard deviation. Conceptually, this step attempts to identify the lattice point closest to the noisy target after the basis has been sufficiently reduced. The search radius $d_i$ controls how far the algorithm explores at each layer, and the error function appears because the success probability is determined by the Gaussian statistics of the noise.

Thus, the overall attack has two main ingredients: lattice reduction, which transforms the basis into a more favorable geometric form, and a subsequent decoding step, which uses that reduced basis to recover the hidden secret. The success of this process depends both on the quality of the reduction and on the magnitude of the underlying noise. These definitions are used in Section~\ref{sec:PQC attack} to construct a time dependent classical attack model.

\section{QRQT Framework and Security Analysis under Quantum-Memory Constraints}
\label{sec:qrqt}
\textbf{QRQT Framework.} The proposed \textit{Quantum-Resistant Quantum Teleportation (QRQT)} framework extends the standard teleportation protocol by integrating post-quantum cryptography into the classical channel. As illustrated in Fig.~\ref{fig:qrqt}, the system consists of two parties, Alice and Bob, sharing an entangled Bell pair and communicating via a classical control link secured by a PQC primitive $\mathcal{E}_{\mathrm{PQC}}$. Let $|\psi\rangle$ denote the input state, $\{M_1, M_2\}$ the measurement outcomes. This section formalizes the QRQT operations and the corresponding timing constraints.

In QRQT, the teleportation process can be described as a quantum classical hybrid map:
\[
\mathcal{T}_{\text{QRQT}} =
\mathcal{R}_{\text{Bob}} \circ
\mathcal{E}_{\mathrm{PQC}} \circ
\mathcal{M}_{\text{Alice}} \circ
\mathcal{U}_{\text{Bell}} \circ
\mathcal{P}_{\text{Bell}},
\]
where $\mathcal{T}_{\text{QRQT}}$ represents the overall map from Alice's input quantum state to the reconstructed state on Bob's side under PQC secured teleportation. Specifically, $\mathcal{P}_{\text{Bell}}$ denotes the preparation of the shared Bell pair
$|\beta_{00}\rangle = \tfrac{1}{\sqrt{2}}(|00\rangle + |11\rangle)$,
$\mathcal{U}_{\text{Bell}}$ denotes the Bell basis transformation implemented by a \textsc{CNOT} gate and a Hadamard gate,
$\mathcal{M}_{\text{Alice}}$ denotes Alice's measurement, which produces the two classical bits $(M_1, M_2)$,
$\mathcal{E}_{\mathrm{PQC}}$ denotes the PQC based encryption and decryption of these bits, and
$\mathcal{R}_{\text{Bob}}$ denotes Bob's reconstruction by applying the Pauli corrections $Z^{M_2}X^{M_1}$.

For successful teleportation, Bob’s quantum memory must preserve coherence throughout the entire delay of the classical channel, including both cryptographic and transmission latencies:
\begin{equation}
\tau_m=T_{\mathcal{P}_{\mathrm{Bell}}}+T_{\mathcal{U}_{\mathrm{Bell}}}+T_{\mathcal{E}_{\mathrm{PQC}}} +T_{\mathcal{M}_{\mathrm{Alice}}}+ T_{\mathrm{comm}}+T_{\mathcal{R}_{\mathrm{Bob}}}
\end{equation}
where $\tau_m$ denotes the effective \textit{quantum memory lifetime limit}, representing the minimum coherence time required to preserve the stored qubit until the classical correction bits are received. Thus,

\begin{equation}
    T_{\mathrm{coh}} > \tau_m
    \label{time equation}.
\end{equation}

The individual timing components are: $T_{\mathcal{P}_{\mathrm{Bell}}}$, the Bell-pair preparation and distribution time; $T_{\mathcal{U}_{\mathrm{Bell}}}$, the Bell-basis transformation latency; $T_{\mathcal{M}_{\mathrm{Alice}}}$, Alice's measurement time; $T_{\mathcal{E}_{\mathrm{PQC}}}$, the PQC encryption and decryption delay; $T_{\mathrm{comm}} = d/v_{\mathrm{fiber}}$, the distance-dependent classical communication delay over optical fiber ($v_{\mathrm{fiber}} \approx 2 \times 10^5$~km/s); and $T_{\mathcal{R}_{\mathrm{Bob}}}$, the time for Bob's local Pauli corrections. Among these, $T_{\mathrm{comm}}$ dominates at long distances (${\approx}\,5\;\mu$s/km), while all local operations remain in the nanosecond range (see Appendix~\ref{appendix: Latency notification} for platform-specific estimates). 
$T_{\mathcal{E}_{\mathrm{PQC}}}$ captures the computational delay introduced by post-quantum encryption and decryption, while $T_{\mathrm{comm}}$ represents the classical communication time. The interplay between these classical-layer delays and quantum-layer decoherence defines the operational feasibility region of QRQT, as quantified below and formalized in Section~\ref{sec:joint-threat}. 

 In the QRQT protocol, Alice secures the two correction bits $(M_1,M_2)$ using a hybrid PQC scheme: a lattice-based key-encapsulation mechanism (e.g., CRYSTALS-Kyber) establishes a one-time session key~$K$, which is then used to symmetrically encrypt the correction bits via AES, yielding ciphertext $(C_1,C_2)$. Bob decapsulates~$C_1$, decrypts~$C_2$, and applies Pauli corrections $Z^{M_2}X^{M_1}$ to recover the teleported state. The full protocol specification is provided in Appendix~\ref{app:pqc protocol}.

\textbf{Memory lifetime limit analysis.} Having established $\tau_m$ as the lower bound on quantum memory coherence required for successful QRQT,
we now demonstrate how this constraint governs the operational feasibility of the QRQT framework. By quantifying this relationship across different PQC schemes and communication distances, we identify the minimal physical conditions (e.g., communication distance) under which QRQT remains both secure and realizable. The PQC computing time is estimated using the JEDI BullSequana XH3000 supercomputer ($R_{\mathrm{max}} = 4.50$~PFlop/s) benchmark~\cite{green500_2024_11}; the details of the estimation method are provided in Appendix~\ref{appendix:JEDI}. We assume spontaneous parametric down-conversion (SPDC) Bell-pair generation~\cite{Kwiat1995,Burnham1970} and fiber-optic delivery for the quantum teleportation experiment; the data used for estimating the feasibility boundary are given in Appendix~\ref{appendix:latency}.

\begin{figure*}[htb!]
\centering
\begin{subfigure}[t]{0.5\textwidth}
    \centering
    \includegraphics[width=\textwidth]{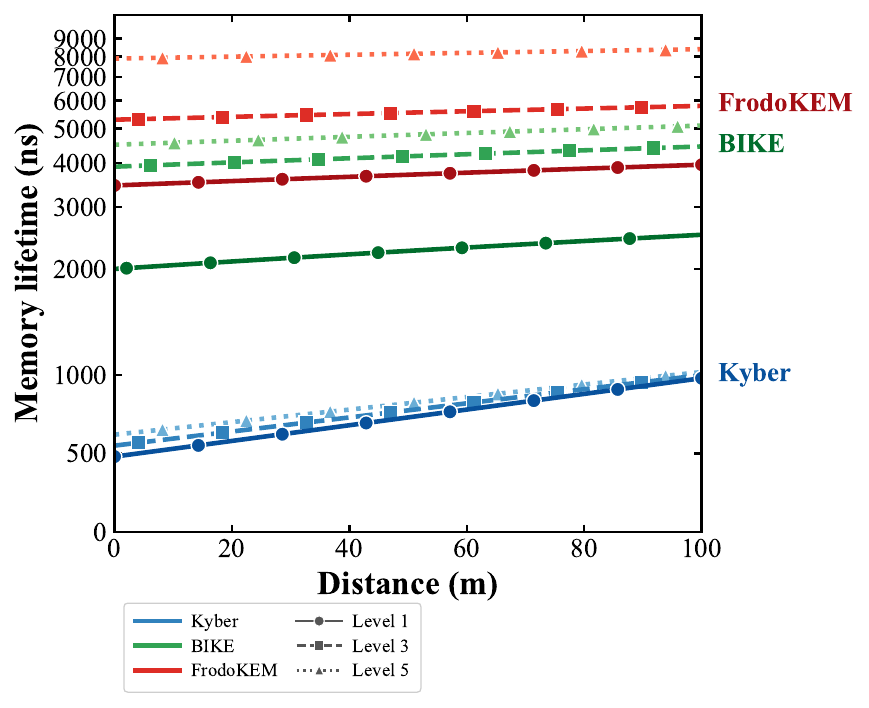}
    \caption{}
    \label{fig:memory_limit_a}
\end{subfigure}
\hfill
\begin{subfigure}[t]{0.48\textwidth}
    \centering
    \includegraphics[width=\textwidth]{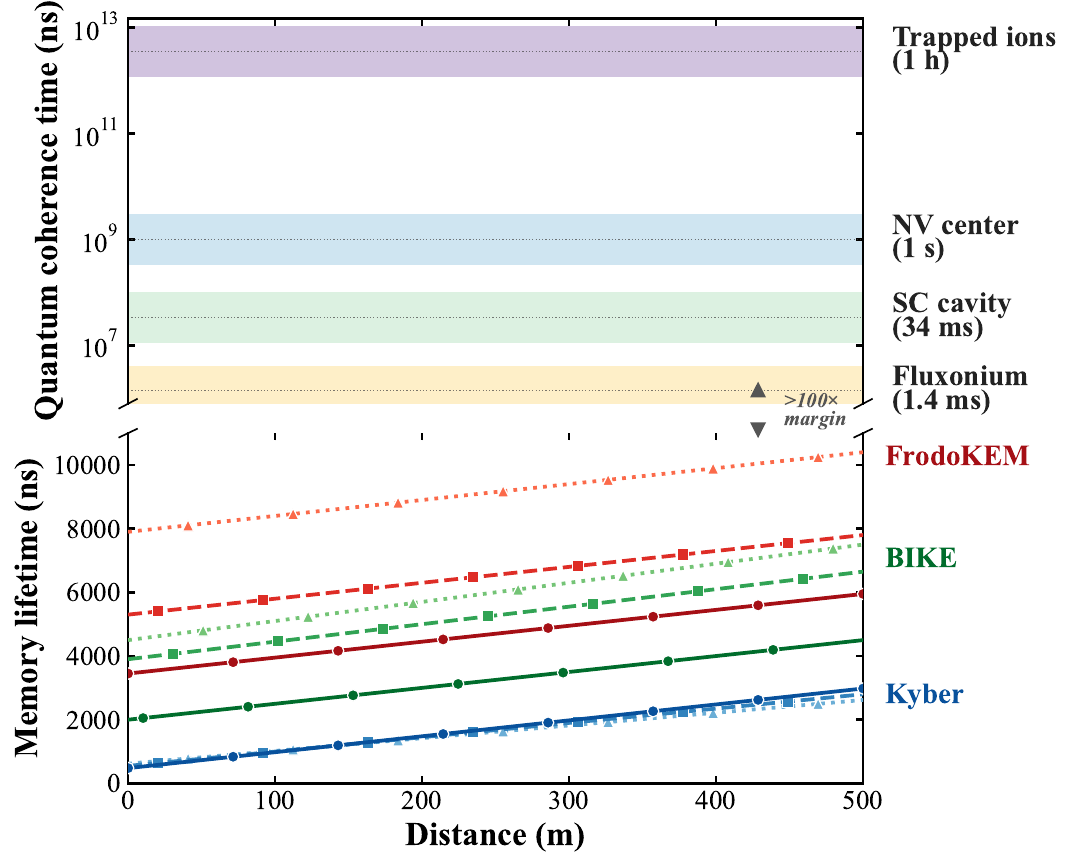}
    \caption{}
    \label{fig:memory_limit_b}
\end{subfigure}
\caption{Quantum memory lifetime vs.\ communication distance for different PQC schemes. (a)~Short-range view (0--100\,m) on a symlog scale. (b)~Long-range view (0--500\,m) with quantum-memory coherence benchmarks (top, log scale) and PQC requirements (bottom, linear); coherence times exceed PQC needs by $>$100$\times$. The slope ($\approx 5~\mu$s/km) reflects fiber propagation delay; the intercept reflects PQC processing time. See Appendix~\ref{appendix: Memory lifetime limit} for the full derivation.}
\label{fig:memory_limit}
\end{figure*}

\begin{table}[htbp]
\caption{PQC overhead and maximum distance ($T_{\mathrm{coh}}=1$~ms).}
\label{tab:pqc_overhead}
\centering
\scriptsize
\setlength{\tabcolsep}{3pt}
\renewcommand{\arraystretch}{0.8}
\begin{tabular}{lccc}
\toprule
Algorithm & Sec. & $T_{\text{enc}}$ ($\mu$s) & Max Dist. (km) \\
\midrule
Kyber512 & 128 & 2.3 & 198.5 \\
Kyber768 & 192 & 3.1 & 198.3 \\
Kyber1024 & 256 & 4.2 & 198.1 \\
FrodoKEM-640 & 128 & 15.7 & 195.9 \\
FrodoKEM-976 & 192 & 28.4 & 194.3 \\
FrodoKEM-1344 & 256 & 45.1 & 191.0 \\
\bottomrule
\end{tabular}
\end{table}

Our results show the tradeoff between PQC robustness and quantum memory requirements: stronger PQC algorithms impose heavier processing delays, which shrink the effective distance window for teleportation. PQC security levels should therefore be chosen in light of physical memory limits and network distance targets.

\textbf{Distance–security tradeoffs.} The maximum feasible communication distance is determined by subtracting fixed delays from the coherence window and multiplying by the propagation speed in fiber:
\begin{equation}
d_{\max} = \left(T_{\mathrm{coh}} - T_{\text{fixed}}\right) \cdot v_{\text{fiber}},
\label{eq:max_distance}
\end{equation}
with $v_{\text{fiber}} \approx 2\times10^5$ km/s.Here $T_{\text{fixed}} = T_{\mathcal{P}_{\mathrm{Bell}}} + T_{\mathcal{U}_{\mathrm{Bell}}} + T_{\mathcal{E}_{\mathrm{PQC}}} + T_{\mathcal{M}_{\mathrm{Alice}}} + T_{\mathcal{R}_{\mathrm{Bob}}}$ collects all distance-independent delays, so that the distance-dependent contribution reduces to $T_{\mathrm{comm}} = d/v_{\mathrm{fiber}}$.

For example, with a practical coherence time of $T_{\mathrm{coh}}=1$ ms (consistent with state-of-the-art trapped-ion and rare-earth doped solid-state memories~\cite{Hedges2010, Bruzewicz2019, Liu2021}), the maximum distance drops from $\sim 200$ km with lightweight schemes such as Kyber512 to below 192 km for heavy-weight schemes such as FrodoKEM-1344. This shows that memory lifetime, rather than raw cryptographic strength, sets the fundamental limit on secure teleportation distances. The derived linear relation provides a design guideline for balancing PQC selection with physical-layer capabilities. For a given quantum memory technology, we can invert Eq.~\eqref{eq:max_distance} to determine the strongest PQC scheme that maintains real-time feasibility.

 The analysis above establishes the physical feasibility region of QRQT by quantifying how PQC overhead and communication distance jointly constrain the available coherence window. A complementary and equally important question is whether the protocol remains \textit{secure} within this feasible region: specifically, can an adversary targeting both the classical and quantum channels simultaneously succeed before decoherence destroys the intercepted quantum information? We address this question in the following section by developing a joint classical--quantum threat model.

\section{Joint Threat Model}
\label{sec:joint-threat}
To evaluate the overall security of the Quantum-Resistant Quantum Teleportation (QRQT)
framework, we establish a joint threat model that unifies the quantum-side and classical-side
adversarial assumptions. The goal is to capture the temporal coupling between physical
quantum coherence and post-quantum computational hardness that determines whether
an attack can succeed before decoherence destroys useful information.

 We consider an adversary, Eve, who can target both the quantum and classical channels of the teleportation protocol. On the \textit{quantum channel}, Eve is assumed capable of coherently interacting with qubits distributed over optical fiber and, in the strongest scenario, performing a SWAP-based interception that redirects Bob's entangled qubit into her own quantum memory (detailed in Sec.~\ref{subsec:swap-attack}). We do not assume that legitimate parties perform entanglement verification or active monitoring; this deliberate omission establishes an upper bound on adversarial capability.

On the \textit{classical channel}, which is treated as public and unauthenticated~\cite{Bennett1993,Nielsen2012, Barnum,Yin2020,Portmann2022}, Eve may eavesdrop on, intercept, or delay the PQC-protected correction bits. Her classical attack capability is modeled in terms of lattice-reduction effort against the underlying LWE instance (detailed in Sec.~\ref{sec:PQC attack}). The joint model couples these two attack surfaces through a shared time parameter: classical cryptanalytic progress improves with time, whereas the quantum information available to Eve degrades due to decoherence, creating the temporal competition analyzed below.

\subsection{Quantum Side Security: SWAP Attack}
\label{subsec:swap-attack}

\begin{figure}[htb!]
\centering
\resizebox{\columnwidth}{!}{%
\begin{tikzpicture}[
    x=1cm,
    y=1.25cm,
    >=Stealth,
    thick,
    gate/.style={
        draw, rounded corners=2pt,
        minimum width=1.05cm, minimum height=0.78cm,
        fill=white, inner sep=2pt, font=\normalsize
    },
    membox/.style={
        draw=teal!70!black, rounded corners=3pt,
        minimum width=1.6cm, minimum height=0.78cm,
        fill=teal!8, inner sep=2pt, font=\small,
        text=teal!70!black
    },
    meas/.style={
        draw, minimum width=0.88cm, minimum height=0.88cm,
        fill=white
    }
]

\node[anchor=east, font=\small] at (-0.15, 3)   {$|\psi\rangle$};
\node[anchor=east, font=\small, text=textdark] at (-0.55, 2) {$A$};
\node[anchor=east, font=\small, text=textdark] at (-0.55, 1) {$B$};
\node[anchor=east, font=\small] at (-0.15, 0)   {$|0\rangle_E$};

\node[font=\scriptsize, text=gray!60!black] at (-0.85, 3) {$C$};

\draw[black, line width=0.8pt] (0, 3) -- (8.0, 3);
\draw[black, line width=0.8pt] (0, 2) -- (8.0, 2);
\draw[black, line width=0.8pt] (0, 1) -- (12.6, 1);
\draw[black, line width=0.8pt] (0, 0) -- (12.6, 0);

\draw[blue!70!black, line width=0.85pt] (1.3, 2) circle (0.10);
\draw[blue!70!black, line width=0.85pt] (1.3, 1) circle (0.10);
\draw[blue!70!black, line width=0.85pt] (1.3, 1.90) -- (1.3, 1.10);
\node[blue!70!black, font=\footnotesize, anchor=west] at (1.5, 1.50)
    {$|\Phi^+\rangle_{AB}$};

\begin{scope}[on background layer]
    \fill[red!6, rounded corners=5pt]
        (2.65, 1.35) rectangle (3.75, -0.35);
    \draw[red!50!black, rounded corners=5pt,
        line width=0.6pt, dashed]
        (2.65, 1.35) rectangle (3.75, -0.35);
\end{scope}
\draw[red!65!black, line width=1pt] (3.02, 1.16) -- (3.38, 0.84);
\draw[red!65!black, line width=1pt] (3.38, 1.16) -- (3.02, 0.84);
\draw[red!65!black, line width=1pt] (3.02, 0.16) -- (3.38,-0.16);
\draw[red!65!black, line width=1pt] (3.38, 0.16) -- (3.02,-0.16);
\draw[red!65!black, line width=0.8pt] (3.20, 0.82) -- (3.20, 0.18);
\node[font=\footnotesize\bfseries, text=red!65!black]
    at (3.20, -0.60) {SWAP};

\node[membox] (qmem) at (5.2, 0) {QM};
\node[font=\tiny, text=teal!60!black, align=center] at (5.2, -0.55)
    {$T_{\mathrm{coh}}$};

\begin{scope}[on background layer]
    \fill[blue!4, rounded corners=5pt]
        (4.3, 3.85) rectangle (8.1, 1.55);
\end{scope}
\node[blue!50!black, font=\scriptsize\bfseries] at (6.2, 4.10)
    {Bell measurement on $(C,\!A)$};

\filldraw[black] (5.0, 3) circle (2.1pt);
\draw[black, line width=0.8pt] (5.0, 3) -- (5.0, 2);
\draw[black, line width=0.8pt] (5.0, 2) circle (0.14);

\node[gate] (H) at (6.2, 3) {$H$};

\node[meas] (M2box) at (7.5, 3) {};
\node[meas] (M1box) at (7.5, 2) {};
\draw[line width=0.6pt] (7.28, 2.87)
    arc[start angle=200, end angle=340, radius=0.21];
\draw[->, line width=0.6pt] (7.43, 2.91) -- (7.65, 3.11);
\draw[line width=0.6pt] (7.28, 1.87)
    arc[start angle=200, end angle=340, radius=0.21];
\draw[->, line width=0.6pt] (7.43, 1.91) -- (7.65, 2.11);
\node[above=2pt, font=\footnotesize, text=textdark]
    at (M2box.north) {$M_2$};
\node[below=2pt, font=\footnotesize, text=textdark]
    at (M1box.south) {$M_1$};

\draw[classwire, dashed, line width=0.9pt, ->]
    (7.95, 3.0) -- (9.6, 3.0) -- (9.6, 0);
\draw[classwire, dashed, line width=0.9pt, ->]
    (7.95, 2.0) -- (9.3, 2.0) -- (9.3, 0);
\node[text=classwire, font=\scriptsize\bfseries, align=center]
    at (10.1, 1.5) {PQC\\attack};

\node[gate, font=\small] (Xcorr) at (10.3, 0) {$X^{M_1}$};
\node[gate, font=\small] (Zcorr) at (11.7, 0) {$Z^{M_2}$};

\draw[->, black, line width=0.8pt] (12.6, 1) -- (13.0, 1);
\node[anchor=west, font=\footnotesize, text=textdark]
    at (13.1, 1) {$|0\rangle$ (Bob)};
\draw[->, black, line width=0.8pt] (12.23, 0) -- (13.0, 0);
\node[anchor=west, font=\footnotesize, text=red!65!black]
    at (13.1, 0) {$|\psi\rangle$ (Eve)};

\draw[gray!50, dotted, rounded corners=7pt, line width=0.7pt]
    (-0.3, 3.85) rectangle (8.1, 1.55);
\node[gray!50!black, font=\footnotesize\bfseries] at (1.5, 4.10) {Alice};

\draw[red!30, dotted, rounded corners=7pt, line width=0.7pt]
    (2.4, 0.48) rectangle (12.8, -0.50);
\node[red!50!black, font=\footnotesize\bfseries]
    at (8.0, -0.78) {Eve};

\draw[teal!60!black, <->, line width=0.6pt]
    (5.95, -0.25) -- (9.75, -0.25);
\node[font=\tiny, text=teal!60!black] at (7.85, -0.42)
    {decoherence window};

\end{tikzpicture}%
}
\caption{Quantum circuit schematic of the SWAP attack on QRQT.
 Initially, a Bell pair $|\Phi^+\rangle_{AB}$ is shared between Alice (qubit~$A$) and Bob's traveling qubit~$B$.
Eve intercepts qubit~$B$ and swaps it with her ancilla $|0\rangle_E$, thereby transferring the remote half of the entangled pair from Bob to Eve and leaving Bob with the dummy state $|0\rangle$.
Alice then measures her input $|\psi\rangle$ (qubit~$C$) and qubit~$A$ in the Bell basis, obtaining outcomes $(M_1,M_2)$.
 If Eve also breaks the PQC-protected classical channel, she can apply $Z^{M_2}X^{M_1}$ to recover~$|\psi\rangle$; Bob still receives the protocol signals (a qubit and two correction bits) but, because his Bell-half has been replaced by a dummy state, the standard teleportation correction no longer reconstructs $|\psi\rangle$ at his side.}
\label{fig:swap attack}
\end{figure}
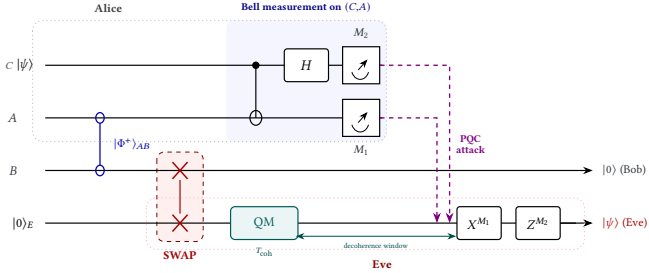

On the quantum channel, we assume a physically powerful adversary (Eve) capable of
performing coherent interactions with transmitted qubits. The strongest practical attack is
modeled as a \textit{SWAP attack}, where Eve replaces Bob's half of the entangled pair with
her own ancilla and stores the legitimate qubit in her quantum memory~(see Fig.~\ref{fig:swap attack} for the circuit-level description).
Operationally, Eve inserts a unitary $U_{\text{SWAP}}$ between Bob's qubit $B$ and her ancilla $E$, effecting the transformation $U_{\mathrm{SWAP}}|x\rangle_B|y\rangle_E = |y\rangle_B|x\rangle_E$. This redirects the entanglement from the $(A,B)$ pair to the $(A,E)$ pair, leaving Bob with the dummy state $|0\rangle$ while Eve inherits the full teleportation correlation~\cite{Nielsen2012}. Such intercept-and-resend strategies on entangled links have been studied in the context of quantum network routing security~\cite{Scarani2009,Caleffi2017} and unreliable entanglement assistance~\cite{Lederman2024}.

 To enable quantitative evaluation, we make the following implementation assumptions:
\begin{itemize}
    \item \textit{Physical access:} Eve can physically access or replace the fiber transmission path (or compromise an intermediate relay node) and apply an approximately ideal SWAP unitary $U_{\text{SWAP}}$ between the traveling photonic mode and a local ancilla, transferring the legitimate qubit into her quantum memory with sufficient fidelity~\cite{Pirandola2015}.
    \item \textit{Classical attack capability:} Concurrently, Eve can intercept, delay, or attempt to break the PQC-protected classical correction information $(M_1,M_2)$, with a classical cryptanalytic budget denoted $T_{\mathrm{BKZ}}$ (see Section~\ref{sec:PQC attack}), and perform the necessary Pauli corrections $Z^{M_2}X^{M_1}$ within the coherence-limited time window.
    \item \textit{Absence of active monitoring:} The legitimate parties do not deploy frequent randomized entanglement verification, high-sensitivity timing or optical-statistics checks, or decoy-state probing, so that the substitution remains undetected.
\end{itemize}
These assumptions collectively define a worst-case adversarial model; practical countermeasures that relax them are discussed after the formal analysis below.

\subsubsection*{SWAP attack model for quantum teleportation}

The SWAP attack can be formalized as a completely positive trace-preserving (CPTP) map. Let the initial state be
\begin{equation}
|\Psi_0\rangle_{CABE} = |\varphi\rangle_C \otimes |\Phi^+\rangle_{AB} \otimes |0\rangle_E,
\end{equation}
where $|\varphi\rangle_C=a|0\rangle+b|1\rangle$ is Alice’s input, $|\Phi^+\rangle_{AB}=(|00\rangle+|11\rangle)/\sqrt{2}$ is the shared entangled state, and $|0\rangle_E$ is Eve’s ancilla. Eve applies a SWAP unitary $U_{\mathrm{SWAP}}$ between Bob’s qubit $B$ and her ancilla $E$,
\[
U_{\mathrm{SWAP}}|x\rangle_B|y\rangle_E = |y\rangle_B|x\rangle_E,
\]
which is a general exchange of two qubit registers. The physical significance in the present context is that Bob's qubit~$B$ is not an isolated state but one half of the entangled pair $|\Phi^+\rangle_{AB}$. The SWAP operation transfers that entanglement from Bob to Eve yielding
\begin{equation}
|\Psi_1\rangle_{CAE B} = |\varphi\rangle_C \otimes |\Phi^+\rangle_{AE} \otimes |0\rangle_B.
\end{equation}
 Eve now holds the teleportation resource which was originally Bob's. When Alice performs her Bell measurement on $(C,A)$, Eve’s qubit collapses to a state related to $|\varphi\rangle$ by a Pauli operator, exactly as in the standard teleportation protocol. By intercepting Alice’s classical correction bits, Eve can apply the appropriate Pauli correction to her qubit and perfectly reconstruct $|\varphi\rangle$. Bob still receives a qubit and the two classical correction bits through the normal protocol flow, but because his Bell-half has been replaced by the dummy state $|0\rangle_B$, the standard Pauli correction $Z^{M_2}X^{M_1}$ no longer reconstructs $|\varphi\rangle$ at his side.

\textbf{Time-dependent success probability. }
Eve’s success depends on her ability to preserve coherence in memory until Alice’s classical bits are received.  To model the finite memory lifetime without committing to a hardware-specific noise process, we adopt a phenomenological depolarizing parametrization for the recovered-state fidelity:
\begin{equation}
F(t)=\frac{1}{2}\bigl(1+e^{-t/T_{\mathrm{coh}}}\bigr),
\end{equation}
where $T_{\mathrm{coh}}$ is the quantum memory coherence time. This form captures the essential physics: $F(0)=1$ for a freshly stored state and $F\to 1/2$ as $t\to\infty$, reflecting complete loss of quantum information to the maximally mixed state.  We emphasize that this is a conservative phenomenological model, not a first-principles derivation tied to a specific platform; a full treatment would require platform-dependent noise operators (e.g., $T_1$/$T_2$ relaxation for superconducting qubits, or spectral diffusion for solid-state memories). The depolarizing-channel form is chosen because it provides a single-parameter upper bound on the fidelity decay and is widely adopted in analyses of optical and solid-state quantum memories~\cite{Holevo1998,Wilde2013,Horodecki1999,Pirandola2015},
where it accurately describes short- to intermediate-time coherence loss under weak coupling
to a thermal bath.

Operationally, $F(\rho,|\psi\rangle)=\langle\psi|\rho|\psi\rangle$ equals the probability that the decohered state $\rho$ passes the projective verification test $|\psi\rangle\!\langle\psi|$~\cite{Wilde2013}.
Under optimal Pauli correction, Eve's attack success probability equals:
\begin{equation}
P_{\mathrm{SWAP}}(t) = F(t) = \frac{1}{2}\bigl(1+e^{-t/T_{\mathrm{coh}}}\bigr)
\label{PSWAP}
\end{equation}

High fidelity ($F \approx 1$) means Eve preserves the quantum information
well; low fidelity ($F \to 1/2$) means decoherence has destroyed the coherent information, reducing Eve's state to the maximally mixed qubit.


Quantum teleportation is known to be intrinsically secure against passive eavesdropping on the quantum channel, yet it still requires authentication of the accompanying classical messages to prevent man-in-the-middle attacks~\cite{Barnum, Curty2004}.
Previous studies have examined teleportation fidelity under various noise and loss mechanisms~\cite{Horodecki1999,Pirandola2015} and how the Holevo bound constrains the accessible information in noisy quantum channels~\cite{Holevo1998,Wilde2013}.
Meanwhile, the classical control channel remains susceptible to side-channel leakage and timing-based inference attacks~\cite{Pantoja2024,Qi2007,Gisin2006}. Building on these results, we formalize the SWAP-style interception as a CPTP map and couple it with a time-dependent quantum-memory fidelity model.
By linking the coherence-limited attack window on the quantum side with the cryptanalytic runtime on the classical side, we obtain a joint model of adversarial success in PQC-protected teleportation.


The SWAP attack formalized above assumes an idealized adversary who can execute a perfect unitary swap on the traveling qubit without introducing detectable disturbances. In practice, several verification mechanisms available to the legitimate parties would significantly constrain such an attack.

\textit{Entanglement verification:} Alice and Bob can periodically perform entanglement witness measurements or Bell-inequality (CHSH) tests on a randomly selected subset of shared pairs~\cite{Brunner2014}. Any SWAP interception breaks the correlations of the tested pair, producing a detectable drop in the Bell parameter $S$ below the classical bound $S\leq 2$.

\textit{Timing-based detection:} Inserting a SWAP gate and rerouting the photonic mode introduces additional optical-path delay and loss. Statistical monitoring of arrival-time jitter and channel transmittance can flag anomalous deviations from the expected profile~\cite{Qi2007,Gisin2006}.

\textit{Decoy-state probing:} Borrowing from QKD practice, Alice can inject decoy states at random intervals to probe channel integrity. Discrepancies in the measured error rate or photon-number statistics reveal the presence of an active interception device~\cite{Lo2005}.

Our model deliberately omits these countermeasures to establish an \emph{upper bound} on adversarial capability: the SWAP success probability $P_{\mathrm{SWAP}}(t)$ quantifies the worst-case scenario in which Eve is undetected.
If a detection probability $P_{\mathrm{det}}$ is incorporated, the effective joint attack probability reduces to
$P_{\mathrm{joint}}^{\mathrm{eff}}(t)=(1 - P_{\mathrm{det}})\,P_{\mathrm{LWE}}(t)\,P_{\mathrm{SWAP}}(t)$,
which strictly tightens the security guarantees derived in this work. Thus, our analysis provides a conservative baseline; real-world deployments with active monitoring will enjoy stronger protection than the bounds reported here.

\subsection{Classical Side Attack Analysis}
\label{sec:PQC attack}
On the classical channel, the adversary targets the post-quantum primitive
$\mathcal{E}_{\mathrm{PQC}}$ used to protect the teleportation correction bits. To model the classical attack, we focus on PQC and adopt the minimal assumption that the
classical-side protection reduces to an underlying Learning-With-Errors (LWE) instance,
and we model adversarial effort in terms of lattice-reduction attacks against LWE.
In particular, we do not explicitly model protocol-specific constructions (such as Kyber
or FrodoKEM) nor do we perform separate cost estimates of symmetric-key primitives
(e.g., AES). In practice, the lattice-reduction stage required to recover the LWE secret dominates the overall cost of classical attacks; therefore, it is reasonable to characterize attack cost by the hardness of LWE alone. Additional overheads, such as brute-force work against AES or Kyber group operations, are smaller and can typically be offset by linearly scaling the number of compute resources. Consequently, whether the PQC attack is solvable or merely requires more machines is determined primarily by the complexity of solving LWE. For this reason, we simplify the classical attack model in this work to the problem of breaking LWE and focus on this dominant complexity.

 \textbf{BKZ Attack Model. }
Building on the LWE formulation and BKZ lattice-reduction framework introduced in Section~\ref{subsec:lwe-prelim}, we now specify the attack model used in Section~\ref{subsec:joint-attack}. The detailed cryptographic parameter choices and communication security assumptions are provided in Appendix~\ref{appendix: LWE parameter}; the derivation of the relationship between $n$, $m$, and $\|\tilde{b}_1\|$ is given in Appendix~\ref{appendix: m,n,b relationship}. Recall that BKZ reduction combined with Nearest-Planes enumeration yields a runtime $T_{\mathrm{BKZ}}$ governed by the empirical fit in Eq.~\eqref{eq:bkz_time_prelim}. In our analysis, the coefficients $(a,b)$ are treated as tunable parameters; representative calibration values are reported in Appendix~\ref{appendix:BKZ_params}.

In this work, the decoding success probability \(P_{\mathrm{LWE}}\) represents the overall success probability of a classical lattice-reduction attack against the LWE instance. Specifically, it models the probability that a BKZ-reduced basis, when followed by a Nearest-Planes decoding step, successfully recovers the secret. Under the Geometric Series Assumption (GSA), this probability can be approximated using the product-of-error-function model~\cite{Lindner2011,Chen2011}. Combined with the BKZ runtime estimate in Eq.~\eqref{eq:bkz_time_prelim}, this formulation provides a direct mapping between the runtime of lattice reduction and the expected success rate of the corresponding attack.

\textbf{Attack Success Probability. }Under the Geometric Series Assumption (GSA), the attack success probability is~\cite{Lindner2011}:
\begin{equation}
P_{\mathrm{LWE}} =
\prod_{i=1}^{m}
\operatorname{erf}\!\left(
\frac{d_i \|\tilde{b}_i\| \sqrt{\pi}}{2s}
\right),
\label{eq:lwe_probability}
\end{equation}
where $\tilde{b}_i$ denotes the $i$-th Gram--Schmidt vector of the BKZ-reduced basis, $s$ denotes the standard deviation as noise magnitude of Gaussian error distribution used in the LWE error terms $\mathrm{e}$, $d_i$ represents the integer search radius used in the NearestPlanes enumeration for the $i$-th search layer,
and $\delta_{\mathrm{root}}$ is the root-Hermite factor characterizing the reduction quality.

By combining the BKZ runtime model with the NearestPlanes success function under the
GSA, we derive a closed-form expression linking the attack success probability
$P_{\mathrm{LWE}}$ to the lattice-reduction runtime $T_{\mathrm{BKZ}}$:
\begin{align}
P_{\mathrm{LWE}} &=
\prod_{i=1}^{m}
\operatorname{erf}\!\left(
\frac{d_i \,\|\tilde{b}_1\|\,\sqrt{\pi}}{2s}
2^{-2(i-1)\frac{a}{\log_2 T_{\mathrm{BKZ}} + b}}
\right).
\label{classical attack}
\end{align}
\noindent
 This derivation is detailed in Appendix~\ref{appendix:PLWE_derivation}. It turns the qualitative notion of attack feasibility into a time-dependent probability model parameterized by computational power and algorithmic complexity.
Here $\|\tilde{b}_i\|$ denotes the length of the $i$-th orthogonalized basis vector,
governed by the reduction quality $\delta_{\mathrm{root}}$ under the GSA relation
$\|\tilde{b}_i\| = \|\tilde{b}_1\| \delta_{\mathrm{root}}^{-2(i-1)}$.
Note that the parameters $(a,b)$ in the BKZ runtime model are unrelated to the basis vectors; they are empirical constants reflecting computational capability.

For efficient numerical evaluation, we adopt a Padé-type rational approximation of the error function~\cite{winitzki2008handy}, which yields a closed-form expression for $\log_2 P_{\mathrm{LWE}}$ with sub-$10^{-3}$ global accuracy while preserving analytic differentiability. The derivation, explicit formula, and numerical validation are provided in Appendix~\ref{appendix:pade_approximation}.
\subsection{Joint Classical-Quantum Attack Probability Model}
\label{subsec:joint-attack}

The key insight is the opposite time dependence: $P_{\mathrm{LWE}}(t)$ increases as Eve gains computation, while $P_{\mathrm{SWAP}}(t)$ decreases due to quantum decoherence. For a joint attack to succeed,
Eve must break LWE \textit{and} preserve quantum coherence within the same time budget $t$ (where $t$ denotes the BKZ attack time, i.e., $t = T_{\mathrm{BKZ}}$).

We model the joint success probability as the product $P_{\mathrm{LWE}} \cdot P_{\mathrm{SWAP}}$, reflecting the statistical independence of the two constituent events conditioned on the shared time parameter~$t$. This factorization is justified on the following grounds.
\textit{First}, the LWE attack on the encrypted classical messages $M_1$ and $M_2$ is fundamentally a cryptanalytic task: recovering the private key from the publicly available PQC public key via lattice reduction. This is a purely computational process that can be initiated at any time: before, during, or after the quantum measurements, which proceeds entirely independently of the quantum channel. In our model, we assume that Eve launches both attacks concurrently at $t=0$, so that the elapsed time~$t$ parameterizes both processes simultaneously.
\textit{Second}, the two attacks operate on physically disjoint channels and consume non-competing resources: the SWAP interception requires coherent quantum hardware to capture and store a traveling photon, whereas the LWE cryptanalysis requires classical (or quantum) computational resources for lattice reduction. Neither resource expenditure constrains or assists the other.
\textit{Third}, the decoherence of Eve's quantum memory is a Markovian physical process governed by the coupling between her storage device and its thermal environment; it proceeds independently of whether the BKZ algorithm running on a separate classical processor has converged.
\textit{Fourth}, there is no information coupling between the two sub-attacks: partial progress on the LWE instance does not reveal any property of the stored quantum state (or vice versa), so the conditional probability of one event given the other equals its marginal probability.
The only coupling between the two channels is \emph{temporal}: both probabilities are parameterized by the same elapsed time~$t$, which is already captured by their explicit time dependence rather than by a statistical correlation.

\begin{figure}[!tb]
\centering
\includegraphics[width=\columnwidth]{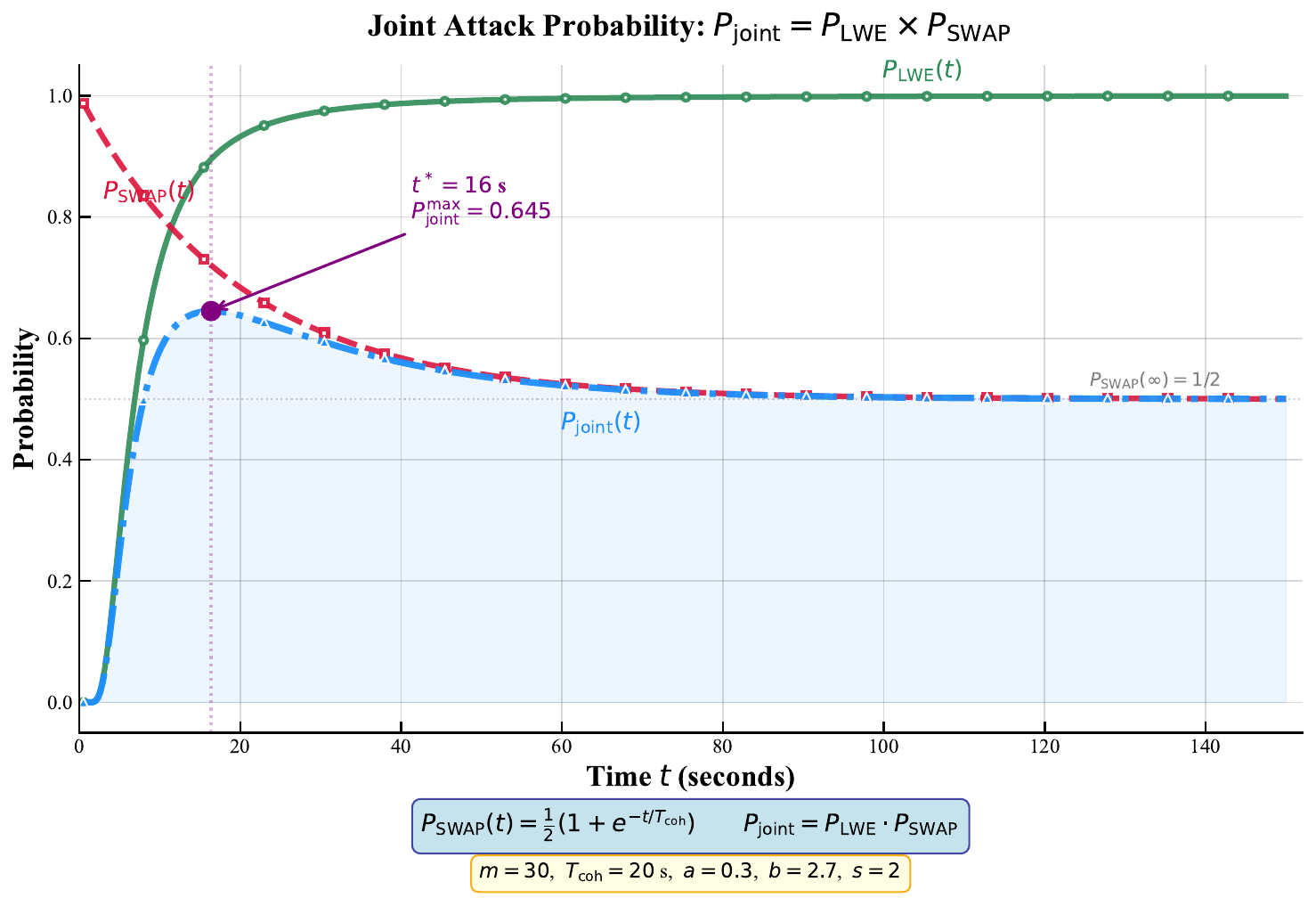}
\caption{Joint attack probability vs.\ time.
$P_{\mathrm{LWE}}$ (green) grows with BKZ runtime; $P_{\mathrm{SWAP}}$ (red, dashed) decays toward~$1/2$ as Eve's quantum memory decoheres ($T_{\mathrm{coh}}=20\,$s).
Their product $P_{\mathrm{joint}}$ (blue, dash-dotted) peaks at $t^*\approx16\,$s, forming a bounded attack window.
Parameters: $m=30$, $s=2$, $a=0.3$, $b=2.7$.
Weak LWE parameters are used for visual clarity; realistic dimensions (e.g., $m \geq 256$) shift $t^*$ by orders of magnitude while preserving the non-monotonic structure.}
\label{fig:p_joint}
\end{figure}

Combining Eq.~\eqref{classical attack} and Eq.~\eqref{PSWAP}, the joint classical-quantum attack probability is:
\begin{equation}
    P_{\mathrm{joint}}= P_{\mathrm{LWE}}\cdot P_{\mathrm{SWAP}}
\end{equation}
\begin{equation}
    P_{\mathrm{joint}}=\prod\limits_{i=1}^{m} \operatorname{erf}\!\left(\frac{d_{i}  2^{\frac{a}{\log_{2} t+b}\cdot(-2)(i-1)} \|\tilde{b}_1\|\sqrt{\pi}}{2s}\right) \cdot \frac{1}{2}\bigl(1+e^{-t/T_{\mathrm{coh}}}\bigr)
\end{equation}
This product exhibits a characteristic bell-shaped curve (Fig.~\ref{fig:p_joint}).
The maximum occurs at an optimal attack time $t^* = \arg\max_t P_{\mathrm{joint}}(t)$,
beyond which quantum decoherence dominates and the attack becomes infeasible.

This unified model reveals a fundamental security principle distinct from
classical cryptography: neither computational hardness nor quantum coherence
alone determines security. Instead, security emerges from their \textit{temporal
competition}. Systems with longer quantum memory coherence times possess wider
vulnerability windows, creating a counterintuitive tradeoff: quantum advantage
comes at the cost of extended attack opportunities during PQC processing. To ensure $P_{\mathrm{joint}}(t) < \epsilon$ for security parameter $\epsilon$,
both $P_{\mathrm{LWE}}(t) < \sqrt{\epsilon}$ and $P_{\mathrm{SWAP}}(t) <
\sqrt{\epsilon}$ must hold. This couples cryptographic parameter selection
(affecting $P_{\mathrm{LWE}}$) with quantum memory characteristics (affecting
$P_{\mathrm{SWAP}}$), establishing a new design constraint for QRQT systems.

The joint threat model developed above quantifies the probability that a specific attack strategy succeeds within a given time budget. However, it does not address a more fundamental question: \textit{how much information is inherently accessible to an adversary}, regardless of the particular attack she employs? An adversary with partial knowledge of the classical correction bits, whether obtained through side-channel leakage, timing analysis, or partial cryptanalytic progress, may extract information from Bob's quantum state without fully breaking the PQC scheme. To establish a complementary, strategy-independent security benchmark, we turn to an information-theoretic analysis based on the Holevo bound. This approach provides an upper limit on the classical information extractable from the quantum ensemble under any measurement, thereby capturing the worst-case information leakage as a function of both quantum decoherence and classical key compromise.



\section{Information-Theoretic Security Analysis}
\label{sec:info-theoretic-security}

While quantum teleportation is often analyzed under the assumption of perfect quantum resources and secure classical channels, practical implementations face a broader set of challenges. In addition to noise and imperfections affecting the entangled states, vulnerabilities can arise through leakage of the classical communication used to convey the correction information. Since the classical bits effectively serve as a ``Pauli one-time pad''~\cite{Nielsen2012}, concealing the quantum state until Bob applies the correction, any partial or full compromise of these bits can lead to significant information leakage. Moreover, if Eve gains access to Bob's quantum system in addition to the classical correction data, this poses a serious threat to the entire teleportation process.

Understanding the impact of classical information leakage requires careful analysis of how much knowledge about the quantum state can be inferred from the compromised classical bits. A natural tool for this analysis is the Holevo Information~\cite{Holevo1973Bounds}, which provides an upper bound on the accessible classical information that can be extracted from a quantum ensemble. Prior to receiving the correction bits, Bob's qubit remains in a maximally mixed state, encoding hidden information about Alice's measurement outcome. The arrival of the classical bits effectively ``decrypts'' this hidden information, restoring the original state.

In this work, we investigate how classical information leakage over time affects the security and fidelity of teleported quantum state in quantum teleportation. We consider scenarios where an eavesdropper, Eve, gains partial access both to the classical communication channels and to Bob's quantum system, potentially through device tampering, side-channel attacks, or physical theft~\cite{Koeune2005,Lucamarini2015,Ghosh2025,Gangopadhyay2022}. Our goal is to model the dynamics of leakage and quantify, at each moment in time, the residual amount of information about the correction bits that remains concealed.

To capture a range of plausible time dependent leakage behaviors, we introduce four stochastic models motivated by temporal patterns commonly studied in stochastic processes, reliability theory, and event dynamics~\cite{Ruijters2019,Zeng2023,Karr1984,2002},

\begin{itemize}
    \item \textbf{Independent exponential leaks}: the two classical correction bits leak independently over time, each according to its own exponential leakage rate.
    \item \textbf{Sequential leaks}: leakage proceeds in stages, with one correction bit becoming vulnerable only after the other has been compromised.
    \item \textbf{Burst leaks}: both correction bits are leaked simultaneously following a single discrete compromise event.
    \item \textbf{Correlated leaks}: leakage of the two bits occurs with nontrivial statistical correlations parameterized by a tunable correlation factor.
\end{itemize}

For each leakage model, we derive closed-form expressions for the expected residual Holevo information, $\mathbb{E}[\chi(t)]$, and the corresponding optimal average fidelity, $F(t)$, achievable by an adversary at a given time $t$. These expressions allow us to quantitatively track how information hidden within the quantum system degrades over time as classical leakage progresses.

Our analysis also accounts for quantum decoherence by modeling imperfections in the entangled resource through an amplitude damping channel acting on one qubit of the Bell pair. Imperfections in the quantum resource degrade the fidelity of the teleported state~\cite{Oh2002,Im2021}, while vulnerabilities in the classical communication channel threaten its security. In particular, the protocol depends on the integrity and latency of the classical correction channel relative to the adversary's ability to access or disrupt it. Interception, delay, or modification of the correction bits can leak information to an eavesdropper and can also affect the reconstructed quantum state. By deriving explicit time dependent bounds on the accessible information under several leakage models, our work provides practical guidance for setting critical Holevo information thresholds and designing teleportation protocols that can help counter such attacks.

\subsection{Role of Holevo Information in Teleportation Security}

In the quantum teleportation protocol~\cite{Bennett1993}, the outcome of Alice's Bell state measurement, specified by two classical bits \((m_1,m_2)\in\{0,1\}^2\), determines the Pauli operation \footnote{Throughout this paper, $m_1 \equiv M_1$ and $m_2 \equiv M_2$. These symbols denote the same classical bits and are used interchangeably in some sections for notational convenience and clarity.} \(U_{m_1m_2}=Z^{m_2}X^{m_1}\) that Bob must apply in order to reconstruct the original quantum state \(\ket{\psi}\). Until Bob receives this classical information, his qubit is described by a statistical mixture over all possible Pauli corrections. Thus, Bob's reduced state prior to correction is,
\begin{equation}
\rho_{\text{Bob}}=\frac{1}{2}I,
\end{equation}
where \(I\) denotes the identity operator.

This maximally mixed state reflects complete local uncertainty: without access to the classical bits \((m_1,m_2)\), Bob's quantum system carries no locally accessible information about the input state or Alice's measurement outcome. Any measurement performed on Bob's qubit before the classical communication is received yields outcomes that are statistically independent of Alice's measurement results.

From an information theoretic perspective, the teleportation protocol distributes the information needed for recovery across two resources: the classical correction bits \((m_1,m_2)\) and Bob's quantum system. Neither resource alone is sufficient to reconstruct the input state; only their combination allows successful recovery of \(\ket{\psi}\).

For an ensemble of states \(\{p_{(m_1,m_2)},\rho^{(m_1,m_2)}\}\), the Holevo information is defined as
\begin{equation}
\chi=S\!\left(\rho_{\text{avg}}\right)-\sum_{(m_1,m_2)}p_{m_1m_2}S\!\left(\rho^{(m_1,m_2)}\right),
\end{equation}
where the ensemble averaged state is
\begin{equation}
\rho_{\text{avg}}=\sum_{(m_1,m_2)}p_{m_1m_2}\rho^{(m_1,m_2)}.
\end{equation}

In the ideal teleportation protocol, each conditional state \(\rho^{(m_1,m_2)}\) is pure and therefore satisfies \(S(\rho^{(m_1,m_2)})=0\). The Holevo information then reduces to,
\begin{equation}
\chi=S(\rho_{\text{avg}}).
\end{equation}

Since \(\rho_{\text{avg}}=\dfrac{I}{2}\), the von Neumann entropy evaluates to \(S(\rho_{\text{avg}})=1\) bit. This quantifies the ensemble level information associated with the unknown correction data, even though Bob's reduced state alone does not reveal Alice's measurement outcome. This separation between quantum and classical information in teleportation motivates the analysis of how partial or delayed leakage of the correction bits affects an adversary's ability to reconstruct the teleported state, which is one of the main focuses of this work.

\subsection{Security Implications for Quantum Communication}

To analyze quantum teleportation in the presence of leakage of the classical correction bits, the disclosure of \((m_1,m_2)\) can be modeled as a stochastic process that captures partial, delayed, or time dependent information exposure. Under such leakage, the adversary's incomplete knowledge of the measurement outcomes induces a classical quantum ensemble of conditional states on Bob's system. The Holevo information provides an information theoretic upper bound for analyzing how much information can be inferred from this ensemble under partial knowledge of the correction bits. This formalism therefore enables a quantitative assessment of how stochastic leakage of the correction bits affects the adversary's information about the teleported quantum state.

This perspective highlights that the security of quantum teleportation is not determined solely by the quantum resources employed, such as shared entanglement, but also by the assumptions made regarding the confidentiality of the accompanying classical communication. Consequently, protecting classical communication channels and mitigating classical side channel leakage are essential for limiting an adversary's information about the input state in teleportation based quantum communication systems.

\section{Holevo Information with an Amplitude-Damped Bell Pair}
\label{sec:holevo-damped}

We analyze the Holevo information associated with quantum teleportation when the shared entangled resource is affected by amplitude damping on Bob's subsystem. Throughout this section, qubit~1 denotes the unknown input state held by Alice, qubit~2 denotes Alice's share of the Bell pair, and qubit~3 denotes Bob's share. The amplitude damping channel acts locally on qubit~3 prior to Alice's Bell measurement and classical communication. Our objective is to characterize how noise in the entangled resource modifies the ensemble of conditional output states on Bob's system and the resulting information-theoretic bounds on accessible classical information.

\subsection{Effect of Amplitude-Damping Channel}

The shared entangled state between qubits~2 and~3 is initially prepared as,
\begin{equation}
\ket{\Phi^+}_{23}=\frac{1}{\sqrt{2}}\bigl(\ket{00}+\ket{11}\bigr).
\end{equation}

An amplitude damping channel with parameter \(\gamma\in[0,1]\) acts on qubit~3 and is described by the Kraus operators,
\begin{equation}
E_0=
\begin{pmatrix}
1 & 0\\
0 & \sqrt{1-\gamma}
\end{pmatrix},
\qquad
E_1=
\begin{pmatrix}
0 & \sqrt{\gamma}\\
0 & 0
\end{pmatrix}.
\end{equation}

Applying the channel yields the mixed two-qubit state
\begin{align}
\rho'_{23}
&=(I\otimes E_0)\ket{\Phi^+}\bra{\Phi^+}(I\otimes E_0)
 +(I\otimes E_1)\ket{\Phi^+}\bra{\Phi^+}(I\otimes E_1) \notag\\
&=\tfrac{1}{2}\Bigl(
\ket{00}\bra{00}
+\sqrt{1-\gamma}\ket{00}\bra{11}
+\sqrt{1-\gamma}\ket{11}\bra{00} \notag\\
&\qquad\quad
+(1-\gamma)\ket{11}\bra{11}
+\gamma\ket{10}\bra{10}
\Bigr).
\end{align}
This noisy entangled state serves as the quantum resource for teleportation.

\subsection{Bell Measurement and Conditional Output States}

Alice performs a Bell basis measurement on qubits~1 and~2, with outcomes labeled by \((m_1,m_2)\in\{0,1\}^2\). Let
\[
\rho'_{123}=\ket{\psi}\bra{\psi}_1\otimes\rho'_{23},
\qquad
\ket{\psi}=\alpha\ket{0}+\beta\ket{1}.
\]
The unnormalized conditional state on Bob's system is then
\begin{equation}
\rho_3^{(m_1,m_2)}
=
\mathrm{Tr}_{12}\!\left[
\left(\ket{\Phi_{m_1m_2}}\bra{\Phi_{m_1m_2}}_{12}\otimes I_3\right)\rho'_{123}
\right].
\end{equation}

All conditional states are related by Pauli conjugation. For the reference outcome \((m_1,m_2)=(0,0)\), the normalized conditional state is
\begin{equation}
\rho_{00}=
\begin{pmatrix}
a & c\\
c^{*} & b
\end{pmatrix},
\end{equation}
with
\begin{equation}
a=|\alpha|^2+\gamma|\beta|^2,\qquad
b=(1-\gamma)|\beta|^2,\qquad
c=\alpha\beta^{*}\sqrt{1-\gamma},
\end{equation}
and \(a+b=1\). This state is the conditional teleported output corresponding to the reference Bell measurement outcome in the presence of amplitude damping on Bob's half of the Bell pair.

For a general Bell outcome, the unnormalized conditional state is
\begin{equation}
\rho_3^{(m_1,m_2)}=\frac{1}{4}\,
X^{m_1}Z^{m_2}\,\rho_{00}\,Z^{m_2}X^{m_1}.
\end{equation}

Since the noise acts only on Bob's subsystem, all Bell outcomes remain equiprobable, with \(p_{m_1m_2}=1/4\). Dividing by these probabilities gives the normalized conditional states
\begin{equation}
\rho_{3,\mathrm{norm}}^{(m_1,m_2)}
=
X^{m_1}Z^{m_2}\,\rho_{00}\,Z^{m_2}X^{m_1}.
\end{equation}

The ensemble \(\{p_{m_1m_2},\rho_{3,\mathrm{norm}}^{(m_1,m_2)}\}\) therefore characterizes Bob's quantum system before the classical correction bits are received.

\subsection{Spectral Properties of the Conditional States}

All conditional states are unitarily equivalent and therefore share the same eigenvalue spectrum. The determinant of the reference state \(\rho_{00}\) is given by,
\begin{equation}
D = ab - |c|^2 = \gamma(1-\gamma)|\beta|^4.
\end{equation}
As a result, each conditional state has eigenvalues,
\begin{equation}
\lambda_{\pm} = \tfrac{1}{2}(1 \pm \delta),
\qquad
\delta = \sqrt{1 - 4D}.
\label{eigenval}
\end{equation}

The von Neumann entropy of each conditional state depends only on this eigenvalue spectrum. Substituting the eigenvalues into the definition of the entropy yields,
\begin{align}
S_{\mathrm{indiv}}
&= -\lambda_{+}\log_2 \lambda_{+} - \lambda_{-}\log_2 \lambda_{-} \notag\\
&= -\frac{1+\delta}{2}\log_2\!\left(\frac{1+\delta}{2}\right)
   -\frac{1-\delta}{2}\log_2\!\left(\frac{1-\delta}{2}\right).
\end{align}
This can be written compactly as,
\begin{equation}
S_{\mathrm{indiv}} = h(\delta),
\end{equation}
where
\begin{equation}
h(x) = -\frac{1+x}{2}\log_2\!\left(\frac{1+x}{2}\right)
       -\frac{1-x}{2}\log_2\!\left(\frac{1-x}{2}\right).
\end{equation}

\subsection{Ensemble Average State and Holevo Information}

The ensemble average state of Bob's qubit, prior to receiving the classical correction bits, is defined as,
\begin{equation}
\rho_{\mathrm{avg}}=\sum_{m_1,m_2}p_{m_1m_2}\,\rho_{3,\mathrm{norm}}^{(m_1,m_2)},
\end{equation}
where \(\rho_{3,\mathrm{norm}}^{(m_1,m_2)}\) denotes the normalized conditional state associated with the Bell measurement outcome \((m_1,m_2)\), and \(p_{m_1m_2}\) is the corresponding outcome probability.

For general outcome probabilities, the ensemble average state can be written in the form,
\begin{equation}
\rho_{\mathrm{avg}}=
\begin{pmatrix}
J & K\\
K^{*} & 1-J
\end{pmatrix},
\end{equation}
where
\begin{equation}
J=ap_0+bp_1,
\qquad
K=c\,\Delta_0+c^{*}\,\Delta_1.
\end{equation}
Here,
\begin{equation}
p_0=p_{00}+p_{01}, \qquad p_1=p_{10}+p_{11},
\end{equation}
and
\begin{equation}
\Delta_0=p_{00}-p_{01}, \qquad \Delta_1=p_{10}-p_{11}.
\end{equation}

In the symmetric case where all Bell measurement outcomes are equally likely, \(p_{m_1m_2}=1/4\), we have \(p_0=p_1=1/2\) and \(\Delta_0=\Delta_1=0\), which gives,
\begin{equation}
J=\frac{1}{2}, \qquad K=0,
\end{equation}
and hence,
\begin{equation}
\rho_{\mathrm{avg}}=\frac{1}{2}I.
\end{equation}
This reflects the fact that, under the equiprobable outcome assumption, Bob's reduced state is maximally mixed before the classical correction bits are received.

For general probabilities, the eigenvalues of \(\rho_{\mathrm{avg}}\) are,
\begin{equation}
\lambda_{\pm}^{(\mathrm{avg})}=\tfrac{1}{2}(1\pm\Delta),
\qquad
\Delta=\sqrt{(1-2J)^2+4|K|^2},
\end{equation}
and therefore
\begin{equation}
S(\rho_{\mathrm{avg}})=h(\Delta).
\end{equation}

The Holevo Information associated with the ensemble of conditional states is then given by,
\begin{equation}
\chi=S(\rho_{\mathrm{avg}})-\sum_{m_1,m_2}p_{m_1m_2}\,
S\!\left(\rho_{3,\mathrm{norm}}^{(m_1,m_2)}\right).
\end{equation}
Since all conditional states are related to \(\rho_{00}\) by Pauli conjugation, they have the same entropy \(S_{\mathrm{indiv}}=h(\delta)\). Hence,
\begin{equation}
\chi=h(\Delta)-h(\delta).
\end{equation}
For uniform outcome probabilities, where \(\rho_{\mathrm{avg}}=\frac{1}{2}I\) and \(S(\rho_{\mathrm{avg}})=1\), this reduces to,
\begin{equation}
\chi=1-h(\delta).
\end{equation}

\subsection{Holevo Information under Amplitude Damping}

The Holevo Information for the ensemble of conditional states indexed by \((m_1,m_2)\) takes the closed form,
\begin{equation}
\chi(\gamma,|\alpha|^2)
=
1
-
h\!\left(
\sqrt{1 - 4\gamma(1-\gamma)(1-|\alpha|^2)^2}
\right),
\label{eq:holevo-final}
\end{equation}
where \(|\alpha|^2=|\langle 0|\psi\rangle|^2\) specifies the input state and \(h(x)\) denotes the binary entropy function. This gives an upper bound on the classical information about the Bell measurement outcome that can be extracted from Bob's quantum system alone, prior to receiving the classical correction bits.

\begin{figure}[htbp]
    \centering
    \includegraphics[width=0.48\textwidth]{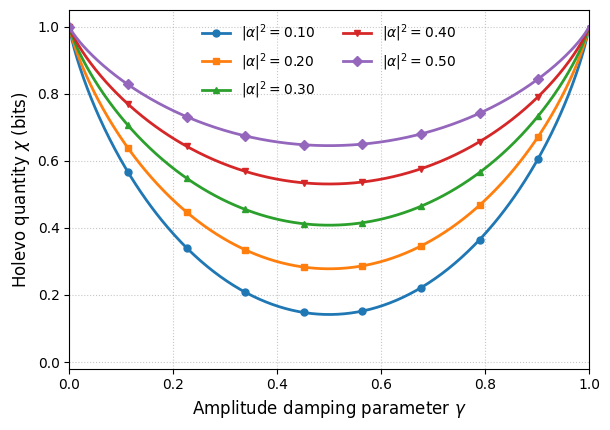}
    \caption{Holevo Information \(\chi(\gamma,|\alpha|^2)\) as a function of the amplitude damping parameter \(\gamma\) for several fixed input amplitudes \(|\alpha|^2\). The curves are symmetric about \(\gamma=0.5\) because \(\chi\) depends on \(\gamma\) only through the factor \(\gamma(1-\gamma)\) under the equiprobable outcome assumption \(p_{m_1m_2}=1/4\).}
    \label{fig:holevo-symmetric}
\end{figure}

In Fig.~\ref{fig:holevo-symmetric}, we plot the Holevo Information \(\chi(\gamma,|\alpha|^2)\) as a function of the amplitude damping parameter \(\gamma\) for several fixed input amplitudes \(|\alpha|^2=|\langle 0|\psi\rangle|^2\). We consider ideal local operations and measurements, with the only imperfection being amplitude damping acting on Bob's half of the shared Bell pair prior to Alice's Bell measurement. Since the noise acts only on Bob's subsystem, the four Bell measurement outcomes \((m_1,m_2)\in\{0,1\}^2\) remain equiprobable, and we take \(p_{m_1m_2}=1/4\).

In this model, \(\chi\) is the Holevo Information of the ensemble \(\{p_{m_1m_2},\rho_{3,\mathrm{norm}}^{(m_1,m_2)}\}\) of conditional states on Bob's qubit indexed by Alice's Bell outcome. It provides an information theoretic upper bound on the amount of classical information about \((m_1,m_2)\) that can be extracted from Bob's quantum system by any measurement performed without access to the classical outcomes.

At \(\gamma=0\), the entangled resource is maximally entangled and the conditional states are pure Pauli rotations of the input state, yielding \(\chi=1\) bit. As \(\gamma\) increases toward \(0.5\), amplitude damping reduces the purity of the conditional states, decreasing their distinguishability and thereby reducing \(\chi\). The symmetry around \(\gamma=0.5\) follows from the eigenvalue gap
\begin{equation}
\delta = \sqrt{1-4\gamma(1-\gamma)(1-|\alpha|^2)^2},
\end{equation}
which depends on \(\gamma\) only through \(\gamma(1-\gamma)\). Because \(\gamma(1-\gamma)\) is maximized at \(\gamma=0.5\), the conditional states are most mixed and \(\chi\) attains its minimum at that point. In contrast, as \(\gamma\to 0\) or \(\gamma\to 1\), the conditional states become pure, and the Holevo Information approaches its maximum value.

\begin{figure}[htbp]
    \centering
    \includegraphics[width=0.48\textwidth]{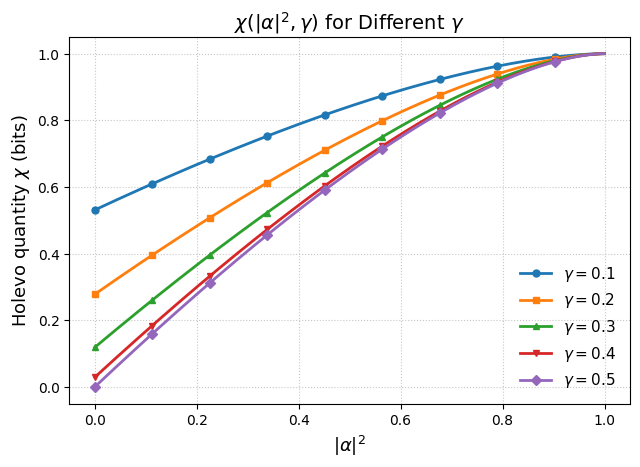}
    \caption{Holevo Information \(\chi(|\alpha|^2,\gamma)\) as a function of the input amplitude \(|\alpha|^2\) for several fixed values of the amplitude damping parameter \(\gamma\), assuming equiprobable Bell outcomes \(p_{m_1m_2}=1/4\).}
    \label{fig:holevo-vs-alpha}
\end{figure}

In Fig.~\ref{fig:holevo-vs-alpha}, we plot \(\chi(|\alpha|^2,\gamma)\) as a function of the input amplitude \(|\alpha|^2\) for several fixed values of \(\gamma\). For each fixed \(\gamma\), the Holevo Information increases monotonically with \(|\alpha|^2\). As \(|\alpha|^2\to 1\), the input approaches the ground state \(\ket{0}\), which is invariant under amplitude damping, and the conditional states remain pure, yielding \(\chi\to 1\). Conversely, for smaller \(|\alpha|^2\), the input has a larger excited-state component, which is more strongly affected by amplitude damping, producing more mixed conditional states and a smaller Holevo Information.

\section{Stochastic Models for Classical Information Leakage}\label{stochastic}

In quantum teleportation, the Pauli correction bits $(m_1,m_2)$ are transmitted over a classical channel. In realistic settings, this channel may be subject to partial interception, delayed compromise, or correlated exposure of the two bits. We model such leakage by introducing a stochastic process $K(t)\subseteq\{1,2\}$ that specifies which subset of correction bits has become available to an adversary by time $t$. The resulting knowledge state $K(t)$ determines the ensemble of quantum states on Bob's system from the adversary's perspective, which in turn determines both the residual hidden information and the adversary's optimal reconstruction performance.

We consider four stochastic leakage models: independent exponential leakage, sequential leakage, burst leakage, and correlated leakage. These models provide phenomenological descriptions of partial and time delayed disclosure of the classical correction bits during quantum teleportation~\cite{Ruijters2019,Zeng2023,Karr1984,2002}. For each model, closed form expressions for the time dependent probabilities of the adversary's knowledge classes $K(t)$ are obtained, enabling explicit evaluation of the expected Holevo Information and the corresponding state reconstruction fidelity.



\subsection{Independent Exponential Leakage Model}
\label{subsec:indep-exp}

We consider a leakage scenario in which the two classical Pauli correction bits
$(m_1,m_2)$ are compromised independently over time.  
This model captures
memoryless side-channel leakage through exponential waiting-time distributions.

The input qubit to be teleported by Alice is in the unknown state,
\begin{equation}
\ket{\psi}
=
\alpha\ket{0}+\beta\ket{1},
\qquad
|\alpha|^2+|\beta|^2=1 .
\end{equation}
As described earlier, Bob’s half of the shared Bell pair undergoes amplitude damping with parameter
$\gamma\in[0,1]$.  
Conditioned on Alice’s Bell measurement outcome $(m_1,m_2)\in\{0,1\}^2$, Bob’s
normalized post-measurement state is given by,
\begin{equation}
\rho_{m_1m_2}
=
X^{m_1} Z^{m_2}\,
\rho_{00}\,
Z^{m_2} X^{m_1},
\end{equation}
where the reference state corresponding to $(m_1,m_2)=(0,0)$ is
\begin{equation}
\rho_{00}
=
\begin{pmatrix}
a & c\\
c^* & b
\end{pmatrix},
\end{equation}
with
\begin{align}
a &= |\alpha|^2+\gamma|\beta|^2, \\
b &= (1-\gamma)|\beta|^2, \\
c &= \alpha\beta^*\sqrt{1-\gamma}.
\end{align}

We introduce the following auxiliary quantities to simplify our calculations (See Appendices \ref{app:holevo-derivation}  and \ref{app:fidelity} for details),
\begin{align}
r &:= \Re(c), \\
D &:= ab-|c|^2, \\
\delta &:= \sqrt{\,1-4D\,}.
\end{align}
The eigenvalues of $\rho_{00}$ are,
\begin{equation}
\lambda_{\pm}
=
\frac{1}{2}(1\pm\delta),
\end{equation}
and since all states $\rho_{m_1m_2}$ are related to $\rho_{00}$ by Pauli
conjugation, they share the same spectrum.  
Consequently, the von Neumann entropy of every conditional state is,
\begin{equation}
S(\rho_{m_1m_2}) = h(\delta),
\end{equation}
where $h(\cdot)$ denotes the binary entropy function.

Classical leakage is modeled by assigning independent exponential waiting times
to the correction bits. Let
\begin{equation}
T_1 \sim \mathrm{Exp}(k_1),
\qquad
T_2 \sim \mathrm{Exp}(k_2),
\end{equation}
be the leakage times of $m_1$ and $m_2$, respectively.  
At time $t$, the adversary’s knowledge set is,
\begin{equation}
K(t)=\{\,i\in\{1,2\}: T_i\le t\,\}.
\end{equation}
The probabilities of the four knowledge classes are,
\begin{align}
P_{\varnothing}(t)
&=
\Pr(K(t)=\varnothing)
=
e^{-(k_1+k_2)t},
\\[4pt]
P_{\{1\}}(t)
&=
\Pr(K(t)=\{1\})
=
(1-e^{-k_1 t})\,e^{-k_2 t},
\\[4pt]
P_{\{2\}}(t)
&=
\Pr(K(t)=\{2\})
=
e^{-k_1 t}\,(1-e^{-k_2 t}),
\\[4pt]
P_{\{1,2\}}(t)
&=
(1-e^{-k_1 t})(1-e^{-k_2 t}).
\end{align}

For each knowledge class, Eve observes an ensemble obtained by averaging over the
unknown correction bits. The corresponding Holevo quantities are (See Appendix~\ref{app:holevo-derivation} for details), 
\begin{align}
\chi_{\varnothing}
&=
1-h(\delta),
\\
\chi_{\{1\}}
&=
h(|1-2a|)-h(\delta),
\\
\chi_{\{2\}}
&=
h(2|r|)-h(\delta),
\\
\chi_{\{1,2\}}
&=
0 .
\end{align}
In this formulation, the Holevo Information measures the residual information about the teleported state that remains concealed from Eve because of uncertainty in the unrevealed correction bits.
The expected Holevo information at time $t$ is therefore,
\begin{align}
\mathbb{E}[\chi(t)]
&=
P_{\varnothing}(t)\,\chi_{\varnothing}
+
P_{\{1\}}(t)\,\chi_{\{1\}}
\nonumber\\
&\quad
+
P_{\{2\}}(t)\,\chi_{\{2\}} .
\label{eq:chi-indep-exp}
\end{align}

We also consider Eve's achievable reconstruction fidelity given partial knowledge of the correction bits together with access to Bob's quantum system (See Appendix~\ref{app:fidelity} ),
\begin{align}
F_{\varnothing}
&=
\frac{1}{2},
\\
F_{\{1\}}
&=
\frac{1}{2}\bigl(1+|1-2a|\bigr),
\\
F_{\{2\}}
&=
\frac{1}{2}+|r|,
\\
F_{\{1,2\}}
&=
\frac{1}{2}(1+\delta).
\end{align}
Averaging over the leakage process yields,
\begin{align}
F(t)
&=
P_{\varnothing}(t)\,F_{\varnothing}
+
P_{\{1\}}(t)\,F_{\{1\}}
\nonumber\\
&\quad
+
P_{\{2\}}(t)\,F_{\{2\}}
+
P_{\{1,2\}}(t)\,F_{\{1,2\}} .
\label{eq:F-indep-exp}
\end{align}

\begin{figure}[h!]
    \centering
    \includegraphics[width=0.48\textwidth]{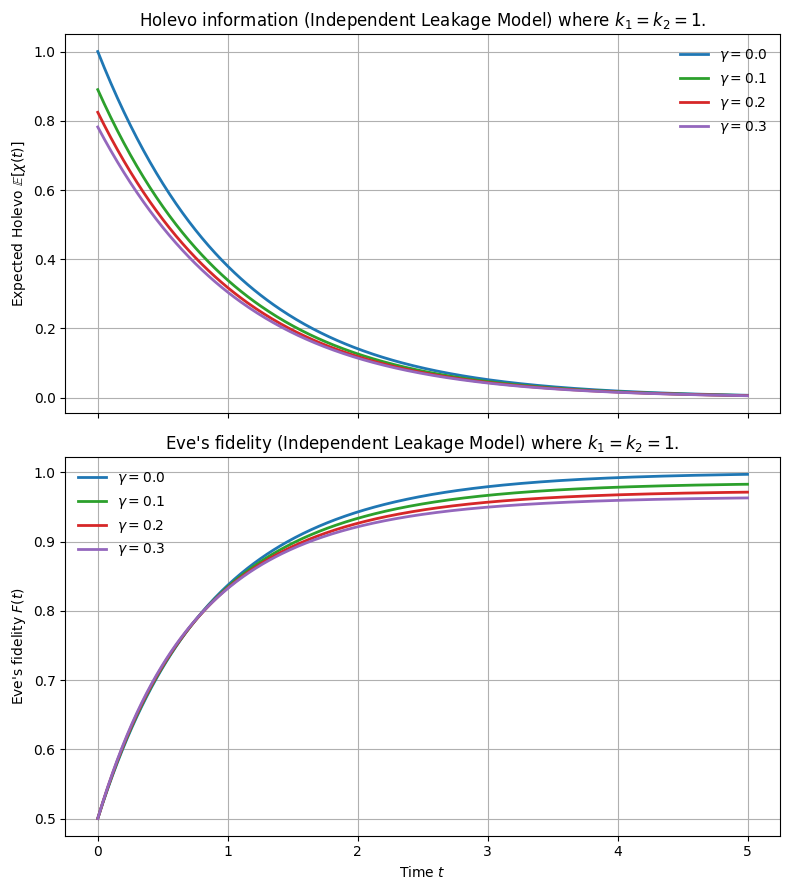}

\caption{Expected Holevo Information \(\mathbb{E}[\chi(t)]\) and Eve's reconstruction fidelity \(F(t)\) under the independent exponential leakage model with symmetric leakage rates \(k_1=k_2=1\),  $\alpha = \sqrt{0.6}$
and
$\beta = \sqrt{0.4}$. As the correction bits are progressively leaked, the residual hidden information decreases with time, while Eve's achievable reconstruction fidelity increases. Different curves correspond to different amplitude damping parameters \(\gamma\).}
    

    \label{fig:ilm}
\end{figure}


Fig.~\ref{fig:ilm} shows that, under independent exponential leakage, the residual hidden information decreases monotonically over time as Eve gains access to more correction data, while her reconstruction fidelity correspondingly increases. Larger values of the amplitude damping parameter \(\gamma\)  degrade the Bob's state and lower the maximum fidelity achievable by Eve.

At any time \(t\), the subset \(K(t)\subseteq\{1,2\}\) of leaked keys determines the ensemble of quantum states accessible to Eve and, consequently, the residual information about the teleported state that remains hidden from her. Because the two leakage processes are independent, partial disclosure events occur without temporal correlation. This model captures side channel scenarios in which the two classical correction bits are compromised at unrelated times.




\subsection{Sequential Leakage Model}
\label{subsec:seq-leak}

In the sequential leakage model, the second correction bit \(m_2\) can be exposed
only after the first correction bit \(m_1\) has leaked. This captures scenarios
in which the disclosure of classical information follows a fixed order, for
example due to protocol structure or layered side-channel vulnerabilities.

Let \(T_1 \sim \mathrm{Exp}(k_1)\) denote the random time at which \(m_1\) leaks,
and let \(T_2 \sim \mathrm{Exp}(k_2)\) denote the additional waiting time required
for \(m_2\) to leak after \(m_1\) has become known. The total time to learn both
keys is therefore \(T_1 + T_2\), where \(T_1\) and \(T_2\) are independent random
variables.

At time \(t\), the probabilities of the possible knowledge classes are:
\begin{align}
P_{\varnothing}(t)
&= \Pr(T_1 > t)
= e^{-k_1 t}, \\[4pt]
P_{\{1\}}(t)
&= \Pr(T_1 \le t < T_1 + T_2) \notag\\
&=
\begin{cases}
\displaystyle
\frac{k_1}{k_2 - k_1}
\bigl(e^{-k_1 t} - e^{-k_2 t}\bigr),
& k_1 \neq k_2, \\[8pt]
k t\, e^{-k t},
& k_1 = k_2 = k,
\end{cases} \\[6pt]
P_{\{1,2\}}(t)
&= 1 - P_{\varnothing}(t) - P_{\{1\}}(t).
\end{align}

Because \(m_2\) cannot leak before \(m_1\), the knowledge class \(\{2\}\) never
occurs in this model. The set of possible knowledge states is therefore
\(\{\varnothing, \{1\}, \{1,2\}\}\).

The Holevo quantities and fidelities associated with each class coincide with
those derived previously, except that the contributions \(\chi_{\{2\}}\) and
\(F_{\{2\}}\) are absent. The expected Holevo Information at time \(t\) is therefore,
\begin{equation}  
\mathbb{E}[\chi(t)]
=
P_{\varnothing}(t)\,\chi_{\varnothing}
+
P_{\{1\}}(t)\,\chi_{\{1\}},
\end{equation}

There is no \(P_{\{1,2\}}(t)\chi_{\{1,2\}}\) term because \(\chi_{\{1,2\}}=0\) (See Appendix~\ref{app:holevo-derivation}).

while the corresponding expected fidelity of Eve’s optimal reconstruction is,
\begin{equation}
F(t)
=
P_{\varnothing}(t)\,F_{\varnothing}
+
P_{\{1\}}(t)\,F_{\{1\}}
+
P_{\{1,2\}}(t)\,F_{\{1,2\}}.
\end{equation}

\begin{figure}[t]
    \centering
    \includegraphics[width=\columnwidth]{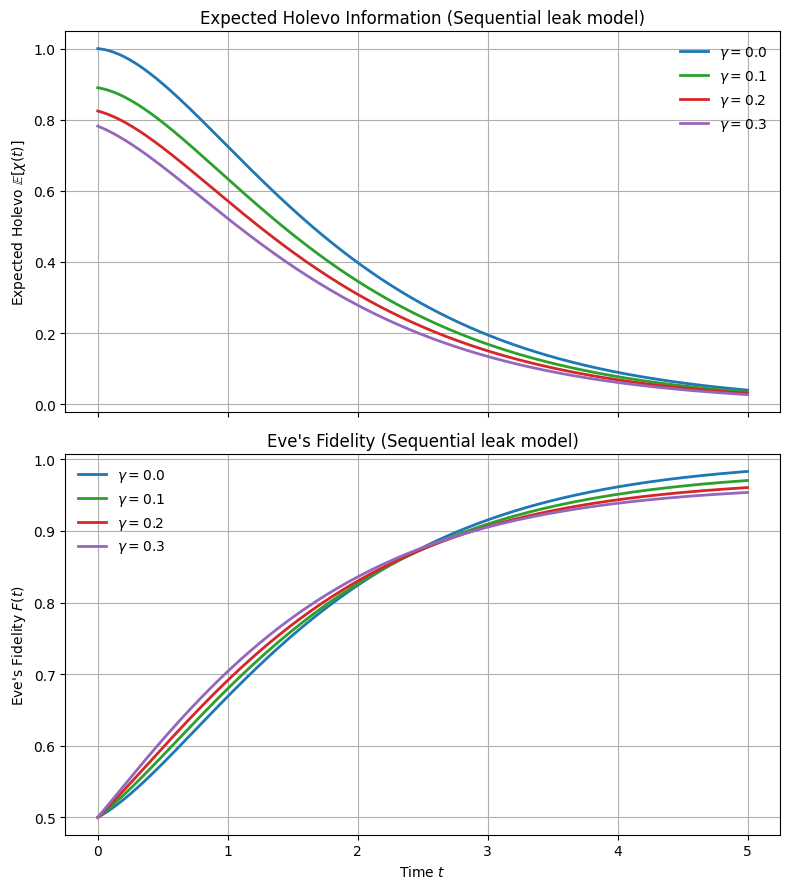}
    \caption{
    Expected Holevo information (top) and Eve’s optimal fidelity (bottom)
    as functions of time in the sequential leakage model with $k_1 = k_2 = 1$,  $\alpha = \sqrt{0.6}$
and
$\beta = \sqrt{0.4}$.
    Curves correspond to different amplitude-damping strengths $\gamma$.}
    \label{fig:seq-leak}
\end{figure}




Fig.~\ref{fig:seq-leak} top subplot shows the expected Holevo information \(\mathbb{E}[\chi(t)]\) in the sequential leakage model. At \(t=0\), Eve has no access to either correction bit, so the Holevo information starts at its maximum value \(\chi_{\varnothing}=1-h(\delta)\). As time increases, leakage of the first key \(m_1\) reduces Eve's uncertainty, causing \(\mathbb{E}[\chi(t)]\) to decrease. Because the second key cannot leak before the first, the decay is constrained by the ordered sequence \(\varnothing \to \{1\} \to \{1,2\}\), in contrast to the independent leakage model. Larger values of the damping parameter \(\gamma\) reduce the initial value of the Holevo information, reflecting the fact that stronger amplitude damping already degrades Bob's state before any classical leakage occurs.

The Fig.~\ref{fig:seq-leak} bottom subplot shows Eve's optimal fidelity \(F(t)\). Initially, Eve's best strategy is random guessing, yielding \(F(0)=1/2\). As the first correction key leaks, partial classical information improves Eve's reconstruction, and the fidelity increases monotonically. At long times, when both correction keys have leaked with high probability, the fidelity approaches the asymptotic value \((1+\delta)/2\). 




We make two observations from Fig.~\ref{fig:seq-leak}. First, the expected Holevo information decreases monotonically with time as more correction information becomes available to Eve. For the chosen input state, increasing \(\gamma\) decreases the Holevo contribution from each relevant knowledge class, so the ordering of the curves is preserved and no crossing occurs.

Second, the fidelity curves exhibit a crossing behavior. In this model, the expected fidelity is determined by a weighted combination of the class-dependent fidelities \(F_{\{1\}}\) and \(F_{\{1,2\}}\), which depend differently on \(\gamma\). Over the plotted range, the partial-leakage fidelity \(F_{\{1\}}\) increases with \(\gamma\), whereas the full-information fidelity \(F_{\{1,2\}}\) decreases. At early times, the \(\{1\}\) knowledge class carries a larger probability weight, while at later times the \(\{1,2\}\) class becomes dominant. The crossing therefore reflects the shift in probability weight from partial leakage to full leakage as time increases.

A similar crossing can also occur in the independent leakage model shown in Fig.~\ref{fig:ilm}. However, in that case Eve's expected fidelity receives contributions from all four knowledge classes, including the additional \(\{2\}\) class, whereas in the sequential leakage model the evolution is restricted to the ordered sequence \(\varnothing \to \{1\} \to \{1,2\}\). Thus, although both models can exhibit fidelity crossings, the behavior in the sequential leakage model is shaped by the enforced order of key disclosure.



\subsection{Burst Leakage Model}
\label{subsec:burst-leak}

In the burst leakage model, both Pauli correction bits are compromised
simultaneously at a single random time. This scenario captures sudden
side-channel failures or catastrophic disclosure events in which all classical
correction information becomes available at once.

Let $T \sim \mathrm{Exp}(k)$ denote the random leakage time at which both bits
$(m_1,m_2)$ are revealed. Prior to this event, Eve has no classical information,
while after $T$ she possesses complete knowledge of the correction keys. The
resulting knowledge class probabilities are,
\begin{align}
P_{\varnothing}(t) &= \Pr(T>t) = e^{-k t}, \\
P_{\{1,2\}}(t) &= \Pr(T\le t) = 1 - e^{-k t}.
\end{align}

Since no intermediate leakage of a single key is possible, the expected Holevo
information simplifies to,
\begin{align}
\mathbb{E}[\chi(t)]
&= P_{\varnothing}(t)\,\chi_{\varnothing} \notag\\
&= e^{-k t}\,\chi_{\varnothing},
\end{align}
where $\chi_{\varnothing}=1-h(\delta)$ denotes the Holevo information when no
correction bits are known.

Similarly, Eve’s optimal fidelity evolves as,
\begin{align}
F(t)
&= P_{\varnothing}(t)F_{\varnothing}
   + P_{\{1,2\}}(t)F_{\{1,2\}} \notag\\
&= e^{-k t}F_{\varnothing}
   + \bigl(1-e^{-k t}\bigr)F_{\{1,2\}},
\end{align}
interpolating between random guessing,
$F_{\varnothing}=\tfrac{1}{2}$, and the maximum achievable fidelity
$F_{\{1,2\}}=\tfrac{1+\delta}{2}$ once both correction keys are disclosed.


\begin{figure}[t]
    \centering
    \includegraphics[width=0.48\textwidth]{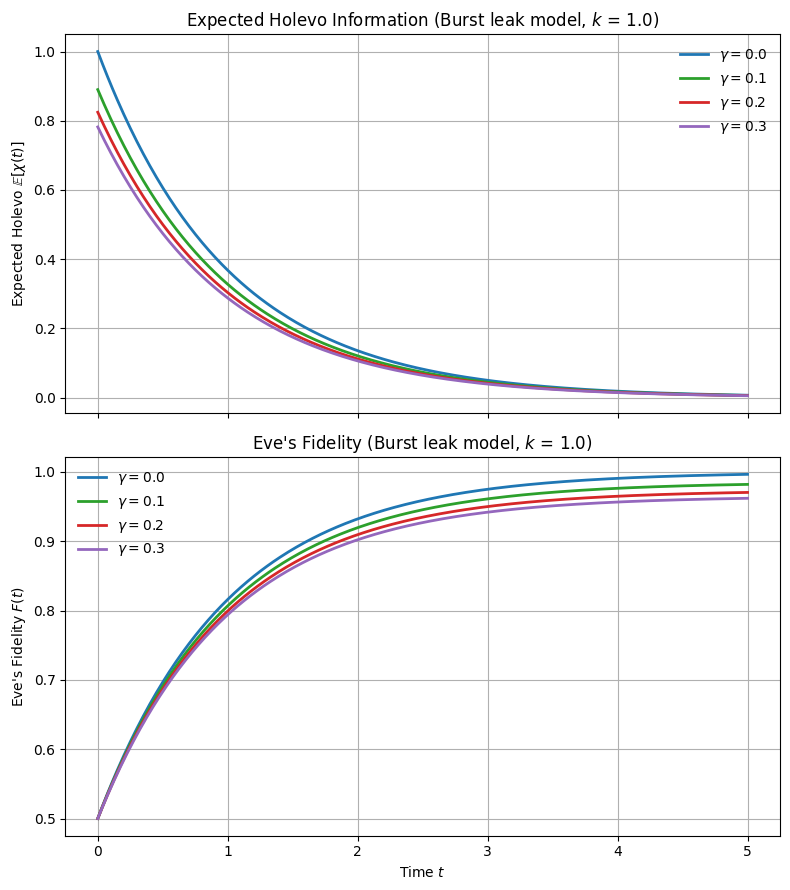}
    \caption{Time evolution of Eve’s expected Holevo information (top) and
    optimal fidelity (bottom) in the burst leakage model with leakage rate
    $k=1.0$,  $\alpha = \sqrt{0.6}$
and
$\beta = \sqrt{0.4}$. Each curve corresponds to a fixed amplitude damping parameter
    $\gamma$. In this model, both Pauli correction bits are revealed simultaneously at a single random time, producing a direct transition from no knowledge to full classical disclosure.}
    \label{fig:burst-leak}
\end{figure}

Figure~\ref{fig:burst-leak} illustrates the temporal behavior of Eve's expected
Holevo information and reconstruction capability in the burst leakage model.
Since both correction bits leak simultaneously at a single exponentially
distributed time, there is no intermediate regime corresponding to partial key
knowledge.

The top panel shows the expected Holevo information
\(\mathbb{E}[\chi(t)] = e^{-k t}\chi_{\varnothing}\), which decays exponentially
from its initial value \(\chi_{\varnothing}=1-h(\delta)\) toward zero. Larger
values of the amplitude damping parameter \(\gamma\) reduce the initial Holevo
information, while the decay rate is determined solely by the leakage rate \(k\).

The bottom panel shows Eve's optimal fidelity,
which rises monotonically from the random guessing value
\(F_{\varnothing}=\tfrac{1}{2}\) toward the saturation value
\(F_{\{1,2\}}=\tfrac{1+\delta}{2}\). Unlike the independent or sequential leakage
models, the burst model has no intermediate partial-leakage phase.

Together, these curves highlight the defining feature of burst leakage: a direct
transition from complete concealment of the correction data to full disclosure,
together with the corresponding increase in Eve's reconstruction fidelity.





\subsection{Correlated Leakage Model}
\label{subsec:corr-leak}

In the correlated leakage model, disclosure of the two Pauli correction bits is neither fully independent nor perfectly simultaneous. Instead, the leakage dynamics consist of a shared component that can reveal both bits at once, together with independent components that can reveal each bit separately. This construction continuously interpolates between the independent leakage and burst leakage limits.

We assume identical marginal leakage rates for the two correction bits, \(k_1=k_2=k\), and introduce a correlation parameter \(\mu\in[0,1]\). Let
\begin{align}
T_{\mathrm{sh}} &\sim \mathrm{Exp}(\mu k), \notag\\
T_1^{\mathrm{ind}} &\sim \mathrm{Exp}((1-\mu)k), \notag\\
T_2^{\mathrm{ind}} &\sim \mathrm{Exp}((1-\mu)k)
\end{align}
be three independent exponential random variables. Here, \(T_{\mathrm{sh}}\) represents a shared leakage mechanism that reveals both correction bits simultaneously, while \(T_i^{\mathrm{ind}}\) represents independent leakage of the individual bit \(m_i\).

The effective leakage time of bit \(m_i\) is defined by
\begin{equation}
T_i=\min\!\bigl(T_{\mathrm{sh}},\,T_i^{\mathrm{ind}}\bigr),
\qquad i\in\{1,2\}.
\end{equation}
Thus, each correction bit may leak either through the shared channel or through its own independent channel, whichever occurs first. By construction, each effective leakage time \(T_i\) has marginal distribution \(\mathrm{Exp}(k)\), while the dependence between \(T_1\) and \(T_2\) is controlled by the shared component \(T_{\mathrm{sh}}\).

When \(\mu=0\), the shared leakage channel is absent and the two bits leak independently at rate \(k\), recovering the independent leakage model. When \(\mu=1\), only the shared leakage channel remains, so both bits are revealed simultaneously at rate \(k\), recovering the burst leakage model.

At time \(t\), the probability that no correction bit has leaked is the probability that none of the three underlying leakage processes has occurred:
\begin{align}
P_{\varnothing}(t)
&=\Pr(T_{\mathrm{sh}}>t)\,
  \Pr(T_1^{\mathrm{ind}}>t)\,
  \Pr(T_2^{\mathrm{ind}}>t) \notag\\
&=e^{-\mu k t}\,e^{-(1-\mu)k t}\,e^{-(1-\mu)k t} \notag\\
&=e^{-(2-\mu)k t}.
\end{align}

The probability that only the \(X\)-correction bit \(m_1\) has leaked is obtained by requiring that the shared leakage has not occurred, that \(m_1\) has leaked through its independent channel, and that \(m_2\) has not leaked:
\begin{align}
P_{\{1\}}(t)
&=\Pr(T_{\mathrm{sh}}>t)\,
  \Pr(T_1^{\mathrm{ind}}\le t)\,
  \Pr(T_2^{\mathrm{ind}}>t) \notag\\
&=e^{-\mu k t}\,
  \bigl(1-e^{-(1-\mu)k t}\bigr)\,
  e^{-(1-\mu)k t}.
\end{align}
By symmetry, the probability that only the \(Z\)-correction bit \(m_2\) has leaked is
\begin{align}
P_{\{2\}}(t)
&=\Pr(T_{\mathrm{sh}}>t)\,
  \Pr(T_1^{\mathrm{ind}}>t)\,
  \Pr(T_2^{\mathrm{ind}}\le t) \notag\\
&=e^{-\mu k t}\,
  e^{-(1-\mu)k t}\,
  \bigl(1-e^{-(1-\mu)k t}\bigr).
\end{align}

Finally, the probability that both correction bits have leaked is obtained by normalization:
\begin{equation}
P_{\{1,2\}}(t)
=
1-P_{\varnothing}(t)-P_{\{1\}}(t)-P_{\{2\}}(t).
\end{equation}
This event includes contributions both from the shared leakage channel and from the case in which the two independent leakage channels have both occurred by time \(t\).

The expected Holevo information and Eve's optimal reconstruction fidelity are obtained by weighting the class dependent quantities \(\chi_K\) and \(F_K\) by the corresponding knowledge class probabilities:
\begin{align}
\mathbb{E}[\chi(t)]
&=
\sum_{K\subseteq\{1,2\}} P_K(t)\,\chi_K, \\
F(t)
&=
\sum_{K\subseteq\{1,2\}} P_K(t)\,F_K .
\end{align}

The correlated leakage model therefore interpolates between the independent and burst leakage limits, allowing the effect of partial correlation in classical side-channel leakage to be quantified in a controlled manner.

\begin{figure}[h!]
    \centering
    \includegraphics[width=0.48\textwidth]{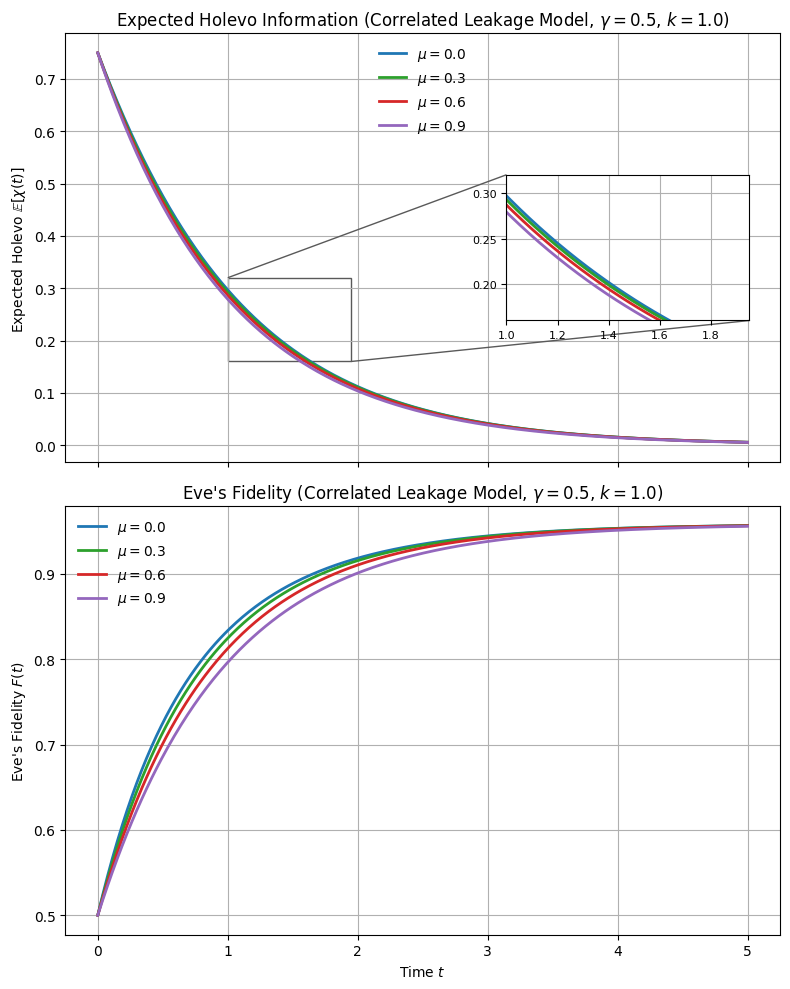}
    \caption{Time evolution of the expected Holevo information
    \(\mathbb{E}[\chi(t)]\) (top) and Eve's optimal fidelity \(F(t)\) (bottom)
    under the correlated leakage model for fixed damping parameter
    \(\gamma=0.5\), leakage rate \(k=1\), \(\alpha=\sqrt{0.6}\), and
    \(\beta=\sqrt{0.4}\). The curves correspond to different values of the
    correlation parameter \(\mu\).}
    \label{fig:correlated-leakage}
\end{figure}

Fig.~\ref{fig:correlated-leakage} illustrates the effect of correlation in classical key leakage on Eve's expected Holevo information and optimal fidelity. The top subplot shows \(\mathbb{E}[\chi(t)]\), which decreases monotonically with time for all values of \(\mu\). The curves remain close, but larger values of \(\mu\) produce a slightly faster decay. The bottom panel shows Eve's optimal fidelity \(F(t)\), which increases monotonically with time for all values of \(\mu\). In this case, larger values of \(\mu\) lead to a slightly slower increase at early and intermediate times, while all curves approach nearly the same asymptotic value at long times.

\section{Conclusion and Future Work}
\label{sec:conclusion}

In this work, we introduced and analyzed the Quantum-Resistant Quantum Teleportation (QRQT) framework, in which the classical correction channel of quantum teleportation is protected using lattice based post quantum cryptographic primitives. We investigated the security of teleportation in the presence of time dependent leakage of the classical correction information and studied how delayed or partial exposure of the correction bits affects both the information available to an adversary and the quality of the reconstructed state. To capture imperfections in the quantum channel, we also incorporated decoherence in the shared entangled resource by modeling an amplitude damping channel acting on Bob's half of the Bell pair.

To study the effect of classical leakage, we introduced four stochastic leakage models: independent exponential leakage, sequential leakage, burst leakage, and correlated leakage. For each model, we derived explicit time dependent expressions for the adversary's knowledge classes and used them to evaluate the expected Holevo information and Eve's reconstruction fidelity. These results show that the temporal structure of leakage plays a central role in determining both the secrecy of the correction data and the adversary's ability to reconstruct the teleported state.

A central conclusion of our analysis is that quantum memory coherence acts as the hidden bottleneck coupling physical and computational security. It simultaneously limits the achievable communication distance, constrains the tolerable post quantum cryptographic overhead, and bounds the adversary's useful attack window. Under realistic parameters, including a 1 ms coherence time and fiber optic propagation, the maximum secure teleportation distance ranges from approximately 191 km for FrodoKEM-1344 to approximately 199 km for Kyber512. We also found that the joint classical and quantum attack probability, $P_{\mathrm{joint}}(t)$, exhibits a characteristic bell shaped profile arising from the competing time dependence of classical cryptanalysis and quantum decoherence, thereby establishing a bounded optimal attack window beyond which adversarial success decays exponentially.

While the present analysis provides useful insights, it also relies on some modeling assumptions that should be considered. First, the quantum noise model is restricted to amplitude damping on one qubit of the Bell pair, although the appropriate noise model may vary depending on the qubit platform and its interaction with the surrounding environment,  and the quantum memory decoherence model adopts a single exponential decay, $F(t)=F_0 e^{-t/T_{\mathrm{coh}}}$, which is appropriate for Markovian noise but may not adequately capture systems with non Markovian memory effects, multi exponential relaxation, or additional device level imperfections. Second, the stochastic leakage models considered here are phenomenological and are intended to capture representative patterns of classical compromise rather than all implementation specific attack mechanisms. Third, our analysis assumes a strong threat model in which the adversary can gain access to Bob's quantum system in addition to leaked classical information, and the SWAP attack model further assumes an idealized adversary capable of performing a perfect unitary swap without introducing detectable disturbance. In practice, entanglement verification, timing analysis, and decoy state style countermeasures may constrain such attacks. Finally, the cryptographic protection in QRQT relies on the conjectured hardness of lattice problems such as LWE and SIS rather than on information theoretic security, so future advances in lattice algorithms or quantum computing could alter the long term security landscape.

Several directions remain for future work. A natural extension is to study more general and realistic quantum noise models, including depolarizing noise, dephasing noise, qubit correlated noise, and non-Markovian memory dynamics, as well as more realistic multi-mode quantum memory architectures~\cite{casey2025multimodequantummemorieshighthroughput}. It would also be valuable to investigate implementation driven leakage models grounded in specific hardware platforms, communication stacks, and side channel assumptions. Another important direction is to extend QRQT to multi-hop teleportation, repeater based quantum networks, and distributed quantum computing architectures, where leakage timing, memory coherence, and cryptographic latency may interact in more complex ways and where post quantum cryptographic overhead may accumulate across intermediate nodes. Experimental validation of QRQT on photonic or trapped ion platforms, even over short distances, would provide useful benchmarks for the latency and coherence tradeoffs predicted here. It is also worth exploring hybrid approaches that combine QKD derived symmetric keys for initial authentication with post quantum cryptography for long term classical channel protection. More broadly, as NIST standardized post quantum schemes mature and quantum memory technologies continue to improve, revisiting the security and distance tradeoff with updated parameters will be essential for assessing the practical deployment of QRQT in future quantum networks.

\vspace{10pt} \noindent \textsf{\textbf{Acknowledgements.}}--- JL is supported in part by the University of Pittsburgh, School of Computing and Information, Department of Computer Science, Pitt Cyber, Pitt Momentum fund, PQI Community Collaboration Awards, John C. Mascaro Faculty Scholar in Sustainability, NSF award 2535915, Thinking Machines Lab and Cisco Research. This research used resources of the Oak Ridge Leadership Computing Facility, which is a DOE Office of Science User Facility supported under Contract DE-AC05-00OR22725 KPS thanks the U.S. Department of Energy, Office of Science, Advanced Scientific Computing Research (ASCR) program, for support under Award Number DE-SC0026264, Cisco Systems, and PQI Community Collaboration Awards.

\bibliography{apssamp}

\begin{thebibliography}{95}%
\makeatletter
\providecommand \@ifxundefined [1]{%
 \@ifx{#1\undefined}
}%
\providecommand \@ifnum [1]{%
 \ifnum #1\expandafter \@firstoftwo
 \else \expandafter \@secondoftwo
 \fi
}%
\providecommand \@ifx [1]{%
 \ifx #1\expandafter \@firstoftwo
 \else \expandafter \@secondoftwo
 \fi
}%
\providecommand \natexlab [1]{#1}%
\providecommand \enquote  [1]{``#1''}%
\providecommand \bibnamefont  [1]{#1}%
\providecommand \bibfnamefont [1]{#1}%
\providecommand \citenamefont [1]{#1}%
\providecommand \href@noop [0]{\@secondoftwo}%
\providecommand \href [0]{\begingroup \@sanitize@url \@href}%
\providecommand \@href[1]{\@@startlink{#1}\@@href}%
\providecommand \@@href[1]{\endgroup#1\@@endlink}%
\providecommand \@sanitize@url [0]{\catcode `\\12\catcode `\$12\catcode
  `\&12\catcode `\#12\catcode `\^12\catcode `\_12\catcode `\%12\relax}%
\providecommand \@@startlink[1]{}%
\providecommand \@@endlink[0]{}%
\providecommand \url  [0]{\begingroup\@sanitize@url \@url }%
\providecommand \@url [1]{\endgroup\@href {#1}{\urlprefix }}%
\providecommand \urlprefix  [0]{URL }%
\providecommand \Eprint [0]{\href }%
\providecommand \doibase [0]{https://doi.org/}%
\providecommand \selectlanguage [0]{\@gobble}%
\providecommand \bibinfo  [0]{\@secondoftwo}%
\providecommand \bibfield  [0]{\@secondoftwo}%
\providecommand \translation [1]{[#1]}%
\providecommand \BibitemOpen [0]{}%
\providecommand \bibitemStop [0]{}%
\providecommand \bibitemNoStop [0]{.\EOS\space}%
\providecommand \EOS [0]{\spacefactor3000\relax}%
\providecommand \BibitemShut  [1]{\csname bibitem#1\endcsname}%
\let\auto@bib@innerbib\@empty
\bibitem [{\citenamefont {Dahlberg}\ \emph {et~al.}(2019)\citenamefont
  {Dahlberg}, \citenamefont {Skrzypczyk}, \citenamefont {Coopmans},
  \citenamefont {Wubben}, \citenamefont {Rozpdek}, \citenamefont {Pompili},
  \citenamefont {Stolk}, \citenamefont {Paweczak}, \citenamefont {Knegjens},
  \citenamefont {de~Oliveira~Filho}, \citenamefont {Hanson},\ and\
  \citenamefont {Wehner}}]{Dahlberg2019}%
  \BibitemOpen
  \bibfield  {author} {\bibinfo {author} {\bibfnamefont {A.}~\bibnamefont
  {Dahlberg}}, \bibinfo {author} {\bibfnamefont {M.}~\bibnamefont
  {Skrzypczyk}}, \bibinfo {author} {\bibfnamefont {T.}~\bibnamefont
  {Coopmans}}, \bibinfo {author} {\bibfnamefont {L.}~\bibnamefont {Wubben}},
  \bibinfo {author} {\bibfnamefont {F.}~\bibnamefont {Rozpdek}}, \bibinfo
  {author} {\bibfnamefont {M.}~\bibnamefont {Pompili}}, \bibinfo {author}
  {\bibfnamefont {A.}~\bibnamefont {Stolk}}, \bibinfo {author} {\bibfnamefont
  {P.}~\bibnamefont {Paweczak}}, \bibinfo {author} {\bibfnamefont
  {R.}~\bibnamefont {Knegjens}}, \bibinfo {author} {\bibfnamefont
  {J.}~\bibnamefont {de~Oliveira~Filho}}, \bibinfo {author} {\bibfnamefont
  {R.}~\bibnamefont {Hanson}},\ and\ \bibinfo {author} {\bibfnamefont
  {S.}~\bibnamefont {Wehner}},\ }\bibfield  {title} {\bibinfo {title} {A link
  layer protocol for quantum networks},\ }in\ \href
  {https://doi.org/10.1145/3341302.3342070} {\emph {\bibinfo {booktitle}
  {Proceedings of the ACM Special Interest Group on Data Communication}}},\
  \bibinfo {series and number} {SIGCOMM -?9}\ (\bibinfo  {publisher} {ACM},\
  \bibinfo {year} {2019})\ pp.\ \bibinfo {pages} {159--?73}\BibitemShut
  {NoStop}%
\bibitem [{\citenamefont {Valivarthi}\ \emph {et~al.}(2020)\citenamefont
  {Valivarthi}, \citenamefont {Davis}, \citenamefont {Pea}, \citenamefont
  {Xie}, \citenamefont {Lauk}, \citenamefont {Narvez}, \citenamefont
  {Allmaras}, \citenamefont {Beyer}, \citenamefont {Gim}, \citenamefont
  {Hussein}, \citenamefont {Iskander}, \citenamefont {Kim}, \citenamefont
  {Korzh}, \citenamefont {Mueller}, \citenamefont {Rominsky}, \citenamefont
  {Shaw}, \citenamefont {Tang}, \citenamefont {Wollman}, \citenamefont {Simon},
  \citenamefont {Spentzouris}, \citenamefont {Oblak}, \citenamefont
  {Sinclair},\ and\ \citenamefont {Spiropulu}}]{Valivarthi2020}%
  \BibitemOpen
  \bibfield  {author} {\bibinfo {author} {\bibfnamefont {R.}~\bibnamefont
  {Valivarthi}}, \bibinfo {author} {\bibfnamefont {S.~I.}\ \bibnamefont
  {Davis}}, \bibinfo {author} {\bibfnamefont {C.}~\bibnamefont {Pea}}, \bibinfo
  {author} {\bibfnamefont {S.}~\bibnamefont {Xie}}, \bibinfo {author}
  {\bibfnamefont {N.}~\bibnamefont {Lauk}}, \bibinfo {author} {\bibfnamefont
  {L.}~\bibnamefont {Narvez}}, \bibinfo {author} {\bibfnamefont {J.~P.}\
  \bibnamefont {Allmaras}}, \bibinfo {author} {\bibfnamefont {A.~D.}\
  \bibnamefont {Beyer}}, \bibinfo {author} {\bibfnamefont {Y.}~\bibnamefont
  {Gim}}, \bibinfo {author} {\bibfnamefont {M.}~\bibnamefont {Hussein}},
  \bibinfo {author} {\bibfnamefont {G.}~\bibnamefont {Iskander}}, \bibinfo
  {author} {\bibfnamefont {H.~L.}\ \bibnamefont {Kim}}, \bibinfo {author}
  {\bibfnamefont {B.}~\bibnamefont {Korzh}}, \bibinfo {author} {\bibfnamefont
  {A.}~\bibnamefont {Mueller}}, \bibinfo {author} {\bibfnamefont
  {M.}~\bibnamefont {Rominsky}}, \bibinfo {author} {\bibfnamefont
  {M.}~\bibnamefont {Shaw}}, \bibinfo {author} {\bibfnamefont {D.}~\bibnamefont
  {Tang}}, \bibinfo {author} {\bibfnamefont {E.~E.}\ \bibnamefont {Wollman}},
  \bibinfo {author} {\bibfnamefont {C.}~\bibnamefont {Simon}}, \bibinfo
  {author} {\bibfnamefont {P.}~\bibnamefont {Spentzouris}}, \bibinfo {author}
  {\bibfnamefont {D.}~\bibnamefont {Oblak}}, \bibinfo {author} {\bibfnamefont
  {N.}~\bibnamefont {Sinclair}},\ and\ \bibinfo {author} {\bibfnamefont
  {M.}~\bibnamefont {Spiropulu}},\ }\bibfield  {title} {\bibinfo {title}
  {Teleportation systems toward a quantum internet},\ }\bibfield  {journal}
  {\bibinfo  {journal} {PRX Quantum}\ }\textbf {\bibinfo {volume} {1}},\ \href
  {https://doi.org/10.1103/prxquantum.1.020317} {10.1103/prxquantum.1.020317}
  (\bibinfo {year} {2020})\BibitemShut {NoStop}%
\bibitem [{\citenamefont {Bennett}\ \emph {et~al.}(1993)\citenamefont
  {Bennett}, \citenamefont {Brassard}, \citenamefont {Crpeau}, \citenamefont
  {Jozsa}, \citenamefont {Peres},\ and\ \citenamefont
  {Wootters}}]{Bennett1993}%
  \BibitemOpen
  \bibfield  {author} {\bibinfo {author} {\bibfnamefont {C.~H.}\ \bibnamefont
  {Bennett}}, \bibinfo {author} {\bibfnamefont {G.}~\bibnamefont {Brassard}},
  \bibinfo {author} {\bibfnamefont {C.}~\bibnamefont {Crpeau}}, \bibinfo
  {author} {\bibfnamefont {R.}~\bibnamefont {Jozsa}}, \bibinfo {author}
  {\bibfnamefont {A.}~\bibnamefont {Peres}},\ and\ \bibinfo {author}
  {\bibfnamefont {W.~K.}\ \bibnamefont {Wootters}},\ }\bibfield  {title}
  {\bibinfo {title} {Teleporting an unknown quantum state via dual classical
  and einstein-podolsky-rosen channels},\ }\href
  {https://doi.org/10.1103/physrevlett.70.1895} {\bibfield  {journal} {\bibinfo
   {journal} {Physical Review Letters}\ }\textbf {\bibinfo {volume} {70}},\
  \bibinfo {pages} {1895} (\bibinfo {year} {1993})}\BibitemShut {NoStop}%
\bibitem [{\citenamefont {Kimble}(2008)}]{Kimble2008}%
  \BibitemOpen
  \bibfield  {author} {\bibinfo {author} {\bibfnamefont {H.~J.}\ \bibnamefont
  {Kimble}},\ }\bibfield  {title} {\bibinfo {title} {The quantum internet},\
  }\href {https://doi.org/10.1038/nature07127} {\bibfield  {journal} {\bibinfo
  {journal} {Nature}\ }\textbf {\bibinfo {volume} {453}},\ \bibinfo {pages}
  {1023} (\bibinfo {year} {2008})}\BibitemShut {NoStop}%
\bibitem [{\citenamefont {Wehner}\ \emph {et~al.}(2018)\citenamefont {Wehner},
  \citenamefont {Elkouss},\ and\ \citenamefont {Hanson}}]{wehner2018quantum}%
  \BibitemOpen
  \bibfield  {author} {\bibinfo {author} {\bibfnamefont {S.}~\bibnamefont
  {Wehner}}, \bibinfo {author} {\bibfnamefont {D.}~\bibnamefont {Elkouss}},\
  and\ \bibinfo {author} {\bibfnamefont {R.}~\bibnamefont {Hanson}},\
  }\bibfield  {title} {\bibinfo {title} {Quantum internet: A vision for the
  road ahead},\ }\bibfield  {journal} {\bibinfo  {journal} {Science}\ }\textbf
  {\bibinfo {volume} {362}},\ \href {https://doi.org/10.1126/science.aam9288}
  {10.1126/science.aam9288} (\bibinfo {year} {2018})\BibitemShut {NoStop}%
\bibitem [{\citenamefont {Sangouard}\ \emph {et~al.}(2011)\citenamefont
  {Sangouard}, \citenamefont {Simon}, \citenamefont {de~Riedmatten},\ and\
  \citenamefont {Gisin}}]{RevModPhys.83.33}%
  \BibitemOpen
  \bibfield  {author} {\bibinfo {author} {\bibfnamefont {N.}~\bibnamefont
  {Sangouard}}, \bibinfo {author} {\bibfnamefont {C.}~\bibnamefont {Simon}},
  \bibinfo {author} {\bibfnamefont {H.}~\bibnamefont {de~Riedmatten}},\ and\
  \bibinfo {author} {\bibfnamefont {N.}~\bibnamefont {Gisin}},\ }\bibfield
  {title} {\bibinfo {title} {Quantum repeaters based on atomic ensembles and
  linear optics},\ }\href {https://doi.org/10.1103/RevModPhys.83.33} {\bibfield
   {journal} {\bibinfo  {journal} {Rev. Mod. Phys.}\ }\textbf {\bibinfo
  {volume} {83}},\ \bibinfo {pages} {33} (\bibinfo {year} {2011})}\BibitemShut
  {NoStop}%
\bibitem [{\citenamefont {Cirac}\ \emph {et~al.}(1999)\citenamefont {Cirac},
  \citenamefont {Ekert}, \citenamefont {Huelga},\ and\ \citenamefont
  {Macchiavello}}]{PhysRevA.59.4249}%
  \BibitemOpen
  \bibfield  {author} {\bibinfo {author} {\bibfnamefont {J.~I.}\ \bibnamefont
  {Cirac}}, \bibinfo {author} {\bibfnamefont {A.~K.}\ \bibnamefont {Ekert}},
  \bibinfo {author} {\bibfnamefont {S.~F.}\ \bibnamefont {Huelga}},\ and\
  \bibinfo {author} {\bibfnamefont {C.}~\bibnamefont {Macchiavello}},\
  }\bibfield  {title} {\bibinfo {title} {Distributed quantum computation over
  noisy channels},\ }\href {https://doi.org/10.1103/PhysRevA.59.4249}
  {\bibfield  {journal} {\bibinfo  {journal} {Phys. Rev. A}\ }\textbf {\bibinfo
  {volume} {59}},\ \bibinfo {pages} {4249} (\bibinfo {year}
  {1999})}\BibitemShut {NoStop}%
\bibitem [{\citenamefont {Van~Meter}(2014)}]{Van2014quantum}%
  \BibitemOpen
  \bibfield  {author} {\bibinfo {author} {\bibfnamefont {R.}~\bibnamefont
  {Van~Meter}},\ }\href {https://doi.org/10.1002/9781118648919} {\emph
  {\bibinfo {title} {Quantum Networking}}}\ (\bibinfo  {publisher} {Wiley},\
  \bibinfo {year} {2014})\BibitemShut {NoStop}%
\bibitem [{\citenamefont {Hermans}\ \emph {et~al.}(2022)\citenamefont
  {Hermans}, \citenamefont {Pompili}, \citenamefont {Beukers}, \citenamefont
  {Baier}, \citenamefont {Borregaard},\ and\ \citenamefont
  {Hanson}}]{Hermans2022}%
  \BibitemOpen
  \bibfield  {author} {\bibinfo {author} {\bibfnamefont {S.~L.~N.}\
  \bibnamefont {Hermans}}, \bibinfo {author} {\bibfnamefont {M.}~\bibnamefont
  {Pompili}}, \bibinfo {author} {\bibfnamefont {H.~K.~C.}\ \bibnamefont
  {Beukers}}, \bibinfo {author} {\bibfnamefont {S.}~\bibnamefont {Baier}},
  \bibinfo {author} {\bibfnamefont {J.}~\bibnamefont {Borregaard}},\ and\
  \bibinfo {author} {\bibfnamefont {R.}~\bibnamefont {Hanson}},\ }\bibfield
  {title} {\bibinfo {title} {Qubit teleportation between non-neighbouring nodes
  in a quantum network},\ }\href {https://doi.org/10.1038/s41586-022-04697-y}
  {\bibfield  {journal} {\bibinfo  {journal} {Nature}\ }\textbf {\bibinfo
  {volume} {605}},\ \bibinfo {pages} {663} (\bibinfo {year}
  {2022})}\BibitemShut {NoStop}%
\bibitem [{\citenamefont {Gangopadhyay}\ \emph {et~al.}(2022)\citenamefont
  {Gangopadhyay}, \citenamefont {Wang}, \citenamefont {Mashatan},\ and\
  \citenamefont {Ghose}}]{Gangopadhyay2022}%
  \BibitemOpen
  \bibfield  {author} {\bibinfo {author} {\bibfnamefont {S.}~\bibnamefont
  {Gangopadhyay}}, \bibinfo {author} {\bibfnamefont {T.}~\bibnamefont {Wang}},
  \bibinfo {author} {\bibfnamefont {A.}~\bibnamefont {Mashatan}},\ and\
  \bibinfo {author} {\bibfnamefont {S.}~\bibnamefont {Ghose}},\ }\bibfield
  {title} {\bibinfo {title} {Controlled quantum teleportation in the presence
  of an adversary},\ }\bibfield  {journal} {\bibinfo  {journal} {Physical
  Review A}\ }\textbf {\bibinfo {volume} {106}},\ \href
  {https://doi.org/10.1103/physreva.106.052433} {10.1103/physreva.106.052433}
  (\bibinfo {year} {2022})\BibitemShut {NoStop}%
\bibitem [{\citenamefont {Tserkis}\ \emph {et~al.}(2020)\citenamefont
  {Tserkis}, \citenamefont {Hosseinidehaj}, \citenamefont {Walk},\ and\
  \citenamefont {Ralph}}]{Tserkis2020}%
  \BibitemOpen
  \bibfield  {author} {\bibinfo {author} {\bibfnamefont {S.}~\bibnamefont
  {Tserkis}}, \bibinfo {author} {\bibfnamefont {N.}~\bibnamefont
  {Hosseinidehaj}}, \bibinfo {author} {\bibfnamefont {N.}~\bibnamefont
  {Walk}},\ and\ \bibinfo {author} {\bibfnamefont {T.~C.}\ \bibnamefont
  {Ralph}},\ }\bibfield  {title} {\bibinfo {title} {Teleportation-based
  collective attacks in gaussian quantum key distribution},\ }\bibfield
  {journal} {\bibinfo  {journal} {Physical Review Research}\ }\textbf {\bibinfo
  {volume} {2}},\ \href {https://doi.org/10.1103/physrevresearch.2.013208}
  {10.1103/physrevresearch.2.013208} (\bibinfo {year} {2020})\BibitemShut
  {NoStop}%
\bibitem [{\citenamefont {Yam}\ \emph {et~al.}(2025)\citenamefont {Yam},
  \citenamefont {Renger}, \citenamefont {Gandorfer}, \citenamefont {Gross},\
  and\ \citenamefont
  {Fedorov}}]{yam2025practicalquantumteleportationfiniteenergy}%
  \BibitemOpen
  \bibfield  {author} {\bibinfo {author} {\bibfnamefont {W.~K.}\ \bibnamefont
  {Yam}}, \bibinfo {author} {\bibfnamefont {M.}~\bibnamefont {Renger}},
  \bibinfo {author} {\bibfnamefont {S.}~\bibnamefont {Gandorfer}}, \bibinfo
  {author} {\bibfnamefont {R.}~\bibnamefont {Gross}},\ and\ \bibinfo {author}
  {\bibfnamefont {K.~G.}\ \bibnamefont {Fedorov}},\ }\href
  {https://arxiv.org/abs/2512.23388} {\bibinfo {title} {Practical quantum
  teleportation with finite-energy codebooks}} (\bibinfo {year} {2025}),\
  \Eprint {https://arxiv.org/abs/2512.23388} {arXiv:2512.23388 [quant-ph]}
  \BibitemShut {NoStop}%
\bibitem [{\citenamefont {Chandra}\ \emph {et~al.}(2026)\citenamefont
  {Chandra}, \citenamefont {Guha},\ and\ \citenamefont
  {Seshadreesan}}]{Chandra2026}%
  \BibitemOpen
  \bibfield  {author} {\bibinfo {author} {\bibfnamefont {N.~K.}\ \bibnamefont
  {Chandra}}, \bibinfo {author} {\bibfnamefont {S.}~\bibnamefont {Guha}},\ and\
  \bibinfo {author} {\bibfnamefont {K.~P.}\ \bibnamefont {Seshadreesan}},\
  }\bibfield  {title} {\bibinfo {title} {Multiplexed bilayered realization of
  fault-tolerant quantum computation over optically networked trapped-ion
  modules},\ }\href {https://doi.org/10.1109/tqe.2025.3649617} {\bibfield
  {journal} {\bibinfo  {journal} {IEEE Transactions on Quantum Engineering}\
  }\textbf {\bibinfo {volume} {7}},\ \bibinfo {pages} {1} (\bibinfo {year}
  {2026})}\BibitemShut {NoStop}%
\bibitem [{\citenamefont {Nielsen}\ and\ \citenamefont
  {Chuang}(2012)}]{Nielsen2012}%
  \BibitemOpen
  \bibfield  {author} {\bibinfo {author} {\bibfnamefont {M.~A.}\ \bibnamefont
  {Nielsen}}\ and\ \bibinfo {author} {\bibfnamefont {I.~L.}\ \bibnamefont
  {Chuang}},\ }\href {https://doi.org/10.1017/cbo9780511976667} {\emph
  {\bibinfo {title} {Quantum Computation and Quantum Information: 10th
  Anniversary Edition}}}\ (\bibinfo  {publisher} {Cambridge University Press},\
  \bibinfo {year} {2012})\BibitemShut {NoStop}%
\bibitem [{\citenamefont {Parakh}(2022)}]{Parakh2022}%
  \BibitemOpen
  \bibfield  {author} {\bibinfo {author} {\bibfnamefont {A.}~\bibnamefont
  {Parakh}},\ }\bibfield  {title} {\bibinfo {title} {Quantum teleportation with
  one classical bit},\ }\bibfield  {journal} {\bibinfo  {journal} {Scientific
  Reports}\ }\textbf {\bibinfo {volume} {12}},\ \href
  {https://doi.org/10.1038/s41598-022-06853-w} {10.1038/s41598-022-06853-w}
  (\bibinfo {year} {2022})\BibitemShut {NoStop}%
\bibitem [{\citenamefont {Ma}\ \emph {et~al.}(2012)\citenamefont {Ma},
  \citenamefont {Herbst}, \citenamefont {Scheidl}, \citenamefont {Wang},
  \citenamefont {Kropatschek}, \citenamefont {Naylor}, \citenamefont
  {Wittmann}, \citenamefont {Mech}, \citenamefont {Kofler}, \citenamefont
  {Anisimova}, \citenamefont {Makarov}, \citenamefont {Jennewein},
  \citenamefont {Ursin},\ and\ \citenamefont {Zeilinger}}]{Ma2012}%
  \BibitemOpen
  \bibfield  {author} {\bibinfo {author} {\bibfnamefont {X.-S.}\ \bibnamefont
  {Ma}}, \bibinfo {author} {\bibfnamefont {T.}~\bibnamefont {Herbst}}, \bibinfo
  {author} {\bibfnamefont {T.}~\bibnamefont {Scheidl}}, \bibinfo {author}
  {\bibfnamefont {D.}~\bibnamefont {Wang}}, \bibinfo {author} {\bibfnamefont
  {S.}~\bibnamefont {Kropatschek}}, \bibinfo {author} {\bibfnamefont
  {W.}~\bibnamefont {Naylor}}, \bibinfo {author} {\bibfnamefont
  {B.}~\bibnamefont {Wittmann}}, \bibinfo {author} {\bibfnamefont
  {A.}~\bibnamefont {Mech}}, \bibinfo {author} {\bibfnamefont {J.}~\bibnamefont
  {Kofler}}, \bibinfo {author} {\bibfnamefont {E.}~\bibnamefont {Anisimova}},
  \bibinfo {author} {\bibfnamefont {V.}~\bibnamefont {Makarov}}, \bibinfo
  {author} {\bibfnamefont {T.}~\bibnamefont {Jennewein}}, \bibinfo {author}
  {\bibfnamefont {R.}~\bibnamefont {Ursin}},\ and\ \bibinfo {author}
  {\bibfnamefont {A.}~\bibnamefont {Zeilinger}},\ }\bibfield  {title} {\bibinfo
  {title} {Quantum teleportation over 143 kilometres using active
  feed-forward},\ }\href {https://doi.org/10.1038/nature11472} {\bibfield
  {journal} {\bibinfo  {journal} {Nature}\ }\textbf {\bibinfo {volume} {489}},\
  \bibinfo {pages} {269} (\bibinfo {year} {2012})}\BibitemShut {NoStop}%
\bibitem [{\citenamefont {Cane}\ \emph {et~al.}(2020)\citenamefont {Cane},
  \citenamefont {Xie},\ and\ \citenamefont {Ng}}]{Cane2020}%
  \BibitemOpen
  \bibfield  {author} {\bibinfo {author} {\bibfnamefont {R.}~\bibnamefont
  {Cane}}, \bibinfo {author} {\bibfnamefont {W.}~\bibnamefont {Xie}},\ and\
  \bibinfo {author} {\bibfnamefont {S.~X.}\ \bibnamefont {Ng}},\ }\bibfield
  {title} {\bibinfo {title} {Turbo-oded secure and reliable quantum
  teleportation},\ }\href {https://doi.org/10.1049/iet-qtc.2020.0004}
  {\bibfield  {journal} {\bibinfo  {journal} {IET Quantum Communication}\
  }\textbf {\bibinfo {volume} {1}},\ \bibinfo {pages} {16} (\bibinfo {year}
  {2020})}\BibitemShut {NoStop}%
\bibitem [{\citenamefont {Liu}\ and\ \citenamefont {Zhou}(2024)}]{Liu2024}%
  \BibitemOpen
  \bibfield  {author} {\bibinfo {author} {\bibfnamefont {Z.}~\bibnamefont
  {Liu}}\ and\ \bibinfo {author} {\bibfnamefont {R.}~\bibnamefont {Zhou}},\
  }\bibfield  {title} {\bibinfo {title} {Asymmetric cyclic controlled quantum
  teleportation in noisy environment},\ }in\ \href
  {https://doi.org/10.1145/3673277.3673363} {\emph {\bibinfo {booktitle}
  {Proceedings of the 2024 3rd International Conference on Cryptography,
  Network Security and Communication Technology}}},\ \bibinfo {series and
  number} {CNSCT 2024}\ (\bibinfo  {publisher} {ACM},\ \bibinfo {year} {2024})\
  pp.\ \bibinfo {pages} {504--?10}\BibitemShut {NoStop}%
\bibitem [{\citenamefont {Renner}\ and\ \citenamefont
  {Wolf}(2023)}]{Renner2023}%
  \BibitemOpen
  \bibfield  {author} {\bibinfo {author} {\bibfnamefont {R.}~\bibnamefont
  {Renner}}\ and\ \bibinfo {author} {\bibfnamefont {R.}~\bibnamefont {Wolf}},\
  }\bibfield  {title} {\bibinfo {title} {Quantum advantage in cryptography},\
  }\href {https://doi.org/10.2514/1.j062267} {\bibfield  {journal} {\bibinfo
  {journal} {AIAA Journal}\ }\textbf {\bibinfo {volume} {61}},\ \bibinfo
  {pages} {1895} (\bibinfo {year} {2023})}\BibitemShut {NoStop}%
\bibitem [{\citenamefont {Rivest}\ \emph {et~al.}(1978)\citenamefont {Rivest},
  \citenamefont {Shamir},\ and\ \citenamefont
  {Adleman}}]{10.1145/359340.359342}%
  \BibitemOpen
  \bibfield  {author} {\bibinfo {author} {\bibfnamefont {R.~L.}\ \bibnamefont
  {Rivest}}, \bibinfo {author} {\bibfnamefont {A.}~\bibnamefont {Shamir}},\
  and\ \bibinfo {author} {\bibfnamefont {L.}~\bibnamefont {Adleman}},\
  }\bibfield  {title} {\bibinfo {title} {A method for obtaining digital
  signatures and public-key cryptosystems},\ }\href
  {https://doi.org/10.1145/359340.359342} {\bibfield  {journal} {\bibinfo
  {journal} {Commun. ACM}\ }\textbf {\bibinfo {volume} {21}},\ \bibinfo {pages}
  {120} (\bibinfo {year} {1978})}\BibitemShut {NoStop}%
\bibitem [{\citenamefont {{National Institute of Standards and
  Technology}}(1994)}]{1994}%
  \BibitemOpen
  \bibfield  {author} {\bibinfo {author} {\bibnamefont {{National Institute of
  Standards and Technology}}},\ }\href {https://doi.org/10.6028/nist.fips.186}
  {\emph {\bibinfo {title} {Digital signature standard (DSS)}}}\ (\bibinfo
  {year} {1994})\BibitemShut {NoStop}%
\bibitem [{\citenamefont {Shor}(1999)}]{Shor1999}%
  \BibitemOpen
  \bibfield  {author} {\bibinfo {author} {\bibfnamefont {P.~W.}\ \bibnamefont
  {Shor}},\ }\bibfield  {title} {\bibinfo {title} {Polynomial-time algorithms
  for prime factorization and discrete logarithms on a quantum computer},\
  }\href {https://doi.org/10.1137/s0036144598347011} {\bibfield  {journal}
  {\bibinfo  {journal} {SIAM Review}\ }\textbf {\bibinfo {volume} {41}},\
  \bibinfo {pages} {303} (\bibinfo {year} {1999})}\BibitemShut {NoStop}%
\bibitem [{\citenamefont {Jin}\ \emph {et~al.}(2025)\citenamefont {Jin},
  \citenamefont {Chandra}, \citenamefont {Azari}, \citenamefont
  {Seshadreesan},\ and\ \citenamefont
  {Liu}}]{jin2025quantumresistantnetworksusingpostquantum}%
  \BibitemOpen
  \bibfield  {author} {\bibinfo {author} {\bibfnamefont {X.}~\bibnamefont
  {Jin}}, \bibinfo {author} {\bibfnamefont {N.~K.}\ \bibnamefont {Chandra}},
  \bibinfo {author} {\bibfnamefont {M.}~\bibnamefont {Azari}}, \bibinfo
  {author} {\bibfnamefont {K.~P.}\ \bibnamefont {Seshadreesan}},\ and\ \bibinfo
  {author} {\bibfnamefont {J.}~\bibnamefont {Liu}},\ }\href
  {https://arxiv.org/abs/2510.24534} {\bibinfo {title} {Quantum-resistant
  networks using post-quantum cryptography}} (\bibinfo {year} {2025}),\ \Eprint
  {https://arxiv.org/abs/2510.24534} {arXiv:2510.24534 [quant-ph]} \BibitemShut
  {NoStop}%
\bibitem [{\citenamefont {Wegman}\ and\ \citenamefont
  {Carter}(1981)}]{Wegman1981}%
  \BibitemOpen
  \bibfield  {author} {\bibinfo {author} {\bibfnamefont {M.~N.}\ \bibnamefont
  {Wegman}}\ and\ \bibinfo {author} {\bibfnamefont {J.}~\bibnamefont
  {Carter}},\ }\bibfield  {title} {\bibinfo {title} {New hash functions and
  their use in authentication and set equality},\ }\href
  {https://doi.org/10.1016/0022-0000(81)90033-7} {\bibfield  {journal}
  {\bibinfo  {journal} {Journal of Computer and System Sciences}\ }\textbf
  {\bibinfo {volume} {22}},\ \bibinfo {pages} {265} (\bibinfo {year}
  {1981})}\BibitemShut {NoStop}%
\bibitem [{\citenamefont {Bennett}\ and\ \citenamefont
  {Brassard}(2014)}]{Bennett2014}%
  \BibitemOpen
  \bibfield  {author} {\bibinfo {author} {\bibfnamefont {C.~H.}\ \bibnamefont
  {Bennett}}\ and\ \bibinfo {author} {\bibfnamefont {G.}~\bibnamefont
  {Brassard}},\ }\bibfield  {title} {\bibinfo {title} {Quantum cryptography:
  Public key distribution and coin tossing},\ }\href
  {https://doi.org/10.1016/j.tcs.2014.05.025} {\bibfield  {journal} {\bibinfo
  {journal} {Theoretical Computer Science}\ }\textbf {\bibinfo {volume}
  {560}},\ \bibinfo {pages} {7} (\bibinfo {year} {2014})}\BibitemShut {NoStop}%
\bibitem [{\citenamefont {Scarani}\ \emph {et~al.}(2009)\citenamefont
  {Scarani}, \citenamefont {Bechmann-Pasquinucci}, \citenamefont {Cerf},
  \citenamefont {Duek}, \citenamefont {L\"{u}tkenhaus},\ and\ \citenamefont
  {Peev}}]{Scarani2009}%
  \BibitemOpen
  \bibfield  {author} {\bibinfo {author} {\bibfnamefont {V.}~\bibnamefont
  {Scarani}}, \bibinfo {author} {\bibfnamefont {H.}~\bibnamefont
  {Bechmann-Pasquinucci}}, \bibinfo {author} {\bibfnamefont {N.~J.}\
  \bibnamefont {Cerf}}, \bibinfo {author} {\bibfnamefont {M.}~\bibnamefont
  {Duek}}, \bibinfo {author} {\bibfnamefont {N.}~\bibnamefont
  {L\"{u}tkenhaus}},\ and\ \bibinfo {author} {\bibfnamefont {M.}~\bibnamefont
  {Peev}},\ }\bibfield  {title} {\bibinfo {title} {The security of practical
  quantum key distribution},\ }\href
  {https://doi.org/10.1103/revmodphys.81.1301} {\bibfield  {journal} {\bibinfo
  {journal} {Reviews of Modern Physics}\ }\textbf {\bibinfo {volume} {81}},\
  \bibinfo {pages} {1301} (\bibinfo {year} {2009})}\BibitemShut {NoStop}%
\bibitem [{\citenamefont {Portmann}\ and\ \citenamefont
  {Renner}(2022)}]{Portmann2022}%
  \BibitemOpen
  \bibfield  {author} {\bibinfo {author} {\bibfnamefont {C.}~\bibnamefont
  {Portmann}}\ and\ \bibinfo {author} {\bibfnamefont {R.}~\bibnamefont
  {Renner}},\ }\bibfield  {title} {\bibinfo {title} {Security in quantum
  cryptography},\ }\bibfield  {journal} {\bibinfo  {journal} {Reviews of Modern
  Physics}\ }\textbf {\bibinfo {volume} {94}},\ \href
  {https://doi.org/10.1103/revmodphys.94.025008} {10.1103/revmodphys.94.025008}
  (\bibinfo {year} {2022})\BibitemShut {NoStop}%
\bibitem [{\citenamefont {Amiri}\ \emph {et~al.}(2016)\citenamefont {Amiri},
  \citenamefont {Wallden}, \citenamefont {Kent},\ and\ \citenamefont
  {Andersson}}]{Amiri2016}%
  \BibitemOpen
  \bibfield  {author} {\bibinfo {author} {\bibfnamefont {R.}~\bibnamefont
  {Amiri}}, \bibinfo {author} {\bibfnamefont {P.}~\bibnamefont {Wallden}},
  \bibinfo {author} {\bibfnamefont {A.}~\bibnamefont {Kent}},\ and\ \bibinfo
  {author} {\bibfnamefont {E.}~\bibnamefont {Andersson}},\ }\bibfield  {title}
  {\bibinfo {title} {Secure quantum signatures using insecure quantum
  channels},\ }\bibfield  {journal} {\bibinfo  {journal} {Physical Review A}\
  }\textbf {\bibinfo {volume} {93}},\ \href
  {https://doi.org/10.1103/physreva.93.032325} {10.1103/physreva.93.032325}
  (\bibinfo {year} {2016})\BibitemShut {NoStop}%
\bibitem [{\citenamefont {Aquina}\ \emph {et~al.}(2025)\citenamefont {Aquina},
  \citenamefont {Cimoli}, \citenamefont {Das}, \citenamefont {H\"{o}velmanns},
  \citenamefont {Weber}, \citenamefont {Okonkwo}, \citenamefont {Rommel},
  \citenamefont {kori}, \citenamefont {Tafur~Monroy},\ and\ \citenamefont
  {Verschoor}}]{Aquina2025}%
  \BibitemOpen
  \bibfield  {author} {\bibinfo {author} {\bibfnamefont {N.}~\bibnamefont
  {Aquina}}, \bibinfo {author} {\bibfnamefont {B.}~\bibnamefont {Cimoli}},
  \bibinfo {author} {\bibfnamefont {S.}~\bibnamefont {Das}}, \bibinfo {author}
  {\bibfnamefont {K.}~\bibnamefont {H\"{o}velmanns}}, \bibinfo {author}
  {\bibfnamefont {F.~J.}\ \bibnamefont {Weber}}, \bibinfo {author}
  {\bibfnamefont {C.}~\bibnamefont {Okonkwo}}, \bibinfo {author} {\bibfnamefont
  {S.}~\bibnamefont {Rommel}}, \bibinfo {author} {\bibfnamefont
  {B.}~\bibnamefont {kori}}, \bibinfo {author} {\bibfnamefont {I.}~\bibnamefont
  {Tafur~Monroy}},\ and\ \bibinfo {author} {\bibfnamefont {S.}~\bibnamefont
  {Verschoor}},\ }\bibfield  {title} {\bibinfo {title} {A critical analysis of
  deployed use cases for quantum key distribution and comparison with
  post-quantum cryptography},\ }\bibfield  {journal} {\bibinfo  {journal} {EPJ
  Quantum Technology}\ }\textbf {\bibinfo {volume} {12}},\ \href
  {https://doi.org/10.1140/epjqt/s40507-025-00350-5}
  {10.1140/epjqt/s40507-025-00350-5} (\bibinfo {year} {2025})\BibitemShut
  {NoStop}%
\bibitem [{\citenamefont {Baseri}\ \emph {et~al.}(2025)\citenamefont {Baseri},
  \citenamefont {Chouhan}, \citenamefont {Ghorbani},\ and\ \citenamefont
  {Chow}}]{Baseri2025}%
  \BibitemOpen
  \bibfield  {author} {\bibinfo {author} {\bibfnamefont {Y.}~\bibnamefont
  {Baseri}}, \bibinfo {author} {\bibfnamefont {V.}~\bibnamefont {Chouhan}},
  \bibinfo {author} {\bibfnamefont {A.}~\bibnamefont {Ghorbani}},\ and\
  \bibinfo {author} {\bibfnamefont {A.}~\bibnamefont {Chow}},\ }\bibfield
  {title} {\bibinfo {title} {Evaluation framework for quantum security risk
  assessment: A comprehensive strategy for quantum-safe transition},\ }\href
  {https://doi.org/10.1016/j.cose.2024.104272} {\bibfield  {journal} {\bibinfo
  {journal} {Computers \& Security}\ }\textbf {\bibinfo {volume} {150}},\
  \bibinfo {pages} {104272} (\bibinfo {year} {2025})}\BibitemShut {NoStop}%
\bibitem [{\citenamefont {Chen}\ \emph {et~al.}(2016)\citenamefont {Chen},
  \citenamefont {Jordan}, \citenamefont {Liu}, \citenamefont {Moody},
  \citenamefont {Peralta}, \citenamefont {Perlner},\ and\ \citenamefont
  {Smith-Tone}}]{Chen2016}%
  \BibitemOpen
  \bibfield  {author} {\bibinfo {author} {\bibfnamefont {L.}~\bibnamefont
  {Chen}}, \bibinfo {author} {\bibfnamefont {S.}~\bibnamefont {Jordan}},
  \bibinfo {author} {\bibfnamefont {Y.-K.}\ \bibnamefont {Liu}}, \bibinfo
  {author} {\bibfnamefont {D.}~\bibnamefont {Moody}}, \bibinfo {author}
  {\bibfnamefont {R.}~\bibnamefont {Peralta}}, \bibinfo {author} {\bibfnamefont
  {R.}~\bibnamefont {Perlner}},\ and\ \bibinfo {author} {\bibfnamefont
  {D.}~\bibnamefont {Smith-Tone}},\ }\href
  {https://doi.org/10.6028/nist.ir.8105} {\emph {\bibinfo {title} {Report on
  Post-Quantum Cryptography}}}\ (\bibinfo {year} {2016})\BibitemShut {NoStop}%
\bibitem [{\citenamefont {Bos}\ \emph {et~al.}(2018)\citenamefont {Bos},
  \citenamefont {Ducas}, \citenamefont {Kiltz}, \citenamefont {Lepoint},
  \citenamefont {Lyubashevsky}, \citenamefont {Schanck}, \citenamefont
  {Schwabe}, \citenamefont {Seiler},\ and\ \citenamefont {Stehle}}]{Bos2018}%
  \BibitemOpen
  \bibfield  {author} {\bibinfo {author} {\bibfnamefont {J.}~\bibnamefont
  {Bos}}, \bibinfo {author} {\bibfnamefont {L.}~\bibnamefont {Ducas}}, \bibinfo
  {author} {\bibfnamefont {E.}~\bibnamefont {Kiltz}}, \bibinfo {author}
  {\bibfnamefont {T.}~\bibnamefont {Lepoint}}, \bibinfo {author} {\bibfnamefont
  {V.}~\bibnamefont {Lyubashevsky}}, \bibinfo {author} {\bibfnamefont {J.~M.}\
  \bibnamefont {Schanck}}, \bibinfo {author} {\bibfnamefont {P.}~\bibnamefont
  {Schwabe}}, \bibinfo {author} {\bibfnamefont {G.}~\bibnamefont {Seiler}},\
  and\ \bibinfo {author} {\bibfnamefont {D.}~\bibnamefont {Stehle}},\
  }\bibfield  {title} {\bibinfo {title} {Crystals - kyber: A cca-secure
  module-lattice-based kem},\ }in\ \href
  {https://doi.org/10.1109/eurosp.2018.00032} {\emph {\bibinfo {booktitle}
  {2018 IEEE European Symposium on Security and Privacy (EuroS\&P)}}}\
  (\bibinfo  {publisher} {IEEE},\ \bibinfo {year} {2018})\ pp.\ \bibinfo
  {pages} {353--?67}\BibitemShut {NoStop}%
\bibitem [{\citenamefont {Ducas}\ \emph {et~al.}(2018)\citenamefont {Ducas},
  \citenamefont {Kiltz}, \citenamefont {Lepoint}, \citenamefont {Lyubashevsky},
  \citenamefont {Schwabe}, \citenamefont {Seiler},\ and\ \citenamefont
  {Stehl}}]{Ducas2018}%
  \BibitemOpen
  \bibfield  {author} {\bibinfo {author} {\bibfnamefont {L.}~\bibnamefont
  {Ducas}}, \bibinfo {author} {\bibfnamefont {E.}~\bibnamefont {Kiltz}},
  \bibinfo {author} {\bibfnamefont {T.}~\bibnamefont {Lepoint}}, \bibinfo
  {author} {\bibfnamefont {V.}~\bibnamefont {Lyubashevsky}}, \bibinfo {author}
  {\bibfnamefont {P.}~\bibnamefont {Schwabe}}, \bibinfo {author} {\bibfnamefont
  {G.}~\bibnamefont {Seiler}},\ and\ \bibinfo {author} {\bibfnamefont
  {D.}~\bibnamefont {Stehl}},\ }\bibfield  {title} {\bibinfo {title}
  {Crystals-dilithium: A lattice-based digital signature scheme},\ }\href
  {https://doi.org/10.46586/tches.v2018.i1.238-268} {\bibfield  {journal}
  {\bibinfo  {journal} {IACR Transactions on Cryptographic Hardware and
  Embedded Systems}\ ,\ \bibinfo {pages} {238}} (\bibinfo {year}
  {2018})}\BibitemShut {NoStop}%
\bibitem [{\citenamefont {Bernstein}\ \emph {et~al.}(2019)\citenamefont
  {Bernstein}, \citenamefont {H\"{u}lsing}, \citenamefont {K\"{o}lbl},
  \citenamefont {Niederhagen}, \citenamefont {Rijneveld},\ and\ \citenamefont
  {Schwabe}}]{Bernstein2019}%
  \BibitemOpen
  \bibfield  {author} {\bibinfo {author} {\bibfnamefont {D.~J.}\ \bibnamefont
  {Bernstein}}, \bibinfo {author} {\bibfnamefont {A.}~\bibnamefont
  {H\"{u}lsing}}, \bibinfo {author} {\bibfnamefont {S.}~\bibnamefont
  {K\"{o}lbl}}, \bibinfo {author} {\bibfnamefont {R.}~\bibnamefont
  {Niederhagen}}, \bibinfo {author} {\bibfnamefont {J.}~\bibnamefont
  {Rijneveld}},\ and\ \bibinfo {author} {\bibfnamefont {P.}~\bibnamefont
  {Schwabe}},\ }\bibfield  {title} {\bibinfo {title} {The sphincs + signature
  framework},\ }in\ \href {https://doi.org/10.1145/3319535.3363229} {\emph
  {\bibinfo {booktitle} {Proceedings of the 2019 ACM SIGSAC Conference on
  Computer and Communications Security}}},\ \bibinfo {series and number} {CCS
  -?9}\ (\bibinfo  {publisher} {ACM},\ \bibinfo {year} {2019})\ pp.\ \bibinfo
  {pages} {2129--?146}\BibitemShut {NoStop}%
\bibitem [{\citenamefont {Moody}\ \emph {et~al.}(2020)\citenamefont {Moody},
  \citenamefont {Alagic}, \citenamefont {Apon}, \citenamefont {Cooper},
  \citenamefont {Dang}, \citenamefont {Kelsey}, \citenamefont {Liu},
  \citenamefont {Miller}, \citenamefont {Peralta}, \citenamefont {Perlner},
  \citenamefont {Robinson}, \citenamefont {Smith-Tone},\ and\ \citenamefont
  {Alperin-Sheriff}}]{Moody2020}%
  \BibitemOpen
  \bibfield  {author} {\bibinfo {author} {\bibfnamefont {D.}~\bibnamefont
  {Moody}}, \bibinfo {author} {\bibfnamefont {G.}~\bibnamefont {Alagic}},
  \bibinfo {author} {\bibfnamefont {D.~C.}\ \bibnamefont {Apon}}, \bibinfo
  {author} {\bibfnamefont {D.~A.}\ \bibnamefont {Cooper}}, \bibinfo {author}
  {\bibfnamefont {Q.~H.}\ \bibnamefont {Dang}}, \bibinfo {author}
  {\bibfnamefont {J.~M.}\ \bibnamefont {Kelsey}}, \bibinfo {author}
  {\bibfnamefont {Y.-K.}\ \bibnamefont {Liu}}, \bibinfo {author} {\bibfnamefont
  {C.~A.}\ \bibnamefont {Miller}}, \bibinfo {author} {\bibfnamefont {R.~C.}\
  \bibnamefont {Peralta}}, \bibinfo {author} {\bibfnamefont {R.~A.}\
  \bibnamefont {Perlner}}, \bibinfo {author} {\bibfnamefont {A.~Y.}\
  \bibnamefont {Robinson}}, \bibinfo {author} {\bibfnamefont {D.~C.}\
  \bibnamefont {Smith-Tone}},\ and\ \bibinfo {author} {\bibfnamefont
  {J.}~\bibnamefont {Alperin-Sheriff}},\ }\href
  {https://doi.org/10.6028/nist.ir.8309} {\emph {\bibinfo {title} {Status
  report on the second round of the NIST post-quantum cryptography
  standardization process}}}\ (\bibinfo {year} {2020})\BibitemShut {NoStop}%
\bibitem [{\citenamefont {Wang}\ \emph
  {et~al.}(2021{\natexlab{a}})\citenamefont {Wang}, \citenamefont {Zhang},
  \citenamefont {Wang}, \citenamefont {Cheng}, \citenamefont {Yang},
  \citenamefont {Tang}, \citenamefont {Yan}, \citenamefont {Tang},
  \citenamefont {Liu}, \citenamefont {Yu}, \citenamefont {Zhang},\ and\
  \citenamefont {Pan}}]{Wang2021}%
  \BibitemOpen
  \bibfield  {author} {\bibinfo {author} {\bibfnamefont {L.-J.}\ \bibnamefont
  {Wang}}, \bibinfo {author} {\bibfnamefont {K.-Y.}\ \bibnamefont {Zhang}},
  \bibinfo {author} {\bibfnamefont {J.-Y.}\ \bibnamefont {Wang}}, \bibinfo
  {author} {\bibfnamefont {J.}~\bibnamefont {Cheng}}, \bibinfo {author}
  {\bibfnamefont {Y.-H.}\ \bibnamefont {Yang}}, \bibinfo {author}
  {\bibfnamefont {S.-B.}\ \bibnamefont {Tang}}, \bibinfo {author}
  {\bibfnamefont {D.}~\bibnamefont {Yan}}, \bibinfo {author} {\bibfnamefont
  {Y.-L.}\ \bibnamefont {Tang}}, \bibinfo {author} {\bibfnamefont
  {Z.}~\bibnamefont {Liu}}, \bibinfo {author} {\bibfnamefont {Y.}~\bibnamefont
  {Yu}}, \bibinfo {author} {\bibfnamefont {Q.}~\bibnamefont {Zhang}},\ and\
  \bibinfo {author} {\bibfnamefont {J.-W.}\ \bibnamefont {Pan}},\ }\bibfield
  {title} {\bibinfo {title} {Experimental authentication of quantum key
  distribution with post-quantum cryptography},\ }\bibfield  {journal}
  {\bibinfo  {journal} {npj Quantum Information}\ }\textbf {\bibinfo {volume}
  {7}},\ \href {https://doi.org/10.1038/s41534-021-00400-7}
  {10.1038/s41534-021-00400-7} (\bibinfo {year}
  {2021}{\natexlab{a}})\BibitemShut {NoStop}%
\bibitem [{\citenamefont {Bar-Gill}\ \emph {et~al.}(2013)\citenamefont
  {Bar-Gill}, \citenamefont {Pham}, \citenamefont {Jarmola}, \citenamefont
  {Budker},\ and\ \citenamefont {Walsworth}}]{BarGill2013}%
  \BibitemOpen
  \bibfield  {author} {\bibinfo {author} {\bibfnamefont {N.}~\bibnamefont
  {Bar-Gill}}, \bibinfo {author} {\bibfnamefont {L.}~\bibnamefont {Pham}},
  \bibinfo {author} {\bibfnamefont {A.}~\bibnamefont {Jarmola}}, \bibinfo
  {author} {\bibfnamefont {D.}~\bibnamefont {Budker}},\ and\ \bibinfo {author}
  {\bibfnamefont {R.}~\bibnamefont {Walsworth}},\ }\bibfield  {title} {\bibinfo
  {title} {Solid-state electronic spin coherence time approaching one second},\
  }\bibfield  {journal} {\bibinfo  {journal} {Nature Communications}\ }\textbf
  {\bibinfo {volume} {4}},\ \href {https://doi.org/10.1038/ncomms2771}
  {10.1038/ncomms2771} (\bibinfo {year} {2013})\BibitemShut {NoStop}%
\bibitem [{\citenamefont {Zhong}\ \emph {et~al.}(2015)\citenamefont {Zhong},
  \citenamefont {Hedges}, \citenamefont {Ahlefeldt}, \citenamefont
  {Bartholomew}, \citenamefont {Beavan}, \citenamefont {Wittig}, \citenamefont
  {Longdell},\ and\ \citenamefont {Sellars}}]{Zhong2015}%
  \BibitemOpen
  \bibfield  {author} {\bibinfo {author} {\bibfnamefont {M.}~\bibnamefont
  {Zhong}}, \bibinfo {author} {\bibfnamefont {M.~P.}\ \bibnamefont {Hedges}},
  \bibinfo {author} {\bibfnamefont {R.~L.}\ \bibnamefont {Ahlefeldt}}, \bibinfo
  {author} {\bibfnamefont {J.~G.}\ \bibnamefont {Bartholomew}}, \bibinfo
  {author} {\bibfnamefont {S.~E.}\ \bibnamefont {Beavan}}, \bibinfo {author}
  {\bibfnamefont {S.~M.}\ \bibnamefont {Wittig}}, \bibinfo {author}
  {\bibfnamefont {J.~J.}\ \bibnamefont {Longdell}},\ and\ \bibinfo {author}
  {\bibfnamefont {M.~J.}\ \bibnamefont {Sellars}},\ }\bibfield  {title}
  {\bibinfo {title} {Optically addressable nuclear spins in a solid with a
  six-hour coherence time},\ }\href {https://doi.org/10.1038/nature14025}
  {\bibfield  {journal} {\bibinfo  {journal} {Nature}\ }\textbf {\bibinfo
  {volume} {517}},\ \bibinfo {pages} {177} (\bibinfo {year}
  {2015})}\BibitemShut {NoStop}%
\bibitem [{\citenamefont {Heshami}\ \emph {et~al.}(2016)\citenamefont
  {Heshami}, \citenamefont {England}, \citenamefont {Humphreys}, \citenamefont
  {Bustard}, \citenamefont {Acosta}, \citenamefont {Nunn},\ and\ \citenamefont
  {Sussman}}]{Heshami2016}%
  \BibitemOpen
  \bibfield  {author} {\bibinfo {author} {\bibfnamefont {K.}~\bibnamefont
  {Heshami}}, \bibinfo {author} {\bibfnamefont {D.~G.}\ \bibnamefont
  {England}}, \bibinfo {author} {\bibfnamefont {P.~C.}\ \bibnamefont
  {Humphreys}}, \bibinfo {author} {\bibfnamefont {P.~J.}\ \bibnamefont
  {Bustard}}, \bibinfo {author} {\bibfnamefont {V.~M.}\ \bibnamefont {Acosta}},
  \bibinfo {author} {\bibfnamefont {J.}~\bibnamefont {Nunn}},\ and\ \bibinfo
  {author} {\bibfnamefont {B.~J.}\ \bibnamefont {Sussman}},\ }\bibfield
  {title} {\bibinfo {title} {Quantum memories: emerging applications and recent
  advances},\ }\href {https://doi.org/10.1080/09500340.2016.1148212} {\bibfield
   {journal} {\bibinfo  {journal} {Journal of Modern Optics}\ }\textbf
  {\bibinfo {volume} {63}},\ \bibinfo {pages} {2005} (\bibinfo {year}
  {2016})}\BibitemShut {NoStop}%
\bibitem [{\citenamefont {Hucul}\ \emph {et~al.}(2014)\citenamefont {Hucul},
  \citenamefont {Inlek}, \citenamefont {Vittorini}, \citenamefont {Crocker},
  \citenamefont {Debnath}, \citenamefont {Clark},\ and\ \citenamefont
  {Monroe}}]{Hucul2014}%
  \BibitemOpen
  \bibfield  {author} {\bibinfo {author} {\bibfnamefont {D.}~\bibnamefont
  {Hucul}}, \bibinfo {author} {\bibfnamefont {I.~V.}\ \bibnamefont {Inlek}},
  \bibinfo {author} {\bibfnamefont {G.}~\bibnamefont {Vittorini}}, \bibinfo
  {author} {\bibfnamefont {C.}~\bibnamefont {Crocker}}, \bibinfo {author}
  {\bibfnamefont {S.}~\bibnamefont {Debnath}}, \bibinfo {author} {\bibfnamefont
  {S.~M.}\ \bibnamefont {Clark}},\ and\ \bibinfo {author} {\bibfnamefont
  {C.}~\bibnamefont {Monroe}},\ }\bibfield  {title} {\bibinfo {title} {Modular
  entanglement of atomic qubits using photons and phonons},\ }\href
  {https://doi.org/10.1038/nphys3150} {\bibfield  {journal} {\bibinfo
  {journal} {Nature Physics}\ }\textbf {\bibinfo {volume} {11}},\ \bibinfo
  {pages} {37} (\bibinfo {year} {2014})}\BibitemShut {NoStop}%
\bibitem [{\citenamefont {Genovese}(2001)}]{Genovese2001}%
  \BibitemOpen
  \bibfield  {author} {\bibinfo {author} {\bibfnamefont {M.}~\bibnamefont
  {Genovese}},\ }\bibfield  {title} {\bibinfo {title} {Proposal of an
  experimental scheme for realizing a translucent eavesdropping on a quantum
  cryptographic channel},\ }\bibfield  {journal} {\bibinfo  {journal} {Physical
  Review A}\ }\textbf {\bibinfo {volume} {63}},\ \href
  {https://doi.org/10.1103/physreva.63.044303} {10.1103/physreva.63.044303}
  (\bibinfo {year} {2001})\BibitemShut {NoStop}%
\bibitem [{\citenamefont {Lederman}\ and\ \citenamefont
  {Pereg}(2024)}]{Lederman2024}%
  \BibitemOpen
  \bibfield  {author} {\bibinfo {author} {\bibfnamefont {M.}~\bibnamefont
  {Lederman}}\ and\ \bibinfo {author} {\bibfnamefont {U.}~\bibnamefont
  {Pereg}},\ }\bibfield  {title} {\bibinfo {title} {Secure communication with
  unreliable entanglement assistance},\ }in\ \href
  {https://doi.org/10.1109/isit57864.2024.10619085} {\emph {\bibinfo
  {booktitle} {2024 IEEE International Symposium on Information Theory
  (ISIT)}}}\ (\bibinfo  {publisher} {IEEE},\ \bibinfo {year} {2024})\ pp.\
  \bibinfo {pages} {1017--?022}\BibitemShut {NoStop}%
\bibitem [{\citenamefont {Shaban}\ and\ \citenamefont
  {Ismail}(2024)}]{Shaban2024}%
  \BibitemOpen
  \bibfield  {author} {\bibinfo {author} {\bibfnamefont {M.}~\bibnamefont
  {Shaban}}\ and\ \bibinfo {author} {\bibfnamefont {M.}~\bibnamefont
  {Ismail}},\ }\bibfield  {title} {\bibinfo {title} {Secured quantum identity
  authentication protocol for quantum networks},\ }in\ \href
  {https://doi.org/10.1109/vtc2024-fall63153.2024.10757654} {\emph {\bibinfo
  {booktitle} {2024 IEEE 100th Vehicular Technology Conference
  (VTC2024-Fall)}}}\ (\bibinfo  {publisher} {IEEE},\ \bibinfo {year} {2024})\
  pp.\ \bibinfo {pages} {1--?}\BibitemShut {Stop}%
\bibitem [{\citenamefont {Klimov}\ \emph {et~al.}(2018)\citenamefont {Klimov},
  \citenamefont {Kelly}, \citenamefont {Chen}, \citenamefont {Neeley},
  \citenamefont {Megrant}, \citenamefont {Burkett}, \citenamefont {Barends},
  \citenamefont {Arya}, \citenamefont {Chiaro}, \citenamefont {Chen},
  \citenamefont {Dunsworth}, \citenamefont {Fowler}, \citenamefont {Foxen},
  \citenamefont {Gidney}, \citenamefont {Giustina}, \citenamefont {Graff},
  \citenamefont {Huang}, \citenamefont {Jeffrey}, \citenamefont {Lucero},
  \citenamefont {Mutus}, \citenamefont {Naaman}, \citenamefont {Neill},
  \citenamefont {Quintana}, \citenamefont {Roushan}, \citenamefont {Sank},
  \citenamefont {Vainsencher}, \citenamefont {Wenner}, \citenamefont {White},
  \citenamefont {Boixo}, \citenamefont {Babbush}, \citenamefont {Smelyanskiy},
  \citenamefont {Neven},\ and\ \citenamefont {Martinis}}]{Klimov2018}%
  \BibitemOpen
  \bibfield  {author} {\bibinfo {author} {\bibfnamefont {P.-a.}\ \bibnamefont
  {Klimov}}, \bibinfo {author} {\bibfnamefont {J.}~\bibnamefont {Kelly}},
  \bibinfo {author} {\bibfnamefont {Z.}~\bibnamefont {Chen}}, \bibinfo {author}
  {\bibfnamefont {M.}~\bibnamefont {Neeley}}, \bibinfo {author} {\bibfnamefont
  {A.}~\bibnamefont {Megrant}}, \bibinfo {author} {\bibfnamefont
  {B.}~\bibnamefont {Burkett}}, \bibinfo {author} {\bibfnamefont
  {R.}~\bibnamefont {Barends}}, \bibinfo {author} {\bibfnamefont
  {K.}~\bibnamefont {Arya}}, \bibinfo {author} {\bibfnamefont {B.}~\bibnamefont
  {Chiaro}}, \bibinfo {author} {\bibfnamefont {Y.}~\bibnamefont {Chen}},
  \bibinfo {author} {\bibfnamefont {A.}~\bibnamefont {Dunsworth}}, \bibinfo
  {author} {\bibfnamefont {A.}~\bibnamefont {Fowler}}, \bibinfo {author}
  {\bibfnamefont {B.}~\bibnamefont {Foxen}}, \bibinfo {author} {\bibfnamefont
  {C.}~\bibnamefont {Gidney}}, \bibinfo {author} {\bibfnamefont
  {M.}~\bibnamefont {Giustina}}, \bibinfo {author} {\bibfnamefont
  {R.}~\bibnamefont {Graff}}, \bibinfo {author} {\bibfnamefont
  {T.}~\bibnamefont {Huang}}, \bibinfo {author} {\bibfnamefont
  {E.}~\bibnamefont {Jeffrey}}, \bibinfo {author} {\bibfnamefont
  {E.}~\bibnamefont {Lucero}}, \bibinfo {author} {\bibfnamefont {J.-.}\
  \bibnamefont {Mutus}}, \bibinfo {author} {\bibfnamefont {O.}~\bibnamefont
  {Naaman}}, \bibinfo {author} {\bibfnamefont {C.}~\bibnamefont {Neill}},
  \bibinfo {author} {\bibfnamefont {C.}~\bibnamefont {Quintana}}, \bibinfo
  {author} {\bibfnamefont {P.}~\bibnamefont {Roushan}}, \bibinfo {author}
  {\bibfnamefont {D.}~\bibnamefont {Sank}}, \bibinfo {author} {\bibfnamefont
  {A.}~\bibnamefont {Vainsencher}}, \bibinfo {author} {\bibfnamefont
  {J.}~\bibnamefont {Wenner}}, \bibinfo {author} {\bibfnamefont {T.-.}\
  \bibnamefont {White}}, \bibinfo {author} {\bibfnamefont {S.}~\bibnamefont
  {Boixo}}, \bibinfo {author} {\bibfnamefont {R.}~\bibnamefont {Babbush}},
  \bibinfo {author} {\bibfnamefont {V.-.}\ \bibnamefont {Smelyanskiy}},
  \bibinfo {author} {\bibfnamefont {H.}~\bibnamefont {Neven}},\ and\ \bibinfo
  {author} {\bibfnamefont {J.-.}\ \bibnamefont {Martinis}},\ }\bibfield
  {title} {\bibinfo {title} {Fluctuations of energy-relaxation times in
  superconducting qubits},\ }\bibfield  {journal} {\bibinfo  {journal}
  {Physical Review Letters}\ }\textbf {\bibinfo {volume} {121}},\ \href
  {https://doi.org/10.1103/physrevlett.121.090502}
  {10.1103/physrevlett.121.090502} (\bibinfo {year} {2018})\BibitemShut
  {NoStop}%
\bibitem [{\citenamefont {Khatri}\ \emph {et~al.}(2020)\citenamefont {Khatri},
  \citenamefont {Sharma},\ and\ \citenamefont {Wilde}}]{Khatri2020}%
  \BibitemOpen
  \bibfield  {author} {\bibinfo {author} {\bibfnamefont {S.}~\bibnamefont
  {Khatri}}, \bibinfo {author} {\bibfnamefont {K.}~\bibnamefont {Sharma}},\
  and\ \bibinfo {author} {\bibfnamefont {M.~M.}\ \bibnamefont {Wilde}},\
  }\bibfield  {title} {\bibinfo {title} {Information-theoretic aspects of the
  generalized amplitude-damping channel},\ }\bibfield  {journal} {\bibinfo
  {journal} {Physical Review A}\ }\textbf {\bibinfo {volume} {102}},\ \href
  {https://doi.org/10.1103/physreva.102.012401} {10.1103/physreva.102.012401}
  (\bibinfo {year} {2020})\BibitemShut {NoStop}%
\bibitem [{\citenamefont {Boykin}\ and\ \citenamefont
  {Roychowdhury}(2003)}]{Boykin2003}%
  \BibitemOpen
  \bibfield  {author} {\bibinfo {author} {\bibfnamefont {P.~O.}\ \bibnamefont
  {Boykin}}\ and\ \bibinfo {author} {\bibfnamefont {V.}~\bibnamefont
  {Roychowdhury}},\ }\bibfield  {title} {\bibinfo {title} {Optimal encryption
  of quantum bits},\ }\bibfield  {journal} {\bibinfo  {journal} {Physical
  Review A}\ }\textbf {\bibinfo {volume} {67}},\ \href
  {https://doi.org/10.1103/physreva.67.042317} {10.1103/physreva.67.042317}
  (\bibinfo {year} {2003})\BibitemShut {NoStop}%
\bibitem [{\citenamefont {Regev}(2009)}]{Regev2009}%
  \BibitemOpen
  \bibfield  {author} {\bibinfo {author} {\bibfnamefont {O.}~\bibnamefont
  {Regev}},\ }\bibfield  {title} {\bibinfo {title} {On lattices, learning with
  errors, random linear codes, and cryptography},\ }\href
  {https://doi.org/10.1145/1568318.1568324} {\bibfield  {journal} {\bibinfo
  {journal} {Journal of the ACM}\ }\textbf {\bibinfo {volume} {56}},\ \bibinfo
  {pages} {1} (\bibinfo {year} {2009})}\BibitemShut {NoStop}%
\bibitem [{\citenamefont {Chen}\ and\ \citenamefont {Nguyen}(2011)}]{Chen2011}%
  \BibitemOpen
  \bibfield  {author} {\bibinfo {author} {\bibfnamefont {Y.}~\bibnamefont
  {Chen}}\ and\ \bibinfo {author} {\bibfnamefont {P.~Q.}\ \bibnamefont
  {Nguyen}},\ }\bibinfo {title} {Bkz 2.0: Better lattice security estimates},\
  in\ \href {https://doi.org/10.1007/978-3-642-25385-0_1} {\emph {\bibinfo
  {booktitle} {Advances in Cryptology -?ASIACRYPT 2011}}}\ (\bibinfo
  {publisher} {Springer Berlin Heidelberg},\ \bibinfo {year} {2011})\ pp.\
  \bibinfo {pages} {1--?0}\BibitemShut {NoStop}%
\bibitem [{\citenamefont {Albrecht}\ \emph {et~al.}(2015)\citenamefont
  {Albrecht}, \citenamefont {Player},\ and\ \citenamefont
  {Scott}}]{Albrecht2015}%
  \BibitemOpen
  \bibfield  {author} {\bibinfo {author} {\bibfnamefont {M.~R.}\ \bibnamefont
  {Albrecht}}, \bibinfo {author} {\bibfnamefont {R.}~\bibnamefont {Player}},\
  and\ \bibinfo {author} {\bibfnamefont {S.}~\bibnamefont {Scott}},\ }\bibfield
   {title} {\bibinfo {title} {On the concrete hardness of learning with
  errors},\ }\href {https://doi.org/10.1515/jmc-2015-0016} {\bibfield
  {journal} {\bibinfo  {journal} {Journal of Mathematical Cryptology}\ }\textbf
  {\bibinfo {volume} {9}},\ \bibinfo {pages} {169} (\bibinfo {year}
  {2015})}\BibitemShut {NoStop}%
\bibitem [{\citenamefont {Schnorr}\ and\ \citenamefont
  {Euchner}(1994)}]{Schnorr1994}%
  \BibitemOpen
  \bibfield  {author} {\bibinfo {author} {\bibfnamefont {C.~P.}\ \bibnamefont
  {Schnorr}}\ and\ \bibinfo {author} {\bibfnamefont {M.}~\bibnamefont
  {Euchner}},\ }\bibfield  {title} {\bibinfo {title} {Lattice basis reduction:
  Improved practical algorithms and solving subset sum problems},\ }\href
  {https://doi.org/10.1007/bf01581144} {\bibfield  {journal} {\bibinfo
  {journal} {Mathematical Programming}\ }\textbf {\bibinfo {volume} {66}},\
  \bibinfo {pages} {181} (\bibinfo {year} {1994})}\BibitemShut {NoStop}%
\bibitem [{\citenamefont {Lindner}\ and\ \citenamefont
  {Peikert}(2011)}]{Lindner2011}%
  \BibitemOpen
  \bibfield  {author} {\bibinfo {author} {\bibfnamefont {R.}~\bibnamefont
  {Lindner}}\ and\ \bibinfo {author} {\bibfnamefont {C.}~\bibnamefont
  {Peikert}},\ }\bibinfo {title} {Better key sizes (and attacks) for lwe-based
  encryption},\ in\ \href {https://doi.org/10.1007/978-3-642-19074-2_21} {\emph
  {\bibinfo {booktitle} {Topics in Cryptology -?CT-RSA 2011}}}\ (\bibinfo
  {publisher} {Springer Berlin Heidelberg},\ \bibinfo {year} {2011})\ pp.\
  \bibinfo {pages} {319--?39}\BibitemShut {NoStop}%
\bibitem [{\citenamefont {{TOP500.org}}(2024)}]{green500_2024_11}%
  \BibitemOpen
  \bibfield  {author} {\bibinfo {author} {\bibnamefont {{TOP500.org}}},\
  }\href@noop {} {\bibinfo {title} {Green500 list - november 2024}},\ \bibinfo
  {howpublished} {\url{https://top500.org/lists/green500/2024/11/}} (\bibinfo
  {year} {2024}),\ \bibinfo {note} {accessed: 2025-08-10}\BibitemShut {NoStop}%
\bibitem [{\citenamefont {Kwiat}\ \emph {et~al.}(1995)\citenamefont {Kwiat},
  \citenamefont {Mattle}, \citenamefont {Weinfurter}, \citenamefont
  {Zeilinger}, \citenamefont {Sergienko},\ and\ \citenamefont
  {Shih}}]{Kwiat1995}%
  \BibitemOpen
  \bibfield  {author} {\bibinfo {author} {\bibfnamefont {P.~G.}\ \bibnamefont
  {Kwiat}}, \bibinfo {author} {\bibfnamefont {K.}~\bibnamefont {Mattle}},
  \bibinfo {author} {\bibfnamefont {H.}~\bibnamefont {Weinfurter}}, \bibinfo
  {author} {\bibfnamefont {A.}~\bibnamefont {Zeilinger}}, \bibinfo {author}
  {\bibfnamefont {A.~V.}\ \bibnamefont {Sergienko}},\ and\ \bibinfo {author}
  {\bibfnamefont {Y.}~\bibnamefont {Shih}},\ }\bibfield  {title} {\bibinfo
  {title} {New high-intensity source of polarization-entangled photon pairs},\
  }\href {https://doi.org/10.1103/physrevlett.75.4337} {\bibfield  {journal}
  {\bibinfo  {journal} {Physical Review Letters}\ }\textbf {\bibinfo {volume}
  {75}},\ \bibinfo {pages} {4337} (\bibinfo {year} {1995})}\BibitemShut
  {NoStop}%
\bibitem [{\citenamefont {Burnham}\ and\ \citenamefont
  {Weinberg}(1970)}]{Burnham1970}%
  \BibitemOpen
  \bibfield  {author} {\bibinfo {author} {\bibfnamefont {D.~C.}\ \bibnamefont
  {Burnham}}\ and\ \bibinfo {author} {\bibfnamefont {D.~L.}\ \bibnamefont
  {Weinberg}},\ }\bibfield  {title} {\bibinfo {title} {Observation of
  simultaneity in parametric production of optical photon pairs},\ }\href
  {https://doi.org/10.1103/physrevlett.25.84} {\bibfield  {journal} {\bibinfo
  {journal} {Physical Review Letters}\ }\textbf {\bibinfo {volume} {25}},\
  \bibinfo {pages} {84} (\bibinfo {year} {1970})}\BibitemShut {NoStop}%
\bibitem [{\citenamefont {Hedges}\ \emph {et~al.}(2010)\citenamefont {Hedges},
  \citenamefont {Longdell}, \citenamefont {Li},\ and\ \citenamefont
  {Sellars}}]{Hedges2010}%
  \BibitemOpen
  \bibfield  {author} {\bibinfo {author} {\bibfnamefont {M.~P.}\ \bibnamefont
  {Hedges}}, \bibinfo {author} {\bibfnamefont {J.~J.}\ \bibnamefont
  {Longdell}}, \bibinfo {author} {\bibfnamefont {Y.}~\bibnamefont {Li}},\ and\
  \bibinfo {author} {\bibfnamefont {M.~J.}\ \bibnamefont {Sellars}},\
  }\bibfield  {title} {\bibinfo {title} {Efficient quantum memory for light},\
  }\href {https://doi.org/10.1038/nature09081} {\bibfield  {journal} {\bibinfo
  {journal} {Nature}\ }\textbf {\bibinfo {volume} {465}},\ \bibinfo {pages}
  {1052} (\bibinfo {year} {2010})}\BibitemShut {NoStop}%
\bibitem [{\citenamefont {Bruzewicz}\ \emph {et~al.}(2019)\citenamefont
  {Bruzewicz}, \citenamefont {Chiaverini}, \citenamefont {McConnell},\ and\
  \citenamefont {Sage}}]{Bruzewicz2019}%
  \BibitemOpen
  \bibfield  {author} {\bibinfo {author} {\bibfnamefont {C.~D.}\ \bibnamefont
  {Bruzewicz}}, \bibinfo {author} {\bibfnamefont {J.}~\bibnamefont
  {Chiaverini}}, \bibinfo {author} {\bibfnamefont {R.}~\bibnamefont
  {McConnell}},\ and\ \bibinfo {author} {\bibfnamefont {J.~M.}\ \bibnamefont
  {Sage}},\ }\bibfield  {title} {\bibinfo {title} {Trapped-ion quantum
  computing: Progress and challenges},\ }\bibfield  {journal} {\bibinfo
  {journal} {Applied Physics Reviews}\ }\textbf {\bibinfo {volume} {6}},\ \href
  {https://doi.org/10.1063/1.5088164} {10.1063/1.5088164} (\bibinfo {year}
  {2019})\BibitemShut {NoStop}%
\bibitem [{\citenamefont {Liu}\ \emph {et~al.}(2021)\citenamefont {Liu},
  \citenamefont {Hu}, \citenamefont {Li}, \citenamefont {Li}, \citenamefont
  {Li}, \citenamefont {Liang}, \citenamefont {Zhou}, \citenamefont {Li},\ and\
  \citenamefont {Guo}}]{Liu2021}%
  \BibitemOpen
  \bibfield  {author} {\bibinfo {author} {\bibfnamefont {X.}~\bibnamefont
  {Liu}}, \bibinfo {author} {\bibfnamefont {J.}~\bibnamefont {Hu}}, \bibinfo
  {author} {\bibfnamefont {Z.-F.}\ \bibnamefont {Li}}, \bibinfo {author}
  {\bibfnamefont {X.}~\bibnamefont {Li}}, \bibinfo {author} {\bibfnamefont
  {P.-Y.}\ \bibnamefont {Li}}, \bibinfo {author} {\bibfnamefont {P.-J.}\
  \bibnamefont {Liang}}, \bibinfo {author} {\bibfnamefont {Z.-Q.}\ \bibnamefont
  {Zhou}}, \bibinfo {author} {\bibfnamefont {C.-F.}\ \bibnamefont {Li}},\ and\
  \bibinfo {author} {\bibfnamefont {G.-C.}\ \bibnamefont {Guo}},\ }\bibfield
  {title} {\bibinfo {title} {Heralded entanglement distribution between two
  absorptive quantum memories},\ }\href
  {https://doi.org/10.1038/s41586-021-03505-3} {\bibfield  {journal} {\bibinfo
  {journal} {Nature}\ }\textbf {\bibinfo {volume} {594}},\ \bibinfo {pages}
  {41} (\bibinfo {year} {2021})}\BibitemShut {NoStop}%
\bibitem [{\citenamefont {Barnum}\ \emph {et~al.}(2002)\citenamefont {Barnum},
  \citenamefont {Crepeau}, \citenamefont {Gottesman}, \citenamefont {Smith},\
  and\ \citenamefont {Tapp}}]{Barnum}%
  \BibitemOpen
  \bibfield  {author} {\bibinfo {author} {\bibfnamefont {H.}~\bibnamefont
  {Barnum}}, \bibinfo {author} {\bibfnamefont {C.}~\bibnamefont {Crepeau}},
  \bibinfo {author} {\bibfnamefont {D.}~\bibnamefont {Gottesman}}, \bibinfo
  {author} {\bibfnamefont {A.}~\bibnamefont {Smith}},\ and\ \bibinfo {author}
  {\bibfnamefont {A.}~\bibnamefont {Tapp}},\ }\bibfield  {title} {\bibinfo
  {title} {Authentication of quantum messages},\ }in\ \href
  {https://doi.org/10.1109/sfcs.2002.1181969} {\emph {\bibinfo {booktitle} {The
  43rd Annual IEEE Symposium on Foundations of Computer Science, 2002.
  Proceedings.}}},\ \bibinfo {series and number} {SFCS-02}\ (\bibinfo
  {publisher} {IEEE Comput. Soc},\ \bibinfo {year} {2002})\ pp.\ \bibinfo
  {pages} {449--?58}\BibitemShut {NoStop}%
\bibitem [{\citenamefont {Yin}\ \emph {et~al.}(2020)\citenamefont {Yin},
  \citenamefont {Li}, \citenamefont {Liao}, \citenamefont {Yang}, \citenamefont
  {Cao}, \citenamefont {Zhang}, \citenamefont {Ren}, \citenamefont {Cai},
  \citenamefont {Liu}, \citenamefont {Li}, \citenamefont {Shu}, \citenamefont
  {Huang}, \citenamefont {Deng}, \citenamefont {Li}, \citenamefont {Zhang},
  \citenamefont {Liu}, \citenamefont {Chen}, \citenamefont {Lu}, \citenamefont
  {Wang}, \citenamefont {Xu}, \citenamefont {Wang}, \citenamefont {Peng},
  \citenamefont {Ekert},\ and\ \citenamefont {Pan}}]{Yin2020}%
  \BibitemOpen
  \bibfield  {author} {\bibinfo {author} {\bibfnamefont {J.}~\bibnamefont
  {Yin}}, \bibinfo {author} {\bibfnamefont {Y.-H.}\ \bibnamefont {Li}},
  \bibinfo {author} {\bibfnamefont {S.-K.}\ \bibnamefont {Liao}}, \bibinfo
  {author} {\bibfnamefont {M.}~\bibnamefont {Yang}}, \bibinfo {author}
  {\bibfnamefont {Y.}~\bibnamefont {Cao}}, \bibinfo {author} {\bibfnamefont
  {L.}~\bibnamefont {Zhang}}, \bibinfo {author} {\bibfnamefont {J.-G.}\
  \bibnamefont {Ren}}, \bibinfo {author} {\bibfnamefont {W.-Q.}\ \bibnamefont
  {Cai}}, \bibinfo {author} {\bibfnamefont {W.-Y.}\ \bibnamefont {Liu}},
  \bibinfo {author} {\bibfnamefont {S.-L.}\ \bibnamefont {Li}}, \bibinfo
  {author} {\bibfnamefont {R.}~\bibnamefont {Shu}}, \bibinfo {author}
  {\bibfnamefont {Y.-M.}\ \bibnamefont {Huang}}, \bibinfo {author}
  {\bibfnamefont {L.}~\bibnamefont {Deng}}, \bibinfo {author} {\bibfnamefont
  {L.}~\bibnamefont {Li}}, \bibinfo {author} {\bibfnamefont {Q.}~\bibnamefont
  {Zhang}}, \bibinfo {author} {\bibfnamefont {N.-L.}\ \bibnamefont {Liu}},
  \bibinfo {author} {\bibfnamefont {Y.-A.}\ \bibnamefont {Chen}}, \bibinfo
  {author} {\bibfnamefont {C.-Y.}\ \bibnamefont {Lu}}, \bibinfo {author}
  {\bibfnamefont {X.-B.}\ \bibnamefont {Wang}}, \bibinfo {author}
  {\bibfnamefont {F.}~\bibnamefont {Xu}}, \bibinfo {author} {\bibfnamefont
  {J.-Y.}\ \bibnamefont {Wang}}, \bibinfo {author} {\bibfnamefont {C.-Z.}\
  \bibnamefont {Peng}}, \bibinfo {author} {\bibfnamefont {A.~K.}\ \bibnamefont
  {Ekert}},\ and\ \bibinfo {author} {\bibfnamefont {J.-W.}\ \bibnamefont
  {Pan}},\ }\bibfield  {title} {\bibinfo {title} {Entanglement-based secure
  quantum cryptography over 1, 120 kilometres},\ }\href
  {https://doi.org/10.1038/s41586-020-2401-y} {\bibfield  {journal} {\bibinfo
  {journal} {Nature}\ }\textbf {\bibinfo {volume} {582}},\ \bibinfo {pages}
  {501} (\bibinfo {year} {2020})}\BibitemShut {NoStop}%
\bibitem [{\citenamefont {Caleffi}(2017)}]{Caleffi2017}%
  \BibitemOpen
  \bibfield  {author} {\bibinfo {author} {\bibfnamefont {M.}~\bibnamefont
  {Caleffi}},\ }\bibfield  {title} {\bibinfo {title} {Optimal routing for
  quantum networks},\ }\href {https://doi.org/10.1109/access.2017.2763325}
  {\bibfield  {journal} {\bibinfo  {journal} {IEEE Access}\ }\textbf {\bibinfo
  {volume} {5}},\ \bibinfo {pages} {22299} (\bibinfo {year}
  {2017})}\BibitemShut {NoStop}%
\bibitem [{\citenamefont {Pirandola}\ \emph {et~al.}(2015)\citenamefont
  {Pirandola}, \citenamefont {Eisert}, \citenamefont {Weedbrook}, \citenamefont
  {Furusawa},\ and\ \citenamefont {Braunstein}}]{Pirandola2015}%
  \BibitemOpen
  \bibfield  {author} {\bibinfo {author} {\bibfnamefont {S.}~\bibnamefont
  {Pirandola}}, \bibinfo {author} {\bibfnamefont {J.}~\bibnamefont {Eisert}},
  \bibinfo {author} {\bibfnamefont {C.}~\bibnamefont {Weedbrook}}, \bibinfo
  {author} {\bibfnamefont {A.}~\bibnamefont {Furusawa}},\ and\ \bibinfo
  {author} {\bibfnamefont {S.~L.}\ \bibnamefont {Braunstein}},\ }\bibfield
  {title} {\bibinfo {title} {Advances in quantum teleportation},\ }\href
  {https://doi.org/10.1038/nphoton.2015.154} {\bibfield  {journal} {\bibinfo
  {journal} {Nature Photonics}\ }\textbf {\bibinfo {volume} {9}},\ \bibinfo
  {pages} {641} (\bibinfo {year} {2015})}\BibitemShut {NoStop}%
\bibitem [{\citenamefont {Holevo}(1998)}]{Holevo1998}%
  \BibitemOpen
  \bibfield  {author} {\bibinfo {author} {\bibfnamefont {A.}~\bibnamefont
  {Holevo}},\ }\bibfield  {title} {\bibinfo {title} {The capacity of the
  quantum channel with general signal states},\ }\href
  {https://doi.org/10.1109/18.651037} {\bibfield  {journal} {\bibinfo
  {journal} {IEEE Transactions on Information Theory}\ }\textbf {\bibinfo
  {volume} {44}},\ \bibinfo {pages} {269} (\bibinfo {year} {1998})}\BibitemShut
  {NoStop}%
\bibitem [{\citenamefont {Wilde}(2013)}]{Wilde2013}%
  \BibitemOpen
  \bibfield  {author} {\bibinfo {author} {\bibfnamefont {M.~M.}\ \bibnamefont
  {Wilde}},\ }\href {https://doi.org/10.1017/cbo9781139525343} {\emph {\bibinfo
  {title} {Quantum Information Theory}}}\ (\bibinfo  {publisher} {Cambridge
  University Press},\ \bibinfo {year} {2013})\BibitemShut {NoStop}%
\bibitem [{\citenamefont {Horodecki}\ \emph {et~al.}(1999)\citenamefont
  {Horodecki}, \citenamefont {Horodecki},\ and\ \citenamefont
  {Horodecki}}]{Horodecki1999}%
  \BibitemOpen
  \bibfield  {author} {\bibinfo {author} {\bibfnamefont {M.}~\bibnamefont
  {Horodecki}}, \bibinfo {author} {\bibfnamefont {P.}~\bibnamefont
  {Horodecki}},\ and\ \bibinfo {author} {\bibfnamefont {R.}~\bibnamefont
  {Horodecki}},\ }\bibfield  {title} {\bibinfo {title} {General teleportation
  channel, singlet fraction, and quasidistillation},\ }\href
  {https://doi.org/10.1103/physreva.60.1888} {\bibfield  {journal} {\bibinfo
  {journal} {Physical Review A}\ }\textbf {\bibinfo {volume} {60}},\ \bibinfo
  {pages} {1888} (\bibinfo {year} {1999})}\BibitemShut {NoStop}%
\bibitem [{\citenamefont {Curty}\ \emph {et~al.}(2004)\citenamefont {Curty},
  \citenamefont {Lewenstein},\ and\ \citenamefont
  {L\"{u}tkenhaus}}]{Curty2004}%
  \BibitemOpen
  \bibfield  {author} {\bibinfo {author} {\bibfnamefont {M.}~\bibnamefont
  {Curty}}, \bibinfo {author} {\bibfnamefont {M.}~\bibnamefont {Lewenstein}},\
  and\ \bibinfo {author} {\bibfnamefont {N.}~\bibnamefont {L\"{u}tkenhaus}},\
  }\bibfield  {title} {\bibinfo {title} {Entanglement as a precondition for
  secure quantum key distribution},\ }\bibfield  {journal} {\bibinfo  {journal}
  {Physical Review Letters}\ }\textbf {\bibinfo {volume} {92}},\ \href
  {https://doi.org/10.1103/physrevlett.92.217903}
  {10.1103/physrevlett.92.217903} (\bibinfo {year} {2004})\BibitemShut
  {NoStop}%
\bibitem [{\citenamefont {Pantoja}\ \emph {et~al.}(2024)\citenamefont
  {Pantoja}, \citenamefont {Bucheli},\ and\ \citenamefont
  {Donaldson}}]{Pantoja2024}%
  \BibitemOpen
  \bibfield  {author} {\bibinfo {author} {\bibfnamefont {J.~J.}\ \bibnamefont
  {Pantoja}}, \bibinfo {author} {\bibfnamefont {V.~A.}\ \bibnamefont
  {Bucheli}},\ and\ \bibinfo {author} {\bibfnamefont {R.}~\bibnamefont
  {Donaldson}},\ }\bibfield  {title} {\bibinfo {title} {Electromagnetic
  side-channel attack risk assessment on a practical quantum-key-distribution
  receiver based on multi-class classification},\ }\bibfield  {journal}
  {\bibinfo  {journal} {EPJ Quantum Technology}\ }\textbf {\bibinfo {volume}
  {11}},\ \href {https://doi.org/10.1140/epjqt/s40507-024-00290-6}
  {10.1140/epjqt/s40507-024-00290-6} (\bibinfo {year} {2024})\BibitemShut
  {NoStop}%
\bibitem [{\citenamefont {Qi}\ \emph {et~al.}(2007)\citenamefont {Qi},
  \citenamefont {Fung}, \citenamefont {Lo},\ and\ \citenamefont {Ma}}]{Qi2007}%
  \BibitemOpen
  \bibfield  {author} {\bibinfo {author} {\bibfnamefont {B.}~\bibnamefont
  {Qi}}, \bibinfo {author} {\bibfnamefont {C.-H.~F.}\ \bibnamefont {Fung}},
  \bibinfo {author} {\bibfnamefont {H.-K.}\ \bibnamefont {Lo}},\ and\ \bibinfo
  {author} {\bibfnamefont {F.-X.}\ \bibnamefont {Ma}},\ }\bibfield  {title}
  {\bibinfo {title} {Time-shift attack in practical quantum cryptosystems},\
  }\href {https://doi.org/10.26421/qic7.1-2-3} {\bibfield  {journal} {\bibinfo
  {journal} {Quantum Information and Computation}\ }\textbf {\bibinfo {volume}
  {7}},\ \bibinfo {pages} {73} (\bibinfo {year} {2007})}\BibitemShut {NoStop}%
\bibitem [{\citenamefont {Gisin}\ \emph {et~al.}(2006)\citenamefont {Gisin},
  \citenamefont {Fasel}, \citenamefont {Kraus}, \citenamefont {Zbinden},\ and\
  \citenamefont {Ribordy}}]{Gisin2006}%
  \BibitemOpen
  \bibfield  {author} {\bibinfo {author} {\bibfnamefont {N.}~\bibnamefont
  {Gisin}}, \bibinfo {author} {\bibfnamefont {S.}~\bibnamefont {Fasel}},
  \bibinfo {author} {\bibfnamefont {B.}~\bibnamefont {Kraus}}, \bibinfo
  {author} {\bibfnamefont {H.}~\bibnamefont {Zbinden}},\ and\ \bibinfo {author}
  {\bibfnamefont {G.}~\bibnamefont {Ribordy}},\ }\bibfield  {title} {\bibinfo
  {title} {Trojan-horse attacks on quantum-key-distribution systems},\
  }\bibfield  {journal} {\bibinfo  {journal} {Physical Review A}\ }\textbf
  {\bibinfo {volume} {73}},\ \href {https://doi.org/10.1103/physreva.73.022320}
  {10.1103/physreva.73.022320} (\bibinfo {year} {2006})\BibitemShut {NoStop}%
\bibitem [{\citenamefont {Brunner}\ \emph {et~al.}(2014)\citenamefont
  {Brunner}, \citenamefont {Cavalcanti}, \citenamefont {Pironio}, \citenamefont
  {Scarani},\ and\ \citenamefont {Wehner}}]{Brunner2014}%
  \BibitemOpen
  \bibfield  {author} {\bibinfo {author} {\bibfnamefont {N.}~\bibnamefont
  {Brunner}}, \bibinfo {author} {\bibfnamefont {D.}~\bibnamefont {Cavalcanti}},
  \bibinfo {author} {\bibfnamefont {S.}~\bibnamefont {Pironio}}, \bibinfo
  {author} {\bibfnamefont {V.}~\bibnamefont {Scarani}},\ and\ \bibinfo {author}
  {\bibfnamefont {S.}~\bibnamefont {Wehner}},\ }\bibfield  {title} {\bibinfo
  {title} {Bell nonlocality},\ }\href
  {https://doi.org/10.1103/revmodphys.86.419} {\bibfield  {journal} {\bibinfo
  {journal} {Reviews of Modern Physics}\ }\textbf {\bibinfo {volume} {86}},\
  \bibinfo {pages} {419} (\bibinfo {year} {2014})}\BibitemShut {NoStop}%
\bibitem [{\citenamefont {Lo}\ \emph {et~al.}(2005)\citenamefont {Lo},
  \citenamefont {Ma},\ and\ \citenamefont {Chen}}]{Lo2005}%
  \BibitemOpen
  \bibfield  {author} {\bibinfo {author} {\bibfnamefont {H.-K.}\ \bibnamefont
  {Lo}}, \bibinfo {author} {\bibfnamefont {X.}~\bibnamefont {Ma}},\ and\
  \bibinfo {author} {\bibfnamefont {K.}~\bibnamefont {Chen}},\ }\bibfield
  {title} {\bibinfo {title} {Decoy state quantum key distribution},\ }\bibfield
   {journal} {\bibinfo  {journal} {Physical Review Letters}\ }\textbf {\bibinfo
  {volume} {94}},\ \href {https://doi.org/10.1103/physrevlett.94.230504}
  {10.1103/physrevlett.94.230504} (\bibinfo {year} {2005})\BibitemShut
  {NoStop}%
\bibitem [{\citenamefont {Winitzki}(2008)}]{winitzki2008handy}%
  \BibitemOpen
  \bibfield  {author} {\bibinfo {author} {\bibfnamefont {S.}~\bibnamefont
  {Winitzki}},\ }\bibfield  {title} {\bibinfo {title} {A handy approximation
  for the error function and its inverse},\ }\href@noop {} {\bibfield
  {journal} {\bibinfo  {journal} {A lecture note obtained through private
  communication}\ } (\bibinfo {year} {2008})}\BibitemShut {NoStop}%
\bibitem [{\citenamefont {Holevo}(1973)}]{Holevo1973Bounds}%
  \BibitemOpen
  \bibfield  {author} {\bibinfo {author} {\bibfnamefont {A.~S.}\ \bibnamefont
  {Holevo}},\ }\bibfield  {title} {\bibinfo {title} {Bounds for the quantity of
  information transmitted by a quantum communication channel},\ }\href
  {http://mi.mathnet.ru/eng/ppi903} {\bibfield  {journal} {\bibinfo  {journal}
  {Problems of Information Transmission}\ }\textbf {\bibinfo {volume} {9}},\
  \bibinfo {pages} {177} (\bibinfo {year} {1973})}\BibitemShut {NoStop}%
\bibitem [{\citenamefont {Koeune}\ and\ \citenamefont
  {Standaert}(2005)}]{Koeune2005}%
  \BibitemOpen
  \bibfield  {author} {\bibinfo {author} {\bibfnamefont {F.}~\bibnamefont
  {Koeune}}\ and\ \bibinfo {author} {\bibfnamefont {F.-X.}\ \bibnamefont
  {Standaert}},\ }\bibinfo {title} {A tutorial on physical security and
  side-channel attacks},\ in\ \href {https://doi.org/10.1007/11554578_3} {\emph
  {\bibinfo {booktitle} {Foundations of Security Analysis and Design III}}}\
  (\bibinfo  {publisher} {Springer Berlin Heidelberg},\ \bibinfo {year}
  {2005})\ pp.\ \bibinfo {pages} {78--?08}\BibitemShut {NoStop}%
\bibitem [{\citenamefont {Lucamarini}\ \emph {et~al.}(2015)\citenamefont
  {Lucamarini}, \citenamefont {Choi}, \citenamefont {Ward}, \citenamefont
  {Dynes}, \citenamefont {Yuan},\ and\ \citenamefont
  {Shields}}]{Lucamarini2015}%
  \BibitemOpen
  \bibfield  {author} {\bibinfo {author} {\bibfnamefont {M.}~\bibnamefont
  {Lucamarini}}, \bibinfo {author} {\bibfnamefont {I.}~\bibnamefont {Choi}},
  \bibinfo {author} {\bibfnamefont {M.-.}\ \bibnamefont {Ward}}, \bibinfo
  {author} {\bibfnamefont {J.-.}\ \bibnamefont {Dynes}}, \bibinfo {author}
  {\bibfnamefont {Z.-.}\ \bibnamefont {Yuan}},\ and\ \bibinfo {author}
  {\bibfnamefont {A.-.}\ \bibnamefont {Shields}},\ }\bibfield  {title}
  {\bibinfo {title} {Practical security bounds against the trojan-horse attack
  in quantum key distribution},\ }\bibfield  {journal} {\bibinfo  {journal}
  {Physical Review X}\ }\textbf {\bibinfo {volume} {5}},\ \href
  {https://doi.org/10.1103/physrevx.5.031030} {10.1103/physrevx.5.031030}
  (\bibinfo {year} {2015})\BibitemShut {NoStop}%
\bibitem [{\citenamefont {Ghosh}\ \emph {et~al.}(2025)\citenamefont {Ghosh},
  \citenamefont {Upadhyay},\ and\ \citenamefont {Ash~Saki}}]{Ghosh2025}%
  \BibitemOpen
  \bibfield  {author} {\bibinfo {author} {\bibfnamefont {S.}~\bibnamefont
  {Ghosh}}, \bibinfo {author} {\bibfnamefont {S.}~\bibnamefont {Upadhyay}},\
  and\ \bibinfo {author} {\bibfnamefont {A.}~\bibnamefont {Ash~Saki}},\
  }\bibfield  {title} {\bibinfo {title} {A primer on security of quantum
  computing hardware},\ }\href {https://doi.org/10.1109/jproc.2025.3630989}
  {\bibfield  {journal} {\bibinfo  {journal} {Proceedings of the IEEE}\
  }\textbf {\bibinfo {volume} {113}},\ \bibinfo {pages} {640} (\bibinfo {year}
  {2025})}\BibitemShut {NoStop}%
\bibitem [{\citenamefont {Ruijters}\ \emph {et~al.}(2019)\citenamefont
  {Ruijters}, \citenamefont {Reijsbergen}, \citenamefont {de~Boer},\ and\
  \citenamefont {Stoelinga}}]{Ruijters2019}%
  \BibitemOpen
  \bibfield  {author} {\bibinfo {author} {\bibfnamefont {E.}~\bibnamefont
  {Ruijters}}, \bibinfo {author} {\bibfnamefont {D.}~\bibnamefont
  {Reijsbergen}}, \bibinfo {author} {\bibfnamefont {P.-T.}\ \bibnamefont
  {de~Boer}},\ and\ \bibinfo {author} {\bibfnamefont {M.}~\bibnamefont
  {Stoelinga}},\ }\bibfield  {title} {\bibinfo {title} {Rare event simulation
  for dynamic fault trees},\ }\href
  {https://doi.org/10.1016/j.ress.2019.02.004} {\bibfield  {journal} {\bibinfo
  {journal} {Reliability Engineering \& System Safety}\ }\textbf {\bibinfo
  {volume} {186}},\ \bibinfo {pages} {220} (\bibinfo {year}
  {2019})}\BibitemShut {NoStop}%
\bibitem [{\citenamefont {Zeng}\ \emph {et~al.}(2023)\citenamefont {Zeng},
  \citenamefont {Barros},\ and\ \citenamefont {Coit}}]{Zeng2023}%
  \BibitemOpen
  \bibfield  {author} {\bibinfo {author} {\bibfnamefont {Z.}~\bibnamefont
  {Zeng}}, \bibinfo {author} {\bibfnamefont {A.}~\bibnamefont {Barros}},\ and\
  \bibinfo {author} {\bibfnamefont {D.}~\bibnamefont {Coit}},\ }\bibfield
  {title} {\bibinfo {title} {Dependent failure behavior modeling for risk and
  reliability: A systematic and critical literature review},\ }\href
  {https://doi.org/10.1016/j.ress.2023.109515} {\bibfield  {journal} {\bibinfo
  {journal} {Reliability Engineering \& System Safety}\ }\textbf {\bibinfo
  {volume} {239}},\ \bibinfo {pages} {109515} (\bibinfo {year}
  {2023})}\BibitemShut {NoStop}%
\bibitem [{\citenamefont {Karr}(1984)}]{Karr1984}%
  \BibitemOpen
  \bibfield  {author} {\bibinfo {author} {\bibfnamefont {A.~F.}\ \bibnamefont
  {Karr}},\ }\bibfield  {title} {\bibinfo {title} {Stochastic processes
  (sheldon m. ross)},\ }\href {https://doi.org/10.1137/1026096} {\bibfield
  {journal} {\bibinfo  {journal} {SIAM Review}\ }\textbf {\bibinfo {volume}
  {26}},\ \bibinfo {pages} {448} (\bibinfo {year} {1984})}\BibitemShut
  {NoStop}%
\bibitem [{200(2002)}]{2002}%
  \BibitemOpen
  \href {https://doi.org/10.1007/978-3-540-24808-8} {\emph {\bibinfo {title}
  {Stochastic Models in Reliability and Maintenance}}}\ (\bibinfo  {publisher}
  {Springer Berlin Heidelberg},\ \bibinfo {year} {2002})\BibitemShut {NoStop}%
\bibitem [{\citenamefont {Oh}\ \emph {et~al.}(2002)\citenamefont {Oh},
  \citenamefont {Lee},\ and\ \citenamefont {Lee}}]{Oh2002}%
  \BibitemOpen
  \bibfield  {author} {\bibinfo {author} {\bibfnamefont {S.}~\bibnamefont
  {Oh}}, \bibinfo {author} {\bibfnamefont {S.}~\bibnamefont {Lee}},\ and\
  \bibinfo {author} {\bibfnamefont {H.-w.}\ \bibnamefont {Lee}},\ }\bibfield
  {title} {\bibinfo {title} {Fidelity of quantum teleportation through noisy
  channels},\ }\bibfield  {journal} {\bibinfo  {journal} {Physical Review A}\
  }\textbf {\bibinfo {volume} {66}},\ \href
  {https://doi.org/10.1103/physreva.66.022316} {10.1103/physreva.66.022316}
  (\bibinfo {year} {2002})\BibitemShut {NoStop}%
\bibitem [{\citenamefont {Im}\ \emph {et~al.}(2021)\citenamefont {Im},
  \citenamefont {Lee}, \citenamefont {Kim}, \citenamefont {Nha}, \citenamefont
  {Kim}, \citenamefont {Lee},\ and\ \citenamefont {Kim}}]{Im2021}%
  \BibitemOpen
  \bibfield  {author} {\bibinfo {author} {\bibfnamefont {D.-G.}\ \bibnamefont
  {Im}}, \bibinfo {author} {\bibfnamefont {C.-H.}\ \bibnamefont {Lee}},
  \bibinfo {author} {\bibfnamefont {Y.}~\bibnamefont {Kim}}, \bibinfo {author}
  {\bibfnamefont {H.}~\bibnamefont {Nha}}, \bibinfo {author} {\bibfnamefont
  {M.~S.}\ \bibnamefont {Kim}}, \bibinfo {author} {\bibfnamefont {S.-W.}\
  \bibnamefont {Lee}},\ and\ \bibinfo {author} {\bibfnamefont {Y.-H.}\
  \bibnamefont {Kim}},\ }\bibfield  {title} {\bibinfo {title} {Optimal
  teleportation via noisy quantum channels without additional qubit
  resources},\ }\bibfield  {journal} {\bibinfo  {journal} {npj Quantum
  Information}\ }\textbf {\bibinfo {volume} {7}},\ \href
  {https://doi.org/10.1038/s41534-021-00426-x} {10.1038/s41534-021-00426-x}
  (\bibinfo {year} {2021})\BibitemShut {NoStop}%
\bibitem [{Note1()}]{Note1}%
  \BibitemOpen
  \bibinfo {note} {Throughout this paper, $m_1 \equiv M_1$ and $m_2 \equiv
  M_2$. These symbols denote the same classical bits and are used
  interchangeably in some sections for notational convenience and
  clarity.}\BibitemShut {Stop}%
\bibitem [{\citenamefont {Casey}\ \emph {et~al.}(2025)\citenamefont {Casey},
  \citenamefont {Williams}, \citenamefont {McCaffrey}, \citenamefont
  {Rotherham},\ and\ \citenamefont
  {Darby}}]{casey2025multimodequantummemorieshighthroughput}%
  \BibitemOpen
  \bibfield  {author} {\bibinfo {author} {\bibfnamefont {C.}~\bibnamefont
  {Casey}}, \bibinfo {author} {\bibfnamefont {A.}~\bibnamefont {Williams}},
  \bibinfo {author} {\bibfnamefont {C.}~\bibnamefont {McCaffrey}}, \bibinfo
  {author} {\bibfnamefont {E.}~\bibnamefont {Rotherham}},\ and\ \bibinfo
  {author} {\bibfnamefont {N.}~\bibnamefont {Darby}},\ }\href
  {https://doi.org/10.48550/ARXIV.2512.00282} {\bibinfo {title} {Multi-mode
  quantum memories for high-throughput satellite entanglement distribution}}
  (\bibinfo {year} {2025})\BibitemShut {NoStop}%
\bibitem [{\citenamefont {Lodahl}(2017)}]{Lodahl2017}%
  \BibitemOpen
  \bibfield  {author} {\bibinfo {author} {\bibfnamefont {P.}~\bibnamefont
  {Lodahl}},\ }\bibfield  {title} {\bibinfo {title} {Quantum-dot based photonic
  quantum networks},\ }\href {https://doi.org/10.1088/2058-9565/aa91bb}
  {\bibfield  {journal} {\bibinfo  {journal} {Quantum Science and Technology}\
  }\textbf {\bibinfo {volume} {3}},\ \bibinfo {pages} {013001} (\bibinfo {year}
  {2017})}\BibitemShut {NoStop}%
\bibitem [{\citenamefont {Krantz}\ \emph {et~al.}(2019)\citenamefont {Krantz},
  \citenamefont {Kjaergaard}, \citenamefont {Yan}, \citenamefont {Orlando},
  \citenamefont {Gustavsson},\ and\ \citenamefont {Oliver}}]{Krantz2019}%
  \BibitemOpen
  \bibfield  {author} {\bibinfo {author} {\bibfnamefont {P.}~\bibnamefont
  {Krantz}}, \bibinfo {author} {\bibfnamefont {M.}~\bibnamefont {Kjaergaard}},
  \bibinfo {author} {\bibfnamefont {F.}~\bibnamefont {Yan}}, \bibinfo {author}
  {\bibfnamefont {T.~P.}\ \bibnamefont {Orlando}}, \bibinfo {author}
  {\bibfnamefont {S.}~\bibnamefont {Gustavsson}},\ and\ \bibinfo {author}
  {\bibfnamefont {W.~D.}\ \bibnamefont {Oliver}},\ }\bibfield  {title}
  {\bibinfo {title} {A quantum engineer- guide to superconducting qubits},\
  }\bibfield  {journal} {\bibinfo  {journal} {Applied Physics Reviews}\
  }\textbf {\bibinfo {volume} {6}},\ \href {https://doi.org/10.1063/1.5089550}
  {10.1063/1.5089550} (\bibinfo {year} {2019})\BibitemShut {NoStop}%
\bibitem [{\citenamefont {Gao}\ \emph {et~al.}(2013)\citenamefont {Gao},
  \citenamefont {Fallahi}, \citenamefont {Togan}, \citenamefont {Delteil},
  \citenamefont {Chin}, \citenamefont {Miguel-Sanchez},\ and\ \citenamefont
  {Imamoglu}}]{Gao2013}%
  \BibitemOpen
  \bibfield  {author} {\bibinfo {author} {\bibfnamefont {W.}~\bibnamefont
  {Gao}}, \bibinfo {author} {\bibfnamefont {P.}~\bibnamefont {Fallahi}},
  \bibinfo {author} {\bibfnamefont {E.}~\bibnamefont {Togan}}, \bibinfo
  {author} {\bibfnamefont {A.}~\bibnamefont {Delteil}}, \bibinfo {author}
  {\bibfnamefont {Y.}~\bibnamefont {Chin}}, \bibinfo {author} {\bibfnamefont
  {J.}~\bibnamefont {Miguel-Sanchez}},\ and\ \bibinfo {author} {\bibfnamefont
  {A.}~\bibnamefont {Imamoglu}},\ }\bibfield  {title} {\bibinfo {title}
  {Quantum teleportation from a propagating photon to a solid-state spin
  qubit},\ }\bibfield  {journal} {\bibinfo  {journal} {Nature Communications}\
  }\textbf {\bibinfo {volume} {4}},\ \href {https://doi.org/10.1038/ncomms3744}
  {10.1038/ncomms3744} (\bibinfo {year} {2013})\BibitemShut {NoStop}%
\bibitem [{\citenamefont {Weinfurter}(1994)}]{Weinfurter1994}%
  \BibitemOpen
  \bibfield  {author} {\bibinfo {author} {\bibfnamefont {H.}~\bibnamefont
  {Weinfurter}},\ }\bibfield  {title} {\bibinfo {title} {Experimental
  bell-state analysis},\ }\href@noop {} {\bibfield  {journal} {\bibinfo
  {journal} {Europhysics Letters}\ }\textbf {\bibinfo {volume} {25}},\ \bibinfo
  {pages} {559} (\bibinfo {year} {1994})}\BibitemShut {NoStop}%
\bibitem [{\citenamefont {Barends}\ \emph {et~al.}(2014)\citenamefont
  {Barends}, \citenamefont {Kelly}, \citenamefont {Megrant}, \citenamefont
  {Veitia}, \citenamefont {Sank}, \citenamefont {Jeffrey}, \citenamefont
  {White}, \citenamefont {Mutus}, \citenamefont {Fowler}, \citenamefont
  {Campbell}, \citenamefont {Chen}, \citenamefont {Chen}, \citenamefont
  {Chiaro}, \citenamefont {Dunsworth}, \citenamefont {Neill}, \citenamefont
  {Roushan}, \citenamefont {Vainsencher}, \citenamefont {Wenner}, \citenamefont
  {Korotkov}, \citenamefont {Cleland},\ and\ \citenamefont
  {Martinis}}]{Barends2014}%
  \BibitemOpen
  \bibfield  {author} {\bibinfo {author} {\bibfnamefont {R.}~\bibnamefont
  {Barends}}, \bibinfo {author} {\bibfnamefont {J.}~\bibnamefont {Kelly}},
  \bibinfo {author} {\bibfnamefont {A.}~\bibnamefont {Megrant}}, \bibinfo
  {author} {\bibfnamefont {A.}~\bibnamefont {Veitia}}, \bibinfo {author}
  {\bibfnamefont {D.}~\bibnamefont {Sank}}, \bibinfo {author} {\bibfnamefont
  {E.}~\bibnamefont {Jeffrey}}, \bibinfo {author} {\bibfnamefont {T.~C.}\
  \bibnamefont {White}}, \bibinfo {author} {\bibfnamefont {J.}~\bibnamefont
  {Mutus}}, \bibinfo {author} {\bibfnamefont {A.~G.}\ \bibnamefont {Fowler}},
  \bibinfo {author} {\bibfnamefont {B.}~\bibnamefont {Campbell}}, \bibinfo
  {author} {\bibfnamefont {Y.}~\bibnamefont {Chen}}, \bibinfo {author}
  {\bibfnamefont {Z.}~\bibnamefont {Chen}}, \bibinfo {author} {\bibfnamefont
  {B.}~\bibnamefont {Chiaro}}, \bibinfo {author} {\bibfnamefont
  {A.}~\bibnamefont {Dunsworth}}, \bibinfo {author} {\bibfnamefont
  {C.}~\bibnamefont {Neill}}, \bibinfo {author} {\bibfnamefont
  {P.}~\bibnamefont {Roushan}}, \bibinfo {author} {\bibfnamefont
  {A.}~\bibnamefont {Vainsencher}}, \bibinfo {author} {\bibfnamefont
  {J.}~\bibnamefont {Wenner}}, \bibinfo {author} {\bibfnamefont {A.~N.}\
  \bibnamefont {Korotkov}}, \bibinfo {author} {\bibfnamefont {A.~N.}\
  \bibnamefont {Cleland}},\ and\ \bibinfo {author} {\bibfnamefont {J.~M.}\
  \bibnamefont {Martinis}},\ }\bibfield  {title} {\bibinfo {title}
  {Superconducting quantum circuits at the surface code threshold for fault
  tolerance},\ }\href {https://doi.org/10.1038/nature13171} {\bibfield
  {journal} {\bibinfo  {journal} {Nature}\ }\textbf {\bibinfo {volume} {508}},\
  \bibinfo {pages} {500} (\bibinfo {year} {2014})}\BibitemShut {NoStop}%
\bibitem [{\citenamefont {O-rien}(2007)}]{OBrien2007}%
  \BibitemOpen
  \bibfield  {author} {\bibinfo {author} {\bibfnamefont {J.~L.}\ \bibnamefont
  {O-rien}},\ }\bibfield  {title} {\bibinfo {title} {Optical quantum
  computing},\ }\href {https://doi.org/10.1126/science.1142892} {\bibfield
  {journal} {\bibinfo  {journal} {Science}\ }\textbf {\bibinfo {volume}
  {318}},\ \bibinfo {pages} {1567} (\bibinfo {year} {2007})}\BibitemShut
  {NoStop}%
\bibitem [{\citenamefont {Wang}\ \emph
  {et~al.}(2021{\natexlab{b}})\citenamefont {Wang}, \citenamefont {Luan},
  \citenamefont {Qiao}, \citenamefont {Um}, \citenamefont {Zhang},
  \citenamefont {Wang}, \citenamefont {Yuan}, \citenamefont {Gu}, \citenamefont
  {Zhang},\ and\ \citenamefont {Kim}}]{wang2021single}%
  \BibitemOpen
  \bibfield  {author} {\bibinfo {author} {\bibfnamefont {P.}~\bibnamefont
  {Wang}}, \bibinfo {author} {\bibfnamefont {C.-Y.}\ \bibnamefont {Luan}},
  \bibinfo {author} {\bibfnamefont {M.}~\bibnamefont {Qiao}}, \bibinfo {author}
  {\bibfnamefont {M.}~\bibnamefont {Um}}, \bibinfo {author} {\bibfnamefont
  {J.}~\bibnamefont {Zhang}}, \bibinfo {author} {\bibfnamefont
  {Y.}~\bibnamefont {Wang}}, \bibinfo {author} {\bibfnamefont {X.}~\bibnamefont
  {Yuan}}, \bibinfo {author} {\bibfnamefont {M.}~\bibnamefont {Gu}}, \bibinfo
  {author} {\bibfnamefont {J.}~\bibnamefont {Zhang}},\ and\ \bibinfo {author}
  {\bibfnamefont {K.}~\bibnamefont {Kim}},\ }\bibfield  {title} {\bibinfo
  {title} {Single ion qubit with estimated coherence time exceeding one hour},\
  }\bibfield  {journal} {\bibinfo  {journal} {Nature Communications}\ }\textbf
  {\bibinfo {volume} {12}},\ \href {https://doi.org/10.1038/s41467-020-20330-w}
  {10.1038/s41467-020-20330-w} (\bibinfo {year}
  {2021}{\natexlab{b}})\BibitemShut {NoStop}%
\bibitem [{\citenamefont {Somoroff}\ \emph {et~al.}(2023)\citenamefont
  {Somoroff}, \citenamefont {Ficheux}, \citenamefont {Mencia}, \citenamefont
  {Xiong}, \citenamefont {Kuzmin},\ and\ \citenamefont
  {Manucharyan}}]{somoroff2023millisecond}%
  \BibitemOpen
  \bibfield  {author} {\bibinfo {author} {\bibfnamefont {A.}~\bibnamefont
  {Somoroff}}, \bibinfo {author} {\bibfnamefont {Q.}~\bibnamefont {Ficheux}},
  \bibinfo {author} {\bibfnamefont {R.~A.}\ \bibnamefont {Mencia}}, \bibinfo
  {author} {\bibfnamefont {H.}~\bibnamefont {Xiong}}, \bibinfo {author}
  {\bibfnamefont {R.}~\bibnamefont {Kuzmin}},\ and\ \bibinfo {author}
  {\bibfnamefont {V.~E.}\ \bibnamefont {Manucharyan}},\ }\bibfield  {title}
  {\bibinfo {title} {Millisecond coherence in a superconducting qubit},\
  }\bibfield  {journal} {\bibinfo  {journal} {Physical Review Letters}\
  }\textbf {\bibinfo {volume} {130}},\ \href
  {https://doi.org/10.1103/physrevlett.130.267001}
  {10.1103/physrevlett.130.267001} (\bibinfo {year} {2023})\BibitemShut
  {NoStop}%
\bibitem [{\citenamefont {Milul}\ \emph {et~al.}(2023)\citenamefont {Milul},
  \citenamefont {Guttel}, \citenamefont {Goldblatt}, \citenamefont {Hazanov},
  \citenamefont {Joshi}, \citenamefont {Chausovsky}, \citenamefont {Kahn},
  \citenamefont {undefinedifty\"{u}rek}, \citenamefont {Lafont},\ and\
  \citenamefont {Rosenblum}}]{milul2023superconducting}%
  \BibitemOpen
  \bibfield  {author} {\bibinfo {author} {\bibfnamefont {O.}~\bibnamefont
  {Milul}}, \bibinfo {author} {\bibfnamefont {B.}~\bibnamefont {Guttel}},
  \bibinfo {author} {\bibfnamefont {U.}~\bibnamefont {Goldblatt}}, \bibinfo
  {author} {\bibfnamefont {S.}~\bibnamefont {Hazanov}}, \bibinfo {author}
  {\bibfnamefont {L.~M.}\ \bibnamefont {Joshi}}, \bibinfo {author}
  {\bibfnamefont {D.}~\bibnamefont {Chausovsky}}, \bibinfo {author}
  {\bibfnamefont {N.}~\bibnamefont {Kahn}}, \bibinfo {author} {\bibfnamefont
  {E.}~\bibnamefont {undefinedifty\"{u}rek}}, \bibinfo {author} {\bibfnamefont
  {F.}~\bibnamefont {Lafont}},\ and\ \bibinfo {author} {\bibfnamefont
  {S.}~\bibnamefont {Rosenblum}},\ }\bibfield  {title} {\bibinfo {title}
  {Superconducting cavity qubit with tens of milliseconds single-photon
  coherence time},\ }\bibfield  {journal} {\bibinfo  {journal} {PRX Quantum}\
  }\textbf {\bibinfo {volume} {4}},\ \href
  {https://doi.org/10.1103/prxquantum.4.030336} {10.1103/prxquantum.4.030336}
  (\bibinfo {year} {2023})\BibitemShut {NoStop}%
\bibitem [{\citenamefont {Ghinea}\ \emph {et~al.}(2023)\citenamefont {Ghinea},
  \citenamefont {Kaczmarczyck}, \citenamefont {Pullman}, \citenamefont
  {Cretin}, \citenamefont {K\"{o}lbl}, \citenamefont {Misoczki}, \citenamefont
  {Picod}, \citenamefont {Invernizzi},\ and\ \citenamefont
  {Bursztein}}]{Ghinea2023}%
  \BibitemOpen
  \bibfield  {author} {\bibinfo {author} {\bibfnamefont {D.}~\bibnamefont
  {Ghinea}}, \bibinfo {author} {\bibfnamefont {F.}~\bibnamefont
  {Kaczmarczyck}}, \bibinfo {author} {\bibfnamefont {J.}~\bibnamefont
  {Pullman}}, \bibinfo {author} {\bibfnamefont {J.}~\bibnamefont {Cretin}},
  \bibinfo {author} {\bibfnamefont {S.}~\bibnamefont {K\"{o}lbl}}, \bibinfo
  {author} {\bibfnamefont {R.}~\bibnamefont {Misoczki}}, \bibinfo {author}
  {\bibfnamefont {J.-M.}\ \bibnamefont {Picod}}, \bibinfo {author}
  {\bibfnamefont {L.}~\bibnamefont {Invernizzi}},\ and\ \bibinfo {author}
  {\bibfnamefont {E.}~\bibnamefont {Bursztein}},\ }\bibinfo {title} {Hybrid
  post-quantum signatures in~hardware security keys},\ in\ \href
  {https://doi.org/10.1007/978-3-031-41181-6_26} {\emph {\bibinfo {booktitle}
  {Applied Cryptography and Network Security Workshops}}}\ (\bibinfo
  {publisher} {Springer Nature Switzerland},\ \bibinfo {year} {2023})\ pp.\
  \bibinfo {pages} {480--?99}\BibitemShut {NoStop}%
\bibitem [{\citenamefont {Babai}(1986)}]{Babai1986}%
  \BibitemOpen
  \bibfield  {author} {\bibinfo {author} {\bibfnamefont {L.}~\bibnamefont
  {Babai}},\ }\bibfield  {title} {\bibinfo {title} {On lovsz-?lattice reduction
  and the nearest lattice point problem},\ }\href
  {https://doi.org/10.1007/bf02579403} {\bibfield  {journal} {\bibinfo
  {journal} {Combinatorica}\ }\textbf {\bibinfo {volume} {6}},\ \bibinfo
  {pages} {1} (\bibinfo {year} {1986})}\BibitemShut {NoStop}%
\bibitem [{\citenamefont {Regev}(2006)}]{regev2006lattice}%
  \BibitemOpen
  \bibfield  {author} {\bibinfo {author} {\bibfnamefont {O.}~\bibnamefont
  {Regev}},\ }\bibinfo {title} {Lattice-based cryptography},\ in\ \href
  {https://doi.org/10.1007/11818175_8} {\emph {\bibinfo {booktitle} {Advances
  in Cryptology - CRYPTO 2006}}}\ (\bibinfo  {publisher} {Springer Berlin
  Heidelberg},\ \bibinfo {year} {2006})\ pp.\ \bibinfo {pages}
  {131--?41}\BibitemShut {NoStop}%
\end{thebibliography}%

\appendix

\clearpage
\section{Table of Notation}
\label{appendix:notation}

\begin{center}
\captionof{table}{Summary of principal symbols and notation used throughout this paper.}
\label{tab:notation}
\renewcommand{\arraystretch}{0.92}
\setlength{\tabcolsep}{4pt}
\scriptsize
\begin{tabular}{@{}lll@{}}
\toprule
\textbf{Symbol} & \textbf{Description} & \textbf{Defined in} \\
\midrule
\multicolumn{3}{@{}l}{\textit{Quantum states and operators}} \\
$|\varphi\rangle = \alpha|0\rangle+\beta|1\rangle$ & Alice's input qubit; $\alpha,\beta$ are the amplitudes ($|\alpha|^2+|\beta|^2=1$) & Sec.~\ref{sec:qrqt} \\
$|\Phi^+\rangle$, $|\beta_{00}\rangle$  & Shared Bell pair $(|00\rangle+|11\rangle)/\sqrt{2}$        & Sec.~\ref{sec:qrqt} \\
$(M_1,M_2)$               & Alice's Bell-measurement outcomes (two classical bits)               & Sec.~\ref{sec:qrqt} \\
$U_{\mathrm{SWAP}}$       & SWAP unitary applied by Eve in the quantum-channel attack            & Sec.~\ref{subsec:swap-attack} \\
$E_0,\,E_1$               & Kraus operators of the amplitude-damping channel                     & App.~\ref{app:holevo-derivation} \\
$\rho_{m_1m_2}$, $\rho_{\mathrm{avg}}$  & Bob's conditional state for outcome $(m_1,m_2)$; ensemble-averaged state & Sec.~\ref{sec:info-theoretic-security} \\
\midrule
\multicolumn{3}{@{}l}{\textit{QRQT protocol maps and timing}} \\
$\mathcal{T}_{\text{QRQT}}$     & End-to-end QRQT teleportation map                              & Sec.~\ref{sec:qrqt} \\
$\mathcal{P}_{\text{Bell}}$, $\mathcal{U}_{\text{Bell}}$, $\mathcal{M}_{\text{Alice}}$, $\mathcal{E}_{\mathrm{PQC}}$, $\mathcal{R}_{\text{Bob}}$
       & Protocol substep operators (preparation, transformation, measurement, PQC, correction) & Sec.~\ref{sec:qrqt} \\
$T_{\mathcal{P}_{\text{Bell}}}$, $T_{\mathcal{U}_{\text{Bell}}}$, $T_{\mathcal{M}_{\text{Alice}}}$, $T_{\mathcal{E}_{\text{PQC}}}$, $T_{\mathcal{R}_{\text{Bob}}}$
       & Latency of each protocol substep                               & Sec.~\ref{sec:qrqt}, App.~\ref{appendix: Latency notification} \\
$T_{\mathrm{comm}}$      & Classical communication delay ($= d/v_{\text{fiber}}$)               & Sec.~\ref{sec:qrqt} \\
$\tau_m$                  & Minimum quantum memory lifetime required for QRQT                   & Sec.~\ref{sec:qrqt} \\
$T_{\mathrm{coh}}$       & Quantum memory coherence time                                       & Sec.~\ref{sec:qrqt} \\
$T_{\text{fixed}}$       & Distance-independent protocol delay                                 & Eq.~\eqref{eq:max_distance}, App.~\ref{appendix:T_fixed} \\
$d$, $d_{\max}$          & Communication distance; maximum feasible distance                    & Sec.~\ref{sec:qrqt}, Eq.~\eqref{eq:max_distance} \\
$v_{\text{fiber}}$       & Speed of light in optical fiber ($\approx 2\times10^5$~km/s)        & Sec.~\ref{sec:qrqt} \\
\midrule
\multicolumn{3}{@{}l}{\textit{SWAP attack and quantum-side security}} \\
$F(t)$, $F_0$            & Time-dependent fidelity of Eve's recovered state; initial fidelity   & Sec.~\ref{subsec:swap-attack} \\
$P_{\mathrm{SWAP}}(t)$   & SWAP attack success probability ($= F_0\,e^{-t/T_{\mathrm{coh}}}$)  & Eq.~\eqref{PSWAP} \\
\midrule
\multicolumn{3}{@{}l}{\textit{Classical-side attack (LWE / BKZ)}} \\
$A$                      & Public matrix $\in\mathbb{Z}_q^{n\times m}$                          & Sec.~\ref{sec:PQC attack} \\
$\mathbf{sk}$, $\mathbf{e}$, $\mathbf{p}$ & Secret key vector, Gaussian error vector, public key / noisy samples & Sec.~\ref{sec:PQC attack} \\
$n$, $m$, $q$            & LWE secret dimension, number of samples / lattice dimension, modulus & Sec.~\ref{sec:PQC attack} \\
$s$                      & Standard deviation of the Gaussian noise                             & Sec.~\ref{sec:PQC attack} \\
$\sigma_e$                 & LWE noise width parameter ($\mathbf{e}\in\mathcal{D}_{\mathbb{Z},\sigma_e q}$) & Sec.~\ref{sec:PQC attack} \\
$\delta_{\mathrm{root}}$ & Root-Hermite factor of BKZ-reduced basis                             & Sec.~\ref{sec:PQC attack} \\
$\tilde{b}_i$, $\|\tilde{b}_1\|$  & $i$-th Gram--Schmidt vector of the reduced basis; length of the first & Sec.~\ref{sec:PQC attack} \\
$d_i$                    & Integer search radius at NearestPlanes level $i$                     & Sec.~\ref{sec:PQC attack} \\
$a$, $b$                 & BKZ runtime scaling coefficients characterizing adversarial computational capability   & Eq.~\eqref{eq:bkz_time_prelim}, App.~\ref{appendix:BKZ_params} \\
$T_{\mathrm{BKZ}}$       & BKZ lattice-reduction runtime                                       & Eq.~\eqref{eq:bkz_time_prelim} \\
$P_{\mathrm{LWE}}$       & LWE attack success probability                                      & Eq.~\eqref{eq:lwe_probability} \\
$\zeta$                  & GSA decay ratio ($\approx\delta_{\mathrm{root}}^{-2}$)               & App.~\ref{appendix:GSA} \\
$w$, $W_i$               & Pad\'e constant ($\approx 0.140012$); scaled argument in log-prob.\ formula & Eq.~\eqref{eq:pade_log_prob} \\
\midrule
\multicolumn{3}{@{}l}{\textit{Joint attack model}} \\
$P_{\mathrm{joint}}(t)$  & Joint classical--quantum attack success probability                  & Sec.~\ref{subsec:joint-attack} \\
$t^*$                    & Optimal attack time $\arg\max_t P_{\mathrm{joint}}(t)$               & Sec.~\ref{subsec:joint-attack} \\
$\epsilon$               & Security threshold parameter                                        & Sec.~\ref{subsec:joint-attack} \\
\midrule
\multicolumn{3}{@{}l}{\textit{Information-theoretic security (Holevo analysis)}} \\
$S(\rho)$                & Von Neumann entropy of density matrix $\rho$                         & Sec.~\ref{sec:info-theoretic-security} \\
$\chi$                   & Holevo quantity (upper bound on extractable information)              & Eq.~\eqref{eq:holevo-def} \\
$\gamma$                 & Amplitude-damping parameter                                         & Sec.~\ref{sec:holevo-damped} \\
$\delta$                 & Eigenvalue gap $\sqrt{1-4\gamma(1{-}\gamma)(1{-}|\alpha|^2)^2}$      & Sec.~\eqref{eigenval} \\
$h(\cdot)$               & Binary entropy function                                              & Sec.~\ref{sec:holevo-damped} \\
$D$                      & Determinant of conditional state $R$; $D=\gamma(1{-}\gamma)|\beta|^4$ & App.~\ref{app:holevo-derivation} \\
$J$, $K$                 & Population bias and residual coherence of $\rho_{\mathrm{avg}}$      & App.~\ref{app:holevo-derivation} \\
$\Delta$                 & Eigenvalue gap of $\rho_{\mathrm{avg}}$; $\sqrt{(1{-}2J)^2+4|K|^2}$ & App.~\ref{app:holevo-derivation} \\
$u$, $r$                 & Effective population $|\alpha|^2{+}\gamma|\beta|^2$; residual coherence $\Re(\alpha\beta^*\!\sqrt{1{-}\gamma})$ & Sec.~\ref{stochastic} \\
$K(t)$                   & Subset of leaked correction-bit indices at time $t$                  & Sec.~\ref{stochastic} \\
$\chi_K$, $F_K$          & Holevo quantity and fidelity for knowledge class $K$                 & Eq.~\ref{stochastic} \\
$k$, $k_1$, $k_2$       & Exponential leakage rates                                           & Sec.~\ref{stochastic} \\
$\mu$                    & Correlation parameter in correlated leakage model                    & Sec.~\ref{stochastic} \\
\midrule
\multicolumn{3}{@{}l}{\textit{PQC protocol and computation}} \\
$\mathrm{Enc}_{\mathrm{PQC}}(\cdot)$, $\mathrm{Enc}_{\mathrm{AES}}(\cdot)$ & PQC and AES encryption operations & App.~\ref{app:pqc protocol} \\
$C_1$, $C_2$             & PQC-encrypted session key; AES-encrypted correction bits             & App.~\ref{app:pqc protocol} \\
$K$ (session key)        & Random session key for hybrid encryption                             & App.~\ref{app:pqc protocol} \\
$pk_{\text{Bob}}$        & Bob's PQC public key                                                & App.~\ref{app:pqc protocol} \\
$N_{\mathrm{ops}}$, $R_{\mathrm{max}}$ & Number of integer operations for PQC; peak throughput (PFlop/s) & App.~\ref{appendix:JEDI} \\
\bottomrule
\end{tabular}
\end{center}
\clearpage

\section{Latency Data for Memory Lifetime Constraint Estimation}
\subsection{Overview of Protocol Timing Components}
\label{appendix: Latency notification}
The total memory lifetime $\tau_m$ (Eq.~\eqref{time equation}) comprises several
physical substeps whose symbols are defined in Table~\ref{tab:notation}.
Representative latencies for each substep are provided in Sec.~\ref{appendix:latency}:
\begin{itemize}
    \item $T_{\mathcal{P}_{\text{Bell}}}$ : Bell-pair preparation and distribution;
    \item $T_{\mathcal{U}_{\text{Bell}}}$, $T_{\mathcal{M}_{\text{Alice}}}$ : Bell-basis transformation and measurement;
    \item $T_{\mathcal{E}_{\text{PQC}}}$ : PQC encryption and decryption (estimated separately in App.~\ref{appendix:JEDI});
    \item $T_{\text{comm}}$ : Classical communication over optical fiber;
    \item $T_{\mathcal{R}_{\text{Bob}}}$ : Local Pauli corrections on Bob's qubit.
\end{itemize}

\subsection{Physical Latency Estimates}
\label{appendix:latency}

\paragraph{State preparation.}
Encoding a quantum state $|\varphi\rangle$ into a photon or superconducting qubit requires $\sim 10$--$500$~ns depending on the platform: single-photon sources typically operate at $\sim 500$~ns~\cite{Lodahl2017}, while superconducting circuits achieve $\sim 10$~ns~\cite{Krantz2019}. This step is local and distance-independent.

\paragraph{Bell-pair generation.}
Entangled photon pairs are produced via spontaneous parametric down-conversion (SPDC, $\sim 1$--$10$~ns\cite{Kwiat1995,Burnham1970}) or quantum dot emitters ($\sim 100$~ps~\cite{Gao2013}). The generation itself is local, but distributing one photon to a distant node adds a distance-dependent propagation delay.

\paragraph{Bell measurement.}
Alice's Bell-basis measurement involves linear optics and photon detection ($\sim 10$--$100$~ns~\cite{Weinfurter1994}) together with basis-changing gates ($\sim 1$--$10$~ns~\cite{Barends2014}). This step is performed locally.

\paragraph{Classical communication.}
The two-bit measurement result must be transmitted to Bob over optical fiber. The delay scales linearly with distance:
\begin{equation}
T_{\mathrm{comm}} = \frac{d}{v_{\mathrm{fiber}}},
\end{equation}
where $v_{\mathrm{fiber}} \approx 2\times10^{5}$~km/s, giving approximately $500~\mu$s at 100~km, 5~ms at 1000~km, and 25~ms at 5000~km. This is the dominant contribution to $\tau_m$.

\paragraph{Local Pauli corrections.}
Upon receiving the classical bits, Bob applies $X$ and/or $Z$ gates ($\sim 1$--$10$~ns in superconducting qubits~\cite{Barends2014}; $\sim 1$~ns in photonic implementations~\cite{OBrien2007}). These operations are local and distance-independent.

\medskip
\noindent Table~\ref{tab:latency_summary} summarizes the above estimates. The dominant latency arises from classical communication, which scales linearly with distance; all other steps operate within nanosecond time scales.

\begin{table}[h]
\centering
\caption{Latency estimates for each step in quantum teleportation.}
\label{tab:latency_summary}
\resizebox{0.5\textwidth}{!}{
\begin{tabular}{@{}lcc@{}}
\toprule
\textbf{Step} & \textbf{Latency} & \textbf{Distance Dep.} \\
\midrule
State Preparation ($|\varphi\rangle$) & $\sim 10{-}500$ ns & No \\
Bell-Pair Generation & $\sim 100$ ps -- 10 ns & No \\
Bell Measurement & $\sim 10{-}100$ ns & No \\
Classical Communication & $\sim 5\,\mu$s/km & Yes \\
Local Pauli Corrections & $\sim 1{-}10$ ns & No \\
\bottomrule
\end{tabular}}
\end{table}
\subsection{Definition of Fixed Delay Term $T_{\text{fixed}}$}
\label{appendix:T_fixed}

The total memory lifetime requirement in the QRQT protocol can be expressed as
\begin{equation}
\tau_m = 
T_{\mathcal{P}_{\text{Bell}}} +
T_{\mathcal{U}_{\text{Bell}}} +
T_{\mathcal{E}_{\text{PQC}}} +
T_{\mathcal{M}_{\text{Alice}}} +
T_{\text{comm}} +
T_{\mathcal{R}_{\text{Bob}}},
\end{equation}

The term $T_{\text{fixed}}$ used in Eq.~\eqref{eq:max_distance} isolates all local and protocol-dependent delays that do \textbf{not} scale with communication distance:
\begin{equation}
T_{\text{fixed}} =
T_{\mathcal{P}_{\text{Bell}}}+
T_{\mathcal{U}_{\text{Bell}}}+
T_{\mathcal{E}_{\text{PQC}}}+
T_{\mathcal{M}_{\text{Alice}}}+
T_{\mathcal{R}_{\text{Bob}}}.
\end{equation}

This separation allows the communication delay to be modeled as 
$T_{\text{comm}} = d / v_{\text{fiber}}$, leading to the distance–lifetime tradeoff 
$d_{\max} = (T_{\mathrm{coh}} - T_{\text{fixed}}) \cdot v_{\text{fiber}}$ used in the main text.
\subsection{Simulation of Memory Lifetime Constraints for PQC-Protected Teleportation}
\label{appendix: Memory lifetime limit}
To quantitatively illustrate the relationship between communication distance and the minimum quantum-memory coherence time required for successful teleportation under post-quantum cryptographic (PQC) latency, we numerically evaluated the total delay
\begin{equation}
\tau_m = T_{\text{fixed}} + T_{\text{comm}}(d),
\end{equation}
where $T_{\text{fixed}}$ represents distance-independent operations including Bell-pair preparation, measurement, PQC encryption/decryption, and Pauli corrections, and
\begin{equation}
T_{\text{comm}}(d) \approx \frac{d}{v_{\text{fiber}}}, \qquad
v_{\text{fiber}} \simeq 2\times10^{5}~\mathrm{km/s}
\ (\approx 5~\mu\mathrm{s/km}).
\end{equation}
Teleportation remains feasible only if the quantum-memory coherence time $T_{\mathrm{coh}}$ exceeds the total latency $\tau_m$.

\noindent To produce Fig.~\ref{fig:memory_limit}, benchmark encryption and decryption latencies (in nanoseconds) for representative NIST-standardized PQC algorithms were used as the baseline offset $T_{\text{fixed}}$:
\begin{table}[htbp]
\caption{Baseline computation time (in ns) for representative PQC schemes.}
\label{tab:pqc_memory_time}
\centering
\small
\setlength{\tabcolsep}{6pt}
\renewcommand{\arraystretch}{1.0}
\begin{tabular}{lccc}
\toprule
\textbf{Scheme Family} & \textbf{Level 1} & \textbf{Level 3} & \textbf{Level 5} \\
\midrule
Kyber (NIST KEM)  & 412 & 432 & 449 \\
BIKE (Code-based) & 2002 & 3932 & 7885 \\
FrodoKEM (LWE)    & 3421 & 5260 & 7955 \\
\bottomrule
\end{tabular}
\end{table}

\textit{Note:} 
Each security level corresponds to the standard NIST parameter sets used in post-quantum KEMs. 
Specifically, Level~1, Level~3, and Level~5 refer respectively to parameter configurations such as 
Kyber512, Kyber768, and Kyber1024 in lattice-based schemes, or BIKE-L1/L3/L5 and FrodoKEM-640/976/1344 
for code-based and LWE-based constructions. 
These mappings reflect equivalent target security levels of approximately 128-bit, 192-bit, and 256-bit 
classical strength.
Each curve represents the minimum memory lifetime requirement
\[
\tau_m(d) = T_{\text{fixed}} + 5d~\text{(ns)},
\]
where the slope of $5~\text{ns/m}$ corresponds to a propagation delay of $5~\mu\text{s/km}$ in optical fiber.
\begin{table}[htbp]
\centering
\caption{Reported quantum memory coherence times for representative physical platforms.}
\label{tab:coh_times}
\begin{tabular}{lcc}
\toprule
\textbf{System} & \textbf{Reported $T_{\mathrm{coh}}$} & \textbf{Reference}\\
\midrule
NV-center & $\sim$1~s ($10^9$~ns) & \cite{BarGill2013}\\
Trapped ion & $\sim$1~h ($3.6\times10^{12}$~ns) & \cite{wang2021single}\\
Fluxonium qubit & $\sim$1.4~ms ($1.4\times10^6$~ns) & \cite{somoroff2023millisecond}\\
Superconducting cavity & $\sim$34~ms ($3.4\times10^7$~ns) & \cite{milul2023superconducting}\\
\bottomrule
\end{tabular}
\end{table}

\section{Estimation of PQC Execution Time}
\label{appendix:JEDI}
To estimate the classical processing delay introduced by post-quantum cryptography (PQC) in the QRQT protocol, 
we approximate the encryption and decapsulation runtimes using the number of integer operations required 
by each PQC primitive and the peak floating-point throughput ($R_{\mathrm{max}}$) of a modern supercomputer. 
Since cryptographic primitives are dominated by integer arithmetic (modular multiplication, polynomial operations) 
rather than floating-point operations, and no standardized integer-throughput benchmark exists for HPC systems, 
we adopt $R_{\mathrm{max}}$ as a proxy; this yields an optimistic (i.e., lower-bound) estimate of the actual latency.
The estimation is based on the high-performance computing system \textbf{JEDI BullSequana XH3000}, 
equipped with NVIDIA GH200 Grace Hopper Superchips and Quad-Rail InfiniBand interconnects, 
ranked \#222 on the November 2024 TOP500 list~\cite{green500_2024_11}. 
The system delivers a sustained performance of $R_{\mathrm{max}} = 4.50~\mathrm{PFlop/s}$ with $19{,}584$ cores 
and an energy efficiency of $72.7~\mathrm{GFlops/W}$.

The decapsulation time is then approximated as:
\begin{equation}
    T_{\mathrm{decap/encap}} \approx 
    \frac{N_{\mathrm{ops}}}{R_{\mathrm{max}}},
    \label{eq:JEDI estimation}
\end{equation}
where $N_{\mathrm{ops}}$ denotes the total number of integer operations required by the PQC and AES128 algorithms.
This estimate assumes ideal parallel execution across all available compute units.
In practice, the actual runtime may be longer due to sequential dependencies and synchronization overheads, 
as certain operations must wait for prior computation events (e.g., key encapsulation preceding decapsulation). 
Hence, the value obtained from Eq.~\eqref{eq:JEDI estimation} provides a theoretical lower bound on PQC computation latency.

\section{PQC-Protected Quantum Teleportation Protocol}
\label{app:pqc protocol}

Alice and Bob first share a maximally entangled Bell state such as
$|\Phi^+\rangle = (|00\rangle + |11\rangle)/\sqrt{2}$, which serves as the quantum resource for teleportation. 
Alice performs a Bell-state measurement on her input qubit 
$|\psi\rangle = \alpha|0\rangle + \beta|1\rangle$ together with her half of the Bell pair, 
projecting the joint system onto one of four Bell-basis states. 
This measurement yields two classical bits $(M_1, M_2) \in \{0,1\}^2$, 
which encode the Pauli corrections required for Bob to recover the original state. 

To secure these correction bits against quantum-capable adversaries, 
Alice employs a hybrid post-quantum cryptographic (PQC) scheme~\cite{Ghinea2023}. 
A random session key $K$ is first encapsulated with a lattice-based key-encapsulation mechanism 
such as CRYSTALS–Kyber, producing 
$C_1 = \mathrm{Enc}_{\mathrm{PQC}}(pk_{\text{Bob}}, K)$. 
The measurement results are then symmetrically encrypted with $K$ using AES, 
yielding $C_2 = \mathrm{Enc}_{\mathrm{AES}}(K, (M_1, M_2))$. 
Alice transmits the ciphertext pair $(C_1, C_2)$ through the PQC-secured classical channel. 
Upon reception, Bob decapsulates $C_1$ to recover $K$, decrypts $C_2$ to obtain $(M_1, M_2)$, 
and applies the corresponding Pauli corrections $Z^{M_2}X^{M_1}$ to his entangled qubit, 
thereby reconstructing the quantum state $|\psi\rangle$. 
The correctness and fidelity of the teleportation process require that 
all classical operations, including PQC encryption, communication, and decryption, complete within 
the quantum memory lifetime $\tau_m$.

In our estimation of the hybrid cryptographic latency, AES-128 is adopted as the symmetric primitive accompanying the post-quantum scheme. However, since the overall classical attack complexity is overwhelmingly dominated by the hardness of the underlying post-quantum cryptographic primitive, the contribution of symmetric encryption is negligible in comparison. Consequently, throughout our security analysis we treat a successful classical-side compromise as equivalent to successfully breaking the PQC scheme, and we focus exclusively on attacks against the post-quantum cryptographic layer.

\section{LWE Attack Model and Analysis}
\label{app:PQC}

\subsection{LWE problem instance}

An LWE instance is given by
\begin{equation}
\mathbf{p} = A^{\top}\mathbf{sk} + \mathbf{e} \pmod q,
\end{equation}
where $A \in \mathbb{Z}_{q}^{n\times m}$ is public, $\mathbf{p}\in\mathbb{Z}_{q}^{m}$ is the public key, $\mathbf{sk}\in\mathbb{Z}_{q}^{n}$ is the secret key, and $\mathbf{e}\in\mathbb{Z}_{q}^{m}$ is an error vector sampled from a discrete Gaussian noise distribution $D_{\mathbb{Z},\sigma_e q}$. Standard assumptions are:
\begin{itemize}
    \item $\mathbf{sk}$ is uniformly random over $\mathbb{Z}_{q}^{n}$.
    \item $A$ is uniformly random over $\mathbb{Z}_{q}^{n\times m}$.
    \item $\mathbf{e}\leftarrow D_{\mathbb{Z},\sigma_e q}$, a discrete Gaussian centered at $0$ with width $\sigma_e q$.
\end{itemize}
Recovering $\mathbf{sk}$ from $(A,\mathbf{p})$ is conjectured to be hard and forms the basis of lattice-based post-quantum cryptography~\cite{Regev2009}.

\subsection{Attack strategy: BKZ + NearestPlanes}

The main practical attack on LWE reduces to two stages~\cite{Lindner2011,Chen2011}:
\begin{enumerate}
    \item Apply the \textbf{BKZ} block lattice reduction algorithm to the lattice basis derived from $(A,\mathbf{p})$, obtaining a shorter and more orthogonal basis.
    \item Use Babai’s \textbf{NearestPlanes} algorithm~\cite{Babai1986} (or enumeration refinements) in the reduced basis to recover the secret vector $\mathbf{sk}$.
\end{enumerate}

The subtlety is that BKZ and NearestPlanes operate in the Euclidean space $\mathbb{R}^m$, while the LWE problem is defined modulo $q$. The attack effectively embeds the finite problem into $\mathbb{R}$ and searches for close lattice points in the Euclidean metric.

\subsection{Parameter Interpretation and Communication Security Assumption}
\label{appendix: LWE parameter}
\paragraph{LWE parameters and key structure.}
In the standard LWE formulation, the matrix \(A\in\mathbb{Z}_q^{m\times n}\) 
consists of \(m\) public sample rows, each corresponding to a noisy linear equation
\(p_i = \langle \mathbf{a}_i, \mathbf{sk}\rangle + e_i \pmod q\).
Here \(n\) denotes the dimension of the secret vector 
\(\mathbf{sk}\in\mathbb{Z}_q^n\), which determines the lattice dimension and hence 
the dominant source of computational hardness.
The parameter \(m\) represents the number of publicly observable LWE samples available 
to an adversary. In standard hardness analyses \(m\ge n\) is often assumed, modeling 
a sample-rich setting.

Accordingly, the LWE-based public and secret keys are defined as
\[
\text{public key: } (A,\,p=A^{\top} \mathbf{sk}+\mathbf{e}\bmod q), \qquad
\text{secret key: } \mathbf{sk}.
\]
The public key thus exposes a set of noisy linear equations parameterized by the secret 
\(\mathbf{sk}\), while the noise vector \(\mathbf{e}\) ensures computational hardness.

\paragraph{Communication-security assumption.}
In our system model, quantum teleportation and the exchange of classical control 
information occur at high speed, whereas solving the underlying LWE instance via 
lattice reduction (e.g., BKZ–Nearest-Planes) requires substantially longer time. 
We therefore adopt a conservative, long-lived-key assumption: the same LWE keying 
material remains in use across many rapid transmissions, allowing an adversary to 
accumulate multiple public LWE samples over time (effectively a large \(m\)). 

Once the LWE secret is eventually recovered, the adversary can decrypt all subsequent 
classical control bits as well as any previously intercepted and stored ciphertexts 
that used the same key. However, recovering quantum states transmitted before the 
LWE compromise would additionally require that these states, or equivalent quantum 
side information have been preserved until the key recovery moment. Such a condition 
imposes significant quantum-memory requirements, which are beyond the scope of this 
study. Our focus is therefore on post-compromise exposure of classical control 
information, rather than on retroactive recovery of past quantum states.

\subsection{NearestPlanes Algorithm and Its Role in LWE Decoding}
\label{appendix:nearestplanes}

The NearestPlanes algorithm, introduced by Gama and Nguyen and used extensively in 
lattice-based cryptanalysis, provides an efficient recursive strategy for 
finding lattice vectors close to a given target point. 
In the context of Learning-With-Errors (LWE) attacks, it forms the core enumeration subroutine 
following BKZ reduction: after obtaining a nearly orthogonal basis $B$, the algorithm 
enumerates lattice points in a narrow parallelepiped centered around the target 
$p = A s + e \pmod q$ in $\mathbb{R}^m$. 
The goal is to recover the lattice vector $v = A s$ whose displacement 
$e = p - v$ is the error term.

\paragraph{Input and output definition.}
The algorithm takes as input:
\begin{itemize}
    \item A lattice basis 
    \[
    B = \{b_1, b_2, \dots, b_k\} \subseteq \mathbb{R}^m,
    \]
    obtained, for instance, from a BKZ-reduced basis of the LWE lattice;
    \item A vector of integer search widths 
    \[
    \mathbf{d} = (d_1, d_2, \dots, d_k) \in (\mathbb{Z}_{>0})^k,
    \]
    where each $d_i$ specifies the number of nearest integer planes considered at depth $i$;
    \item A target vector 
    \[
    p \in \mathbb{R}^m,
    \]
    typically corresponding to the noisy sample to be decoded.
\end{itemize}

\noindent
\textbf{Output:}  
A set of $\prod_{i=1}^{k} d_i$ candidate lattice vectors of the form
\[
v = \sum_{j=1}^{k} c_{j} b_j, \quad c_j \in \mathbb{Z},
\]
which represent points in the lattice $\mathcal{L}(B)$ lying within a bounded 
distance from the projection of $p$ onto $\mathrm{span}(B)$.
Among these, the closest candidate (in Euclidean distance) is the final decoding output.

\paragraph{Recursive formulation.}
The recursive NearestPlanes procedure can be expressed as follows:

\begin{itemize}
    \item If $k=0$, return the zero vector.
    \item Let $v$ be the orthogonal projection of $p$ onto $\mathrm{span}(B)$,
    where $\mathrm{span}(B) = \{\alpha_1 b_1 + \dots + \alpha_k b_k \mid \alpha_i \in \mathbb{R}\}$.
    \item Compute the $d_k$ integers closest to the normalized inner product
    \[
    \frac{\langle \tilde{b}_k, v \rangle}{\langle \tilde{b}_k, \tilde{b}_k \rangle},
    \]
    where $\tilde{b}_k$ is the $k$-th Gram–Schmidt vector.
    \item For each such integer $c_i$, recursively call
    \[
    NearestPlanes(\{b_1, \dots, b_{k-1}\}, (d_1, \dots, d_{k-1}), v - c_i b_k),
    \]
    and add $c_i b_k$ to the returned partial lattice vector.
\end{itemize}

\noindent
The complete output is the union of all recursive branches:
\begin{equation}
\mathcal{S}_k = \bigcup_{i \in [d_k]} \Bigl( c_i b_k + \text{NearestPlanes}(\{b_j\}_{j<k}, \mathbf{d}_{<k}, v - c_i b_k) \Bigr)
\end{equation}
\paragraph{Interpretation.}
Intuitively, the algorithm performs a depth-first enumeration across 
hyperplanes orthogonal to $\tilde{b}_k$, choosing the nearest integer 
projections at each level. 
Each recursion step refines the residual target by subtracting the best multiple of $b_k$,
so that the final set of candidate vectors covers all points 
in a parallelepiped centered around the projected target. 
The correct lattice point $v$ (and hence the true secret $s$) 
appears among the outputs if the cumulative rounding error across all levels 
remains smaller than the Gaussian noise norm.

Under the Geometric Series Assumption (GSA), the lengths 
$\|\tilde b_i\|$ decay geometrically with $i$, which renders the coordinate errors approximately independent.

\subsection{Success attack probability of BKZ-NearestPlanes} 

Suppose $p=v+e$ for some $v\in \mathcal{L}(B)$ and $e$ drawn from a spherical Gaussian of parameter $s$. Then the probability that $v$ is contained in the NearestPlanes output satisfies~\cite{Lindner2011}:
\begin{equation}
Pr[e\in \mathcal{P}_{\frac{1}{2}}(\tilde{B}\cdot diag(d))]=\prod\limits_{i=1}^{m} \operatorname{erf}\!\left(\frac{d_{i}\|\tilde{b}_{i}\|\sqrt{\pi}}{2s}\right)
    \label{eq:nearestplanes-pr}
\end{equation}

When $e$ is drawn from a sufficiently wide discrete Gaussian (width parameter $\geq 6$), this product-form expression provides an extremely accurate approximation to the true success probability. The NearestPlanes enumeration operates in $O(\prod_i d_i)$ time and is typically used in conjunction with BKZ-reduced bases.
Under the Geometric Series Assumption (GSA), the Gram–Schmidt norms 
$\|\tilde b_i\|$ decay geometrically, which makes the error contributions approximately independent. 
This independence justifies the product-form success probability in Eq.~\eqref{eq:nearestplanes-pr},
which is the standard analytical model for estimating LWE decoding success~\cite{Lindner2011,Chen2011}.

\subsection{BKZ reduction and GSA}

BKZ is the standard block reduction algorithm for lattice cryptanalysis. Its effectiveness is characterized by the \emph{root-Hermite factor} $\delta_{\mathrm{root}}$, which controls the geometric decay of Gram–Schmidt lengths. Under the Geometric Series Assumption (GSA)~\cite{Chen2011}:
\begin{equation}
\Vert \tilde{b}_{i}\Vert \;\approx\; \Vert b_{1}\Vert \cdot (\delta_{\mathrm{root}})^{\,-2(i-1)}.
\end{equation}
Attack cost is parameterized by the BKZ block size (or equivalently $\delta_{\mathrm{root}}$). A common runtime model is
\begin{equation}
\log_{2}T_{\mathrm{BKZ}} \;\approx\; \frac{a}{\log_{2}\delta_{\mathrm{root}}} - b,
\end{equation}
where $a,b$ are calibrated to estimator data~\cite{Albrecht2015}. Larger block sizes yield smaller $\delta_{\mathrm{root}}$ but exponentially higher cost.

\subsection{BKZ Runtime and LWE Attack Probability Derivation}
\label{appendix:PLWE_derivation}

Building on the security condition introduced in Eq.~\eqref{classical attack}, 
we now derive the quantitative relationship between the attack runtime 
\(T_{\mathrm{BKZ}}\) and the success probability \(P_{\mathrm{LWE}}\).  
This formulation connects the computational hardness of lattice reduction 
with the physical coherence constraints discussed in the main text, 
thereby defining a unified, time-dependent metric for adversarial feasibility.  
The derivation follows the modeling framework of 
Lindner and Peikert~\cite{Lindner2011}, 
with parameters calibrated according to the empirical runtime relations 
summarized in Appendix~\ref{appendix:BKZ_params}.

\subsubsection{Geometric Series Assumption (GSA)}
\label{appendix:GSA}

The GSA states that in a BKZ-reduced basis 
\(B = \{b_1, b_2, \dots, b_m\}\),
the lengths of the Gram--Schmidt vectors \(\tilde b_i\) 
decay geometrically with \(i\) (as stated in Sec.~E\,6 above):~\cite{Lindner2011}
\begin{equation}
\|\tilde b_i\| = \|\tilde b_1\| \cdot \zeta^{i-1},
\label{eq:gsa_decay}
\end{equation}
where \(\zeta < 1\) is the decay ratio.  

Following Lindner and Peikert~\cite{Lindner2011},
the determinant of the lattice satisfies
$\det(\Lambda) = \prod_{i=1}^{m} \|\tilde{b}_i\|$.
Substituting Eq.~\eqref{eq:gsa_decay} and the Hermite factor
definition $\|b_1\| = \delta_{\mathrm{root}}^m \det(\Lambda)^{1/m}$
gives
\begin{equation}
\prod_{i=1}^{m} \|\tilde{b}_i\|
  = \|b_1\|^{m} \, \zeta^{m(m-1)/2}
  = \delta_{\mathrm{root}}^{m^2} \, \det(\Lambda) \, \zeta^{m(m-1)/2}.
\end{equation}
Equating this with $\det(\Lambda)$ and cancelling yields
\begin{equation}
\zeta = \delta_{\mathrm{root}}^{-2m/(m-1)} \approx \delta_{\mathrm{root}}^{-2}.
\label{eq:zeta_def}
\end{equation}
This approximation (valid for large \(m\)) bridges the decay factor \(\zeta\) 
and the root-Hermite factor \(\delta_{\mathrm{root}}\), 
forming the foundation for the GSA-based norm distribution model 
in BKZ-reduced lattices.

Consequently, each level of the basis satisfies
\begin{equation}
\zeta^{i-1} \approx \delta_{\mathrm{root}}^{-2(i-1)}.
\label{eq:zeta_substitution}
\end{equation}

\paragraph{Empirical BKZ runtime fit.}
The empirical relation between the achieved root-Hermite factor 
and the BKZ runtime (rearranging Eq.~\eqref{eq:bkz_time_prelim} from the main text)~\cite{Lindner2011,Albrecht2015} is modeled as
\begin{equation}
\log_2 \delta_{\mathrm{root}} = \frac{a}{\log_2 T_{\mathrm{BKZ}} + b},
\label{eq:delta_from_T}
\end{equation}
where the coefficients \(a\) and \(b\) depend on computational power 
(see Appendix~\ref{appendix:BKZ_params}).  
This can be rewritten as
\begin{equation}
\delta_{\mathrm{root}} = 2^{\frac{a}{\log_2 T_{\mathrm{BKZ}} + b}},
\label{eq:delta_exponential}
\end{equation}
and thus
\begin{equation}
\zeta^{i-1} = 
2^{-2(i-1)\frac{a}{\log_2 T_{\mathrm{BKZ}} + b}}.
\label{eq:zeta_T_relation}
\end{equation}

\subsubsection{NearestPlanes Success Probability under GSA}
\label{appendix:NearestPlanes}

The NearestPlanes algorithm enumerates lattice candidates in the neighborhood of 
a target vector \(p = v + e\), where \(v \in \mathcal{L}(B)\) and 
\(e\) follows a Gaussian noise distribution with standard deviation \(s\).
The success probability that the correct lattice vector \(v\) is found
in the enumeration output is given by the product-form approximation:
\begin{equation}
P_{\mathrm{LWE}}(\{\tilde b_i\},\mathbf{d},s)
\;=\;
\prod_{i=1}^{m}
\operatorname{erf}\!\left(
\frac{d_i\,\|\tilde b_i\|\,\sqrt{\pi}}{2s}
\right),
\end{equation}
where \(\mathbf{d} = (d_1,\dots,d_m)\) denotes the integer search widths 
(number of nearest integer planes per level).  
Substituting Eqs.~\eqref{eq:gsa_decay}--\eqref{eq:zeta_T_relation} into the product-form probability gives
\begin{equation}
P_{\mathrm{LWE}} =
\prod_{i=1}^{m}
\operatorname{erf}\!\left(
\frac{d_i\,\|\tilde b_1\|\,\sqrt{\pi}}{2s}
\cdot
2^{-2(i-1)\frac{a}{\log_2 T_{\mathrm{BKZ}} + b}}
\right),
\label{eq:PLWE_final}
\end{equation}
which explicitly links the attack success probability to the BKZ runtime 
and thereby to computational capability.

\paragraph{Practical interpretation.}
Equation~\eqref{eq:PLWE_final} shows that the success probability increases 
as the achievable reduction quality (i.e., smaller \(\delta_{\mathrm{root}}\)) improves, 
or equivalently, as the runtime \(T_{\mathrm{BKZ}}\) and computing power increase. 
In most practical lattice attacks, the search widths are chosen as 
\(d_i = 1\) for nearly all dimensions, representing enumeration along a single
nearest plane per level.  
Only in high-dimensional or low-noise regimes are a few deeper layers extended to 
\(d_i \in \{2,3,4\}\) to improve coverage, since the total enumeration cost scales 
as \(\prod_i d_i\).  
This tradeoff defines the practical boundary between feasible and infeasible 
LWE instance attacks under realistic hardware constraints.

\subsection{Empirical BKZ Runtime Fit from Lindner and Peikert (2010)}
\label{appendix:BKZ_params}

This appendix details the empirical calibration of the BKZ runtime model adopted in Eq.~\eqref{eq:bkz_time_prelim}. 
Lindner and Peikert~\cite{Lindner2011} performed systematic BKZ lattice-reduction experiments on random $q$-ary lattices with parameters 
$n=72$, $q=1021$, and $m=\sqrt{n\log(q)/\log(\delta_{\mathrm{root}})}$. 
Their experiments were executed on a 2.3~GHz AMD Opteron processor 
(comparable in performance to an Intel Xeon server core) 
and measured the runtime (in seconds)
required to obtain a target root-Hermite factor~$\delta_{\mathrm{root}}$. 
A least-squares regression yielded the empirical relation~\cite{Lindner2011}:
\[
\log_2 T_{\mathrm{BKZ}}(\delta_{\mathrm{root}}) = \frac{1.806}{\log_{2}(\delta_{\mathrm{root}})} - 91,
\]
and a conservative lower-bound estimate~\cite{Lindner2011}:
\[
\log_2 T_{\mathrm{BKZ}}(\delta_{\mathrm{root}}) = \frac{1.8}{\log_{2}(\delta_{\mathrm{root}})} - 110.
\]
These values correspond to measurements on commodity Intel-class hardware 
and have since become the de facto reference for estimating the cost of BKZ-based lattice reduction. 

In summary, the BKZ runtime model takes the form
\[
\log_2 T_{\mathrm{BKZ}} = \frac{a}{\log_2 \delta_{\mathrm{root}}} - b,
\]
where $\delta_{\mathrm{root}}$ denotes the achieved root-Hermite factor after reduction. 
The coefficients $a$ and $b$ jointly capture the effective computational capability of the attacker:
smaller~$a$ indicates steeper efficiency scaling (stronger algorithmic or hardware performance), 
while larger~$b$ implies higher baseline throughput. 
Hence, $a$ and $b$ together form an abstract metric of attack power. 

For reference, the 2.3~GHz baseline platform corresponds to $(a,b)=(1.8,110)$,
which we adopt as the conservative default in our simulations. 
Stronger adversaries can be modeled by decreasing $a$ or increasing $b$. 
Representative settings are summarized in Table~\ref{tab:bkz_params}.

\begin{table}[htbp]
\caption{Representative calibration of BKZ parameters $(a,b)$ under different hardware assumptions.}
\label{tab:bkz_params}
\centering
\scriptsize
\setlength{\tabcolsep}{3pt}
\begin{tabular}{@{}lcrl@{}}
\toprule
\textbf{Hardware} & $a$ & $b$ & \textbf{Description} \\
\midrule
CPU (2.3\,GHz) & 1.8 & 110 & Ref.\ fit~\cite{Lindner2011} \\
GPU cluster     & 1.6 & 115 & Moderate parallelism \\
HPC / quantum   & 1.4 & 120 & Aggressive assumption \\
\bottomrule
\end{tabular}
\end{table}

This calibration enables a quantitative mapping from BKZ runtime 
to physical computational resources, 
forming the bridge between classical cryptanalytic effort 
and the coherence-time-limited quantum teleportation window in our joint threat model.

\section{Complete Derivation: Relationship Between $n$, $\|b_1\|$, and $m$}
\label{appendix: m,n,b relationship}
\subsection{Foundation from Paper}

We use the following established relationships from the paper:

\begin{align}
\|\tilde{b}_i\| &= \|b_1\| \cdot \zeta^{i-1} && \text{(GSA, Eq.~\ref{eq:gsa_decay})} \nonumber\\
\zeta &= \delta^{-2m/(m-1)} \approx \delta^{-2} && \text{(Eq.~\ref{eq:zeta_def}, large $m$)} \nonumber\\
\|b_1\| &= \delta^m \cdot \det(\Lambda)^{1/m} && \text{(Hermite factor~\cite{Schnorr1994})}\label{eq:hermite}\\
\det(\Lambda) &= q^n && \text{($q$-ary lattice~\cite{regev2006lattice})}\label{eq:det}\\
m &= \sqrt{\frac{n \lg(q)}{\lg(\delta)}} && \text{(Optimal subdim.~\cite{Lindner2011})}\label{eq:m-opt}
\end{align}

Equation~\eqref{eq:det} is derived from standard q-ary LWE embedding construct.
The Hermite factor $\delta$ in Eq.~\eqref{eq:hermite} characterizes the quality of the reduced basis~\cite{Lindner2011,Regev2009}.

Equation~\eqref{eq:m-opt} represents the optimal embedding dimension, achieving a balance between basis reduction and the cost of bounded-distance decoding~\cite{Regev2009,Lindner2011}. Inverting this equation yields $\delta \approx q^{n/m^2}$; given $n, q, m$, the corresponding Hermite factor is automatically determined.

\subsection{Step-by-Step Derivation}

\textbf{Step 1:} Express $\|\tilde{b}_m\|$ using GSA
\begin{equation*}
\|\tilde{b}_m\| = \|b_1\| \cdot \zeta^{m-1}
\end{equation*}

\textbf{Step 2:} Substitute $\zeta \approx \delta^{-2}$
\begin{equation*}
\|\tilde{b}_m\| = \|b_1\| \cdot (\delta^{-2})^{m-1} = \|b_1\| \cdot \delta^{-2(m-1)}
\end{equation*}

\textbf{Step 3:} Substitute Hermite factor definition
\begin{align*}
\|b_1\| &= \delta^m \cdot (q^n)^{1/m} = \delta^m \cdot q^{n/m}\\
\|\tilde{b}_m\| &= \delta^m \cdot q^{n/m} \cdot \delta^{-2(m-1)}\\
&= \delta^{m-2(m-1)} \cdot q^{n/m}\\
&= \delta^{-m+2} \cdot q^{n/m}
\end{align*}

\textbf{Step 4:} Use optimal dimension relationship to eliminate $\delta$
From $m = \sqrt{\frac{n \lg(q)}{\lg(\delta)}}$:
\begin{align*}
m^2 &= \frac{n \lg(q)}{\lg(\delta)}\\
\lg(\delta) &= \frac{n \lg(q)}{m^2}\\
\delta &= q^{n/m^2}
\end{align*}

\textbf{Step 5:} Substitute back into $\|\tilde{b}_m\|$
\begin{align*}
\|\tilde{b}_m\| &= (q^{n/m^2})^{-m+2} \cdot q^{n/m}\\
&= q^{-n(m-2)/m^2} \cdot q^{n/m}\\
&= q^{n\left(\frac{1}{m} - \frac{m-2}{m^2}\right)}\\
&= q^{n\left(\frac{m-(m-2)}{m^2}\right)}\\
&= q^{2n/m^2}
\end{align*}

\textbf{Step 6:} Express $\|b_1\|$ in terms of $n$ and $m$
\begin{align*}
\|b_1\| &= \delta^m \cdot q^{n/m}\\
&= (q^{n/m^2})^m \cdot q^{n/m}\\
&= q^{n/m} \cdot q^{n/m}\\
&= q^{2n/m}
\end{align*}
The above equations demonstrate relationships between variables under our assumptions; note that $\|b_1\|$ is not an independent free parameter but is determined by $n$, $m$, and $q$.
\subsection{Final Results}

\textbf{Primary relationship:}
\begin{equation*}
\|b_1\| = q^{2n/m}
\end{equation*}

\textbf{Solving for $n$:}
\begin{equation*}
n = \frac{m \cdot \lg(\|b_1\|)}{2 \cdot \lg(q)}
\end{equation*}

\textbf{Last Gram-Schmidt vector:}
\begin{equation*}
\|\tilde{b}_m\| = q^{2n/m^2}
\end{equation*}

\subsection{Practical Considerations}
The relationship $\|b_1\| = q^{2n/m}$ and its inverse 
$n = \frac{m \cdot \lg\|b_1\|}{2\lg q}$ are exact 
consequences of the GSA and optimal embedding 
dimension assumptions. In practice, BKZ with finite 
block size $\beta$ achieves a root-Hermite factor 
$\delta_{\mathrm{ach}} > \delta_{\mathrm{opt}}$, 
which invalidates the substitution 
$\delta = q^{n/m^2}$ used in Step~4. As a result, 
the actual first basis vector satisfies 
$\|b_1\|_{\mathrm{ach}} = \delta_{\mathrm{ach}}^m 
\cdot q^{n/m} > q^{2n/m}$, and the simple 
closed-form relation between $n$ and $\|b_1\|$ 
no longer holds exactly. 

The expressions derived above should therefore be 
understood as idealized analytical relationships 
illustrating how lattice parameters scale with one 
another, rather than as operational formulas for 
concrete security estimation. For the latter purpose, 
dedicated tools such as the LWE 
Estimator~\cite{Albrecht2015} and the BKZ 
simulator of Chen and Nguyen~\cite{Chen2011}
should be used, as they incorporate empirical 
corrections for finite block sizes and 
non-asymptotic effects.

\subsection{Padé Approximation for the Error Function}
\label{appendix:pade_approximation}

To accelerate the computation of Eq.~\eqref{eq:lwe_probability}, we use 
Winitzki’s Padé-type rational approximation~\cite{winitzki2008handy}, which provides
a globally accurate closed-form for the error function:
\[
\operatorname{erf}(z) 
\approx 
\sqrt{1 - \exp\!\left(-\frac{z^2(4/\pi + w z^2)}{1 + w z^2}\right)},
\]
where $w \approx 0.140012$. 
This approximation achieves a maximum absolute error of less than 
$3.5\times10^{-4}$ for all real~$z$, and has been widely adopted in analytic 
approximations for Gaussian cumulative functions. 

Substituting this form into Eq.~\eqref{eq:lwe_probability} and taking the base-2 
logarithm yields the closed-form log-probability:
\begin{equation}
\log_2 P_{\mathrm{LWE}} \approx
\sum_{i=1}^m
\frac{1}{2}
\log_2\!\left(
1 -
\exp\!\left(
-\frac{W_i^{2}\!\left(\frac{4}{\pi} + w W_i^{2}\right)}{1 + w W_i^{2}}
\right)
\right),
\label{eq:pade_log_prob}
\end{equation}
where
\[
W_i = d_i \frac{\|\tilde b_1\| \sqrt{\pi}}{2s} 2^{- 2 (i-1)\frac{a}{\log_2 T_{\mathrm{BKZ}} + b}}.
\]
This formulation enables 
efficient evaluation of $P_{\mathrm{LWE}}$ with negligible numerical loss 
while maintaining analytic differentiability with respect to 
lattice and attack parameters $(a,b,s,d_i)$.

\subsubsection{Padé Approximation Validation}

To validate the efficiency of our Padé approximation of $\log_2 P_{\mathrm{LWE}}$, 
we compare it with exact values across a range of lattice dimensions.

\begin{figure}[tb]
\centering
\includegraphics[width=\columnwidth]{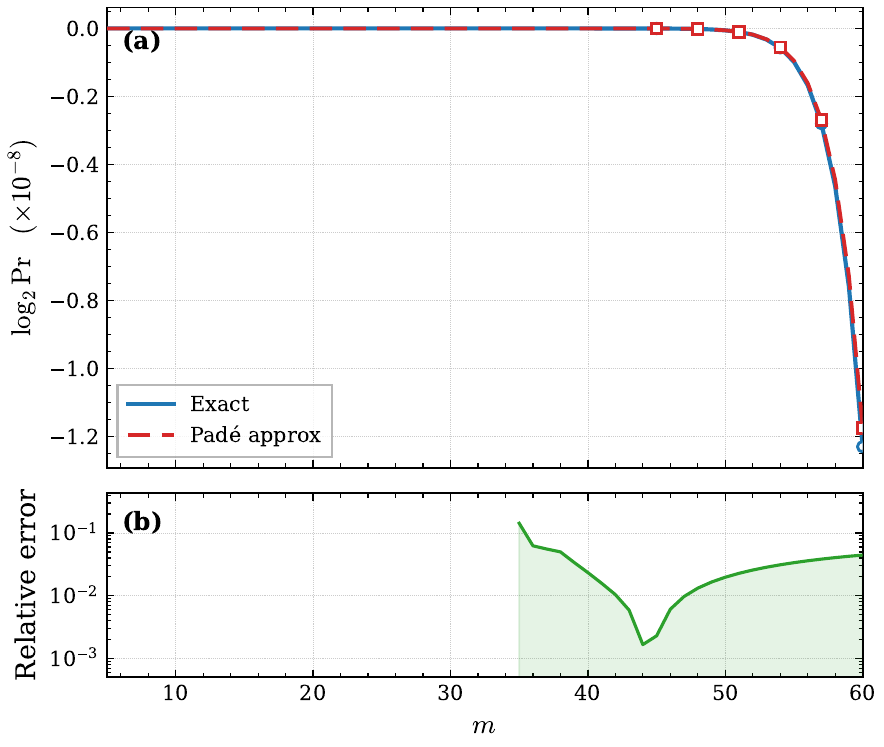}
\caption{Comparison of exact and Padé-approximated $\log_2 P_{\mathrm{LWE}}$ versus lattice dimension $m$. The approximation achieves $<10^{-4}$ relative error while reducing computation time by orders of magnitude. Parameters: $a = 1.8$, $b = 2.7$, $T=2^{30}$~s, $\|b_{1}\|=10$, $d_{i}=2$, $s=2$.}
\label{fig:pade_comparison}
\end{figure}

\begin{figure}[htb!]
\centering
\includegraphics[width=\columnwidth]{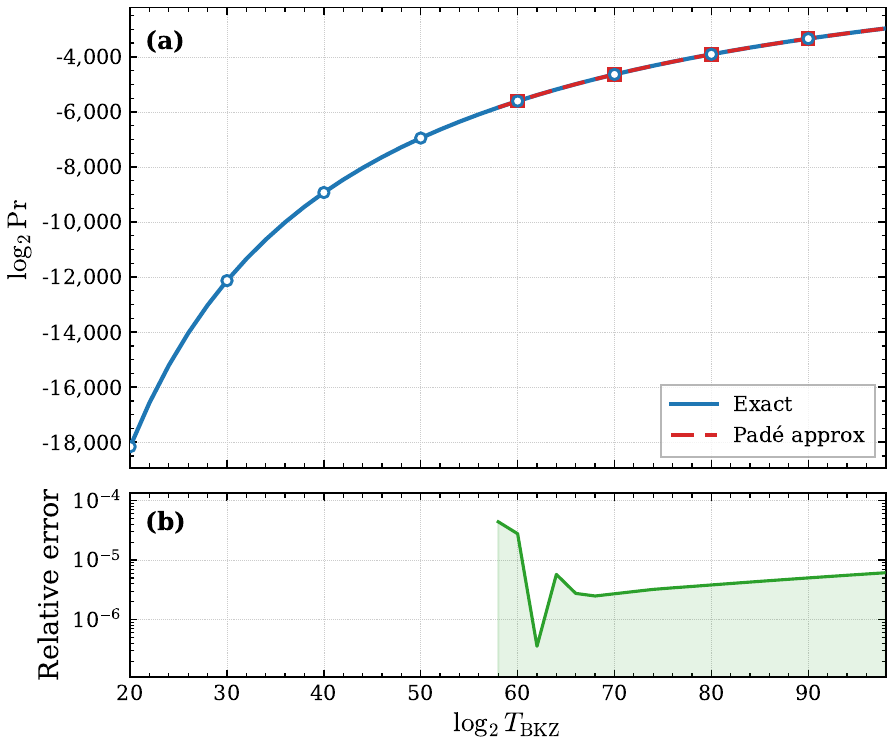}
\caption{Attack success probability versus BKZ computation time for various PQC algorithms. Quantum memory constraints limit attack time, directly impacting achievable security levels. Parameters: $a = 0.3$, $b = 2.7$, $m=30$, $\|b_{1}\|=10$, $d_{i}=1$, $s=2$.}
\label{fig:attack_time}
\end{figure}

Figure~\ref{fig:attack_time} shows attack success probability versus available BKZ time for different PQC parameter sets.

\section{Holevo Information with an Amplitude-Damped Bell Pair}
\label{app:holevo-derivation}
We analyze the the ensemble induced by the adversary's (Eve's) partial knowledge  who learns a subset
$K \subseteq \{1,2\}$ of the classical teleportation correction bits
$(m_1,m_2)\in\{0,1\}^2$. For each leakage pattern, we compute the corresponding Holevo quantity \(\chi_K\), which measures the residual information about the teleported state that remains hidden from Eve because of uncertainty in the unrevealed correction bits. We assume that Eve has access to Bob's quantum system and attempts to infer the teleported state using the leaked classical information.


Throughout this appendix, Bob’s conditional pre-correction state corresponding
to the Bell measurement outcome $(m_1,m_2)$ is denoted by $\rho_{m_1m_2}$. All four
conditional states are related by Pauli conjugation of a reference state
$\rho_{00}$. Since Pauli conjugation is unitary, it preserves the spectrum of the state, and hence all four conditional states have the same von Neumann entropy. 

The four cases below correspond to Eve learning none of the correction bits, only \(m_1\), only \(m_2\), or both bits.

\subsubsection*{Case A: No Keys Leaked ($K=\varnothing$)}

If neither correction bit is leaked, Eve must average uniformly over all four
conditional states. The ensemble-averaged state is,
\begin{equation}
    \rho_{\mathrm{avg}|\varnothing}
=
\frac{1}{4}\sum_{m_1,m_2}\rho_{m_1m_2}
=
\frac{I}{2}.
\end{equation}
The entropy of the averaged state is therefore
$S(\rho_{\mathrm{avg}|\varnothing})=1$.
Since each conditional state $\rho_{m_1m_2}$ has entropy $h(\delta)$, the Holevo
quantity in this case is
\[
\chi_{\varnothing}
=
1-h(\delta).
\]

Here \(h(x)\) denotes the binary entropy associated with eigenvalues \(\tfrac{1}{2}(1\pm x)\), namely
\[
h(x)
=
-\frac{1+x}{2}\log_2\!\frac{1+x}{2}
-\frac{1-x}{2}\log_2\!\frac{1-x}{2}.
\]

\subsubsection*{Case B: $X$ Key Leaked Only ($K=\{1\}$)}

If Eve learns the $X$-correction bit $m_1$ but not the $Z$-correction bit $m_2$,
she averages over the two states that differ by a $Z$ conjugation. Writing
\[
\rho_{00}
=
\begin{pmatrix}
a & c\\
c^* & b
\end{pmatrix},
\qquad
\rho_{01}
=
\begin{pmatrix}
a & -c\\
-c^* & b
\end{pmatrix},
\]
the averaged state for fixed $m_1=0$ is
\[
\rho_{\mathrm{avg}|m_1=0}
=
\frac{1}{2}(\rho_{00}+\rho_{01})
=
\operatorname{diag}(a,b).
\]
Similarly, for $m_1=1$, averaging over $\rho_{10}$ and $\rho_{11}$ yields
$\operatorname{diag}(b,a)$. In both cases, the eigenvalue gap of the averaged
state is $|1-2a|$, and its entropy is therefore $h(|1-2a|)$. Subtracting the
entropy $h(\delta)$ of the individual conditional states gives
\[
\chi_{\{1\}}
=
h(|1-2a|)-h(\delta).
\]

\subsubsection*{Case C: $Z$ Key Leaked Only ($K=\{2\}$)}

If Eve learns the $Z$-correction bit $m_2$ but not $m_1$, she averages over states
that differ by an $X$ conjugation. For fixed $m_2=0$,
\[
\rho_{00}
=
\begin{pmatrix}
a & c\\
c^* & b
\end{pmatrix},
\qquad
\rho_{10}
=
\begin{pmatrix}
b & c^*\\
c & a
\end{pmatrix},
\]
and the averaged state is
\[
\rho_{\mathrm{avg}|m_2=0}
=
\frac{1}{2}(\rho_{00}+\rho_{10})
=
\begin{pmatrix}
\frac{a+b}{2} & \Re(c)\\
\Re(c) & \frac{a+b}{2}
\end{pmatrix}.
\]
Since $a+b=1$, this state has eigenvalue gap
\[
\Delta_2 = 2|\Re(c)|.
\]
The same spectrum is obtained for $m_2=1$. The entropy of the averaged state is
therefore $h(2|\Re(c)|)$, and the corresponding Holevo quantity is
\[
\chi_{\{2\}}
=
h(2|\Re(c)|)-h(\delta).
\]

\subsubsection*{Case D: Both Keys Leaked ($K=\{1,2\}$)}


If both correction bits are leaked, the ensemble collapses to a single known conditional state \(\rho_{m_1m_2}\). No uncertainty associated with the correction-bit ensemble remains, and therefore
\[
\chi_{\{1,2\}}=0.
\]

\medskip
Summarizing the above results, the Holevo quantities associated with the four
leakage patterns are
\begin{align}
\chi_{\varnothing} &= 1-h(\delta), \qquad
\chi_{\{1\}} = h(|1-2a|)-h(\delta), \notag\\
\chi_{\{2\}} &= h(2|\Re(c)|)-h(\delta), \qquad
\chi_{\{1,2\}} = 0 .
\end{align}

\section{ Derivation of Eve’s Optimal Fidelity}
\label{app:fidelity}

In this appendix, we derive Eve’s optimal average fidelity for each possible
subset \(K \subseteq \{1,2\}\) of leaked Pauli correction keys in the teleportation
protocol. Eve’s task is to infer Bob’s post-measurement quantum state based on
partial or complete knowledge of the classical correction bits.

Let Bob’s conditional pre-correction state be
\begin{equation}
\rho_{00} =
\begin{pmatrix}
a & c \\
c^{*} & b
\end{pmatrix},
\end{equation}
with auxiliary parameters
\begin{equation}
r := \Re(c),
\qquad
\delta := \sqrt{1 - 4(ab - |c|^{2})}.
\end{equation}
The parameter \(\delta\) determines the eigenvalue spectrum of \(\rho_{00}\),
whose eigenvalues are \(\tfrac{1}{2}(1 \pm \delta)\).

For a given knowledge class \(K\), Eve observes an ensemble
\(\{p_i, \rho_i\}\) of possible quantum states and selects a pure state
\(\ket{\sigma}\) that maximizes the average overlap with the ensemble. Her
optimal single-shot fidelity is therefore given by
\begin{equation}
F_K
=
\max_{\ket{\sigma}}
\sum_i p_i \bra{\sigma} \rho_i \ket{\sigma},
\end{equation}
which is achieved by projecting onto the eigenvector corresponding to the
largest eigenvalue of the effective averaged state available to her.

\vspace{0.5em}

Table~\ref{tab:fidelity-summary} summarizes the effective states observed by Eve,
their eigenvalue spectra, and the resulting optimal fidelities for each
knowledge class.

\begin{table}[h]
\centering
\renewcommand{\arraystretch}{1.3}
\begin{tabular}{@{}c@{\quad}c@{\quad}c@{\quad}c@{}}
\toprule
\textbf{Key Set \(K\)} &
\textbf{Effective State \(\rho_{\mathrm{avg}|K}\)} &
\textbf{Eigenvalues} &
\textbf{Fidelity \(F_K\)} \\
\midrule
\(\varnothing\) &
\(\tfrac{1}{2} I\) &
\(\tfrac{1}{2}, \tfrac{1}{2}\) &
\(\tfrac{1}{2}\) \\

\(\{1\}\) &
\(\begin{pmatrix} a & 0 \\ 0 & b \end{pmatrix}\) &
\(a, b\) &
\(\tfrac{1}{2}(1 + |1 - 2a|)\) \\

\(\{2\}\) &
\(\begin{pmatrix} \tfrac{1}{2} & r \\ r & \tfrac{1}{2} \end{pmatrix}\) &
\(\tfrac{1}{2} \pm r\) &
\(\tfrac{1}{2} + |r|\) \\

\(\{1,2\}\) &
\(\rho_{m_1 m_2}\) &
\(\tfrac{1}{2}(1 \pm \delta)\) &
\(\tfrac{1}{2}(1 + \delta)\) \\
\bottomrule
\end{tabular}
\caption{Eve’s optimal fidelity for each subset \(K\) of known Pauli correction keys.}
\label{tab:fidelity-summary}
\end{table}

\vspace{0.5em}

When no correction keys are known (\(K=\varnothing\)), Eve observes the maximally
mixed state, yielding a fidelity of \(1/2\), equivalent to random guessing.
Knowledge of only the \(X\)-correction key removes phase coherence and produces a
diagonal effective state, while knowledge of only the \(Z\)-correction key
preserves coherence but symmetrizes the population distribution. In both cases,
Eve’s optimal fidelity is determined by the largest eigenvalue of the
corresponding averaged state. When both correction keys are known, Eve can
identify the exact conditional state, and her fidelity is limited only by the
intrinsic mixedness induced by noise, as quantified by \(\delta\).

This analysis provides a direct operational characterization of partial
classical key leakage, complementing the Holevo information results by
quantifying Eve’s best achievable state reconstruction performance.

\end{document}